\documentclass[a4paper,11pt]{article}
\pdfoutput=1
\usepackage{jheppub}
\usepackage[T1]{fontenc}
\usepackage{amssymb}
\usepackage{amsmath}
\usepackage{graphicx}
\usepackage{subcaption}
\usepackage[colorlinks=true]{hyperref}
\usepackage{appendix}
\usepackage{amsbsy}
\allowdisplaybreaks
\DeclareMathAlphabet\mathbfcal{OMS}{cmsy}{b}{n}
\usepackage{booktabs}
\usepackage{multirow}
\usepackage{tabularx}

\def\beq{\begin{equation}}   
\def\eeq{\end{equation}}
\def\bea{\begin{eqnarray}}  
\def\eea{\end{eqnarray}} 
\def\nn{\nonumber}

\def\doubletilde#1{\widetilde{\vphantom{\raise 1.5pt \hbox{#1}}\smash{\kern -2pt\widetilde{#1}}}}

\def\CA{C_A}
\def\CF{C_F}
\def\TF{T_F}

\def\Re{\mbox{Re}}

\def\glu{{\tilde{g}}}
\def\NG{{\color{blue} N_{\glu}}}
\def\NGSquare{{\color{blue} N_{\glu}^2}}
\def\NGCube{{\color{blue} N_{\glu}^3}}

\def\NF{{\color{magenta} N_F}}
\def\NFSquare{{\color{magenta} N_F^2}}
\def\NFCube{{\color{magenta} N_F^3}}

\def\e{\epsilon}
\def\d{\hbox{d}}

\def\nn{\nonumber}

\def\CA{C_A}
\def\CF{C_F}
\def\TF{T_F}

\def\e{\epsilon}
\def\d{\hbox{d}}

\def\nn{\nonumber}

\newcommand{\code}[1]{\textsc{#1}}

\newcommand{\Zren}[1]{Z_\eta^{(#1)}}

\newcommand{\T}[2]{{\cal T}^{(#1)}_{#2}}

\preprint{{\raggedleft%
ZU-TH 64/23, IPPP/23/57
}}

\title{Radiation from a gluon-gluino colour-singlet dipole at N$^3$LO}
\author[a,b]{Xuan Chen,}
\author[b]{Petr Jakub\v{c}\'{i}k,}
\author[b,c]{Matteo Marcoli}
\author[b,d]{and Giovanni Stagnitto}

\affiliation[a]{School of Physics, Shandong University,
            Jinan, Shandong 250100, China}
\affiliation[b]{Physik-Institut, Universit\"at Z\"urich,
  Winterthurerstrasse 190, CH-8057 Z\"urich, Switzerland}
\affiliation[c]{Institute for Particle Physics Phenomenology, Department of Physics, Durham University, Durham, DH1 3LE, UK}
\affiliation[d]{Universit\`{a} degli Studi di Milano-Bicocca \& INFN, Piazza della Scienza 3, Milano 20126, Italy }
\emailAdd{xuan.chen@sdu.edu.cn}
\emailAdd{petr.jakubcik@physik.uzh.ch}
\emailAdd{matteo.marcoli@durham.ac.uk}
\emailAdd{giovanni.stagnitto@unimib.it
}

\abstract{
We compute the quantum chromodynamics (QCD) corrections to the decay of a neutralino to gluinos and partons at next-to-next-to-next-to-leading order (N$^3$LO) in the strong coupling constant $\alpha_s$, integrated separately over the phase-space of two, three, four or five particles in the final state. The resulting matrix elements are related to the quark-gluon antenna functions, completing the set of integrated antenna functions in final-final kinematics required for the extension of the antenna subtraction scheme to N$^3$LO.
For a model with massless partons (quarks and gluons) and an arbitrary number of adjoint fermions, we obtained the following new results to $\mathcal{O}(\alpha_s^3)$: the inclusive cross section for the decay of a neutralino, the renormalization of the effective coupling of a neutralino to a gluino and gluon(s), the gluino collinear anomalous dimension, the gluino contribution to the parton collinear anomalous dimensions and the neutralino-gluino-gluon(s) vertex form factors. A special case of this model is the $\mathcal{N}=1$ super Yang-Mills theory, where we can relate some of our findings to known results.
}

\begin{document} 
\maketitle
\flushbottom

%%%%%%%%%%%%%%%%%%%%%%%%%%%%%%%%%%%%%%%%%%%%%%%%%%%%%%%%%%%%%%%%%%%%%%%%%%%%%%%%

\section{Introduction}
Precise fixed-order theory predictions in quantum chromodynamics (QCD) play an essential role in the physics programme of the Large Hadron Collider.
The current state of the art for most Standard Model cross sections is
next-to-next-to-leading order (NNLO) in QCD, both for fiducial cross sections and differential distributions. 
In the coming years, particularly during the scheduled high-luminosity phase of the LHC, measurements for many observables will approach  percent-level precision. This will require the extension of the technology for higher-order calculations to next-to-next-to-next-to-leading order (N$^3$LO). Several remarkable steps towards N$^3$LO phenomenology have recently been made, see~\cite{Caola:2022ayt} for a summary.

A prescription for handling the infrared (IR) divergences of matrix elements at a fixed perturbative order, such as a subtraction or slicing method, is one of the main 
missing process-independent ingredients.
The development of a local subtraction method
requires an understanding of the universal
behaviour of N$^3$LO matrix elements in unresolved configurations,
including single-unresolved limits of two-loop
amplitudes, double-unresolved limits of one-loop amplitudes, and
triple-unresolved limits of tree-level amplitudes. Some unresolved limits can be
described with the iteration of single and double unresolved structures, while
others require novel computations, e.g.\ single collinear limits of two-loop
amplitudes~\cite{Badger:2004uk,Duhr:2014nda}; the two-loop current for the emission
of a soft gluon~\cite{Duhr:2013msa,Li:2013lsa,Dixon:2019lnw}; triple collinear
limits of one-loop amplitudes~\cite{Catani:2003vu,Czakon:2022fqi}; double soft
emission at one loop~\cite{Catani:2021kcy,Zhu:2020ftr,Czakon:2022dwk}; quadruple
collinear splitting functions~\cite{DelDuca:2019ggv,DelDuca:2020vst}; and triple
soft emission in tree-level
amplitudes~\cite{Catani:2019nqv,DelDuca:2022noh,Catani:2022hkb}.
At the integrated level, the interplay of the different unresolved limits is not yet understood.

Several subtraction schemes have been proposed in the last two decades for NNLO calculations in QCD. In the antenna subtraction method~\cite{Gehrmann-DeRidder:2005btv,Currie:2013vh},
the matrix elements for several $1\to 2$ processes (a photon decaying into a pair of
quarks~\cite{Gehrmann-DeRidder:2004ttg},
a Higgs boson decaying into a pair of gluons~\cite{Gehrmann-DeRidder:2005alt} and a neutralino decaying
into a gluino-gluon pair~\cite{Gehrmann-DeRidder:2005svg}) are
used to define \textit{antenna functions}, building blocks for subtraction terms 
which remove the singular behaviour of matrix elements in the presence of unresolved radiation between a
pair of hard emitters. On the other hand, the integrals of these
matrix elements over the phase space are used to remove the explicit
singularities present in virtual corrections. A future extension of antenna subtraction to N$^3$LO relies on the computation of these ingredients. In~\cite{Jakubcik:2022zdi}, we presented analytic results for the phase-space integration of the corrections to the $\gamma^*\to q\bar{q}$ decay up to N$^3$LO. We also performed an analogous calculation for the $H\to gg$ decay in the large top mass limit~\cite{Chen:2023fba}. These results represent the integrated quark-antiquark and gluon-gluon antenna functions in final-final kinematics. 

In this work, we follow the strategy of~\cite{Gehrmann-DeRidder:2005svg} to extract the quark-gluon antenna functions from the decay of a neutralino into a gluino and a gluon. 
In other words, we
analytically integrate the tree-level five-particle, the one-loop four-particle, the
two-loop three-particle, and the three-loop two-particle matrix elements over the
respective phase space,
\begin{equation}\label{eqn:matelems}
  \sigma^{(3)} = \int \d\Phi_5\,M_{5}^0 + \int \d\Phi_4\,M_{4}^1 + \int
  \d\Phi_3\,M_{3}^2 + \int \d\Phi_2\,M_{2}^3\,,
\end{equation}
where $M_{n}^{\ell}$ denotes the $\ell$-loop matrix element for the decay of a neutralino
into $n$ partons and gluinos. We refer to the four terms
in~\eqref{eqn:matelems} as the triple-real (RRR), double-real-virtual (VRR),
double-virtual-real (VVR), and triple-virtual (VVV) \textit{layers} of the
calculation. For completeness we recompute also the next-to-leading order (NLO)
and NNLO corrections. 
Our method was first described in~\cite{Jakubcik:2022zdi} in the context of the
decay of a virtual photon into hadrons and it leverages the reverse unitarity
relation~\cite{Cutkosky:1960sp,Anastasiou:2002yz,Anastasiou:2002wq,Anastasiou:2003yy,Anastasiou:2003ds} to gain access to modern multi-loop techniques.

In our previous work, we could validate the sum of the layers with different final states against known inclusive cross sections for the production of jets in photon and Higgs boson decay. As we shall explain in detail, in order to perform the equivalent check, we extend the present calculation to an effective theory of the minimally supersymmetric Standard Model (MSSM) with gluons, $\NF$ light quark flavours and $\NG$ gluino flavours. In this model, we can perform an independent calculation of the total decay width of the neutralino and derive the relevant renormalization constants from higher-order corrections to the neutralino propagator, which themselves constitute a new result. In Section~\ref{sec:QCD_applications}, we subsequently isolate the desired pure QCD radiation by systematically discarding certain final states.

Having control over the ultraviolet (UV) poles of the triple-virtual layer (the three loop amplitude $\tilde{\chi} \to \glu g$ trivially integrated over the two-particle phase space), we can compare its IR poles with well-known factorization formulae. In particular, the prediction in soft-collinear effective theory (SCET)~\cite{Becher:2009cu,Gardi:2009qi} is cast in terms of universal ingredients such as the collinear anomalous dimensions of the partons. We are therefore able to extract the gluino anomalous dimension and the gluino contribution to the cusp and parton anomalous dimensions to $\mathcal{O}(\alpha_s^3)$. We provide several observations on their structure and make comparisons with existing results.

The rest of the paper is organised as follows. In Section~\ref{sec:method}, we explain our method for the computation of the integrated matrix elements in~\eqref{eqn:matelems}. In Section~\ref{sec:renorm}, we describe how we obtained the necessary renormalization constants. The results are presented in Section~\ref{sec:results} and printed in full in Appendices~\ref{app:results} and~\ref{app:exprlower}. We perform several checks on our results and elaborate on the application to higher-order QCD calculations in Section~\ref{sec:disc}, before drawing conclusions in Section~\ref{sec:concl}. Appendix~\ref{app:ren} provides details about the renormalization  of individual matrix elements and in Appendix~\ref{app:coltables} we report tables with the colour factors appearing in the integrated expressions up to N$^3$LO.

%%%%%%%%%%%%%%%%%%%%%%%%%%%%%%%%%%%%%%%%%%%%%%%%%%%%%%%%%%%%%%%%%%%%%%%%%%%%%%%%

\section{Method}\label{sec:method}

We want to investigate the infrared structure of the decay of an off-shell particle into a massless quark (a spin 1/2 particle in the fundamental representation of SU($3$)) and a gluon (a spin 1 particle in the adjoint representation of SU($3$)). It follows that the decaying particle must have spin 1/2 and also transform in the fundamental representation. However, off-shell external states are forbidden in QCD and besides, we want to focus on final-state radiation by studying the decay of a colour singlet.

Nonetheless, after introducing colour-ordering, one can combine matrix elements for two different partons in the fundamental representation and with equal momenta to emulate a particle in the adjoint representation in the final state. This way, one can recover the desired infrared behaviour from the decay to a spin 1/2 octet state and a spin 1 octet state. In this case, the off-shell particle is a spin 1/2 singlet. Such a process can be realized within the minimal supersymmetric Standard Model (MSSM). For the purposes of this paper, we can simply consider the effective theory for the decay of a heavy neutralino originally formulated in~\cite{Haber:1988px} and its generalization to the decay of a heavy neutralino into a gluon and a gluino~\cite{Gehrmann-DeRidder:2005svg}. In order to include the necessary interactions in the calculation, we extend the SM QCD Lagrangian by adding the gluino field $\psi_\glu$~\cite{Kuroda:1999ks},
\begin{equation}\label{eq:glu_Lagr}
	{\cal L}_{{\rm gluino}}=\dfrac{i}{2}\overline{\psi}^a_{\glu}\gamma^{\mu} D_{\mu} \psi^a_{\glu}
\end{equation}
with the covariant derivative of the field given by
\begin{equation}
	D_\mu \psi^a_{\glu}=\partial_\mu\psi_\glu^a-g_sf^{abc}G_\mu^b\psi_\glu^c\,.
\end{equation}
We use $G_\mu^b$ for the gluon field, the indices $a$, $b$ and $c$ transform in the adjoint representation of SU($3$), $f^{abc}$ are the SU($3$) structure constants and $g_s$ is the strong coupling constant. The term~\eqref{eq:glu_Lagr} describes the gluinos' coupling to gluons and their propagation. For the effective coupling of a gluon and a gluino to the external neutralino field $\psi_{\tilde{\chi}}$, we add to the Lagrangian
\begin{align}\label{eq:eff_Lagr}
{\cal L}_{{\rm eff}} = i \eta \overline{\psi}^a_{\glu} \sigma^{\mu\nu} 
\psi_{\tilde{\chi}} F_{\mu\nu}^a  + ({\rm h.c.})\,,
\end{align}
where the commutator of gamma matrices reads
\begin{align}
\sigma^{\mu\nu} = \frac{i}{2}[\gamma^\mu,\gamma^\nu]\,
\end{align}
and the standard gluon field strength is 
\begin{align}
	F^a_{\mu\nu} = \partial_\mu G_\nu^a - \partial_\nu G_\mu^a - g_s f^{abc}G_\mu^b G_\nu^c\,.
\end{align}
The resulting Feynman rules are
\begin{align}
\begin{gathered}
\includegraphics[width=4cm]{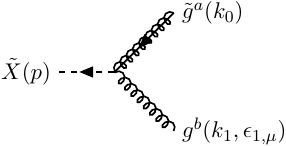}
\end{gathered} &= -i\eta\delta^{ab}\sigma_{\mu\nu}k_1^\nu\,,\\
\begin{gathered}
\includegraphics[width=4cm]{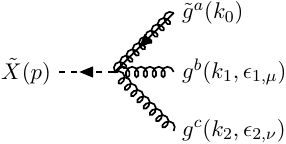}
\end{gathered} &= -g_s\eta f^{abc}\sigma_{\mu\nu}\,,\\
\begin{gathered}
\includegraphics[width=4cm]{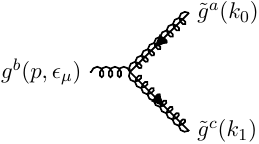}
\end{gathered} &= -g_s f^{abc}\gamma^\mu\,,\\
\begin{gathered}
\includegraphics[width=3cm]{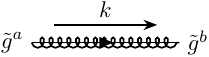}
\end{gathered} &= i\delta^{ab}\frac{1}{k^2+i\epsilon}\,,
\end{align}
where the first two determine the coupling to the external neutralino, while the third one describes the interaction between gluinos and gluons inside the diagram. The essential difference between a gluino and a quark lies in the fact that gluon emissions off a gluino line are weighted with a factor $\CA=N$, instead of $\CF=(N^2-1)/(2\,N)$, where $N=3$ is the number of QCD colours.

In~\cite{Gehrmann-DeRidder:2005svg}, where the equivalent calculation was performed up to NNLO, only QCD emissions from a single gluino line were considered which is the scenario relevant for the definition of quark-gluon antenna functions. However, when we represent  the various layers of the calculation as cuts of the neutralino two-point function, we must allow for \textit{all} physical cuts of a given diagram in order to derive the relevant renormalization constants or define a total width for the decay $\tilde{\chi}\to$ gluinos + partons. For this reason, we perform the calculation with the full set of diagrams allowed by the combination of the QCD Lagrangian with the gluino contribution in~\eqref{eq:glu_Lagr} and the effective Lagrangian in~\eqref{eq:eff_Lagr}. In Section~\ref{sec:QCD_applications} we will illustrate how the presented result can be modified to recover the appropriate quantities for quark-gluon antenna functions, systematically removing undesired gluino emissions.

For the computation of the individual matrix elements in~\eqref{eqn:matelems}, we follow the strategy outlined for the decay $\gamma^{*}\to q\bar{q}$ in~\cite{Jakubcik:2022zdi}, later applied to $H\to gg$ in the large top mass limit and to $H\to b\bar{b}$ in~\cite{Chen:2023fba}.
First we generate the relevant
decay diagrams with QGRAF~\cite{Nogueira:1991ex} as self-energies
of the neutralino with cut internal propagators. We take particular care in treating the supersymmetric particles. As a Majorana fermion, the gluino is its own anti-particle and requires a symmetry factor of 1/2 for every closed loop.
We then apply a procedure which selects only physical cuts and tags loop- and final-state configurations. The selected diagrams are subsequently matched onto the
integral families reported in~\cite{Jakubcik:2022zdi} using
\code{Reduze2}~\cite{vonManteuffel:2012np} and the Feynman rules are inserted
and evaluated in \code{FORM}~\cite{Vermaseren:2000nd}.

The integrals appearing in the matrix elements have up to eleven propagators in
the denominator and a maximum of five scalar products in the numerator.
The integrals are reduced with the help of
\code{Reduze2}~\cite{vonManteuffel:2012np} to a set of 22, 27, 35 and 31 master
integrals for the four terms of~\eqref{eqn:matelems}, respectively. The master
integrals required for the NNLO calculation can be found
in~\cite{Gehrmann-DeRidder:2003pne} and have been extended up to weight 6
in~\cite{Jakubcik:2022zdi,Gituliar:2018bcr,Gehrmann:2010ue}.
The integrals required for the N$^3$LO  calculation were computed
in~\cite{Gituliar:2018bcr,Magerya:2019cvz}.

%%%%%%%%%%%%%%%%%%%%%%%%%%%%%%%%%%%%%%%%%%%%%%%%%%%%%%%%%%%%%%%%%%%%%%%%%%%%%%%%

\section{Renormalization}
\label{sec:renorm}
We work in dimensional regularisation with $d=4-2\epsilon$ dimensions and we present results for the integration of renormalised squared amplitudes
in the $\overline{\text{MS}}$ scheme. We replace the bare coupling $\alpha_0$ with the
renormalised coupling $\alpha_s$ according to
\begin{align}\label{eq:alfaren}
\alpha_0\,\mu_0^{2\e}\,S_\e = \alpha_s\,\mu^{2\e}\Bigg[
1- \frac{\beta_0}{\e}\left(\frac{\alpha_s}{2\pi}\right) 
+\left(\dfrac{\beta_0^2}{\e^2}-\dfrac{\beta_1}{2\e}\right)\left(\frac{\alpha_s}{2\pi}\right)^2  
+{\cal O}(\alpha_s^3) \Bigg]\,,
\end{align}
where $\alpha_0$ is the bare coupling, $S_\e = (4\pi)^\e e^{-\e\gamma}$ with $\gamma$ the Euler constant, and $\mu_0^2$ is the mass parameter
introduced in dimensional regularisation to maintain a dimensionless coupling in
the bare Lagrangian density. We fix the renormalisation scale $\mu^2$ to be the
invariant mass of the decaying particle $q^2$. Note that the coefficients $\beta_i$ in the SM extension considered in this work are not equal to the standard QCD $\beta$-function coefficients.

The effective coupling $\eta_0=Z_{\eta}\eta$ of the neutralino to a gluino and gluon(s) is formally renormalised with
\begin{align}\label{eq:etarenorm}
 Z_\eta &= 1 + Z_\eta^{(1)}\left(\frac{\alpha_s}{2\pi}\right)
 + Z_\eta^{(2)}
 \left(\frac{\alpha_s}{2\pi}\right)^2
 + Z_\eta^{(3)}   \left(\frac{\alpha_s}{2\pi}\right)^3
 +{\cal O}(\alpha_s^4)\,,\nn\\
 Z_\eta^{(1)} &= \dfrac{Z_\eta^{(1,1)}}{\e}\,,\nn\\
 Z_\eta^{(2)} &= \dfrac{Z_\eta^{(2,2)}}{\e^2}+\dfrac{Z_\eta^{(2,1)}}{\e}\,,\nn\\
 Z_\eta^{(3)} &= \dfrac{Z_\eta^{(3,3)}}{\e^3}+\dfrac{Z_\eta^{(3,2)}}{\e^2}
 + \dfrac{Z_\eta^{(3,1)}}{\e}  \,.
\end{align}
To the best of our knowledge, the renormalization constants above are not known. We explain the method for calculating $Z_{\eta}^{(i)}$ and give the results in this section.

A common strategy to obtain the inclusive cross section for the decay of a particle is to use the optical theorem
\begin{align}
2\operatorname{Im}\left[\mathcal{M}(a\to a)\right] = \sum_f \int d\Pi_f \mathcal{M}^{\dagger}(a\to f) \mathcal{M}(a\to f)\,,
\end{align}
which relates the imaginary part of the self-energy of the decaying particle $a$ to the sum over all allowed intermediate final states $f$. In dimensional regularization, the $\epsilon$-poles of the bare self-energy are entirely of ultraviolet origin. Their removal therefore determines the correct renormalization of the relevant couplings of the theory. Note that the sole source of the imaginary part is the expansion of the time-like kinematic prefactor of the self-energy:
\begin{align}
\frac{1}{\pi}\operatorname{Im}\left[(-1)^{\ell\epsilon}\right] =  \sum_{n=0}^{\infty} \frac{(-1)^n \pi^{2n} \ell^{2n+1}}{(2n+1)!} \epsilon^{2n+1} \,.
\end{align} 
at $\ell$ loops. Since this expansion starts at order $\epsilon$, only the poles of the bare expression for $\mathcal{M}(a\to a)$ contribute to the renormalized finite part of the sum of the cuts, which is the total cross section for the production of partons and gluinos in the decay.

We apply this strategy first to the corrections to the photon propagator in order to extract the renormalization of $\alpha_s$, validate our workflow and replicate the results for the well-known QCD beta function and the QCD+gluinos beta function computed in~\cite{Clavelli:1996pz}. After generating the relevant diagrams and evaluating the Feynman rules, we translate the integrand to the notation of \code{FORCER}~\cite{Ueda:2016yjm}, a specialised programme for the IBP reduction of self-energies up to 4 loops to the master integrals calculated in~\cite{Lee:2011jt,Baikov:2010hf}. Up to 3 loops, we also perform the reduction ourselves in~\code{Reduze2} and insert the master integrals in~\cite{Chetyrkin:1981qh,Baikov:2010hf} and find agreement. The bare self-energy is reported in the ancillary files. From~\eqref{eq:alfaren}, one can read off the renormalization of the $\ell$-loop correction to the photon propagator $|\mathcal{M}^{(\ell)}\rangle$:
\begin{align}
|\mathcal{M}^{(1)}\rangle &= |\mathcal{M}^{(1), U}\rangle\,, \\
|\mathcal{M}^{(2)}\rangle &= |\mathcal{M}^{(2), U}\rangle\,, \\
|\mathcal{M}^{(3)}\rangle &= |\mathcal{M}^{(3), U}\rangle - \frac{\beta_0}{\e}|\mathcal{M}^{(2), U}\rangle\,, \\
|\mathcal{M}^{(4)}\rangle &= |\mathcal{M}^{(4), U}\rangle - \frac{2\beta_0}{\e}|\mathcal{M}^{(3), U}\rangle+ \left(\frac{\beta_0^2}{\e^2} - \frac{\beta_1}{2\e}\right)|\mathcal{M}^{(2), U}\rangle\,,
\end{align}
where the superscript $U$ denotes unrenormalised quantities.
Requiring the vanishing of each ultraviolet pole in the total decay cross section yields a system of equations with a unique solution,
\begin{eqnarray}
\beta_0 &=& \frac{1}{6}(11 \CA - 2 \NF)-\frac{1}{3}\NG \CA\,,\label{eq:beta0}\\
\beta_1 &=& \frac{1}{6}(17 \CA^2 - 5 \CA \NF -3 \CF \NF)-\frac{4}{3}\NG\CA^2\,,\label{eq:beta1}
\end{eqnarray}
which is in agreement with the literature~\cite{Clavelli:1996pz}.

\begin{figure}
\centering
\begin{subfigure}[t]{.4\textwidth}
\centering
\includegraphics[width=0.98\linewidth]{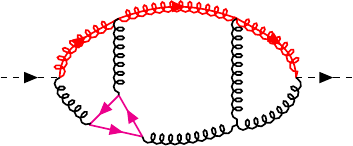}
\caption{Self-energy with a single gluino line coupling to the external neutralino}
\end{subfigure}\hspace{1cm}
\begin{subfigure}[t]{.4\textwidth}
\centering
\includegraphics[width=0.98\linewidth]{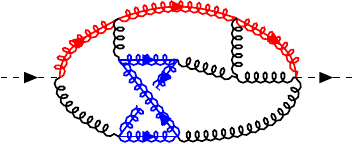}
\caption{Self-energy with a closed gluino loop}
\end{subfigure}
\caption{Example diagrams of N$^3$LO QCD corrections to the neutralino propagator, with and without closed gluino loops (blue). In red we highlight the gluino line which connects to the external neutralinos. All the physical two-, three-, four- and five-particle cuts of these diagrams contribute to the respective layers of the squared matrix element in~\eqref{eqn:matelems}}\label{fig:SE1}
\end{figure}

Next we focus on the corrections to the neutralino propagator (see example diagrams in Figure~\ref{fig:SE1}), where $\alpha_s$ and $\eta$ ought to be renormalized,
\begingroup
\allowdisplaybreaks
\begin{align}
|\mathcal{M}^{(1)}\rangle &= |\mathcal{M}^{(1), U}\rangle\,, \\
|\mathcal{M}^{(2)}\rangle &= |\mathcal{M}^{(2), U}\rangle 
+ 2Z_\eta^{(1)}|\mathcal{M}^{(1), U}\rangle\,, \\
|\mathcal{M}^{(3)}\rangle &= |\mathcal{M}^{(3), U}\rangle 
+ \left(2Z_\eta^{(1)} - \frac{\beta_0}{\e}\right)|\mathcal{M}^{(2), U}\rangle 
+ \left((Z_\eta^{(1)})^2+2Z_\eta^{(2)}\right)|\mathcal{M}^{(1), U}\rangle\,, \\
|\mathcal{M}^{(4)}\rangle &= |\mathcal{M}^{(4), U}\rangle 
+ \left(2Z_\eta^{(1)}- \frac{2\beta_0}{\e}\right)|\mathcal{M}^{(3), U}\rangle \nn\\
&\phantom{\,\,= |\mathcal{M}^{(4), U}\rangle}+ \left((Z_\eta^{(1)})^2+2Z_\eta^{(2)}-\frac{2\beta_0 Z_\eta^{(1)}}{\e} + \frac{\beta_0^2}{\e^2} - \frac{\beta_1}{2\e}\right)|\mathcal{M}^{(2), U}\rangle \nn\\
&\phantom{\,\,= |\mathcal{M}^{(4), U}\rangle}+ \left(2Z_\eta^{(1)}Z_\eta^{(2)}+2Z_\eta^{(3)}\right)|\mathcal{M}^{(1), U}\rangle\,.
\end{align}
\endgroup
Due to the presence of the effective coupling, the general form of the renormalization is the same as for the $H\to gg$ decay. As a validation of our method, we verified that we can with our setup reproduce the known renormalization coefficients of the Higgs effective coupling up to N$^3$LO.

The bare neutralino self-energy corrections up to four loops are given in the ancillary files. Fixing the beta values according to~\eqref{eq:beta0} and~\eqref{eq:beta1}, we can again solve a system of equations with a unique solution:
\begingroup
\allowdisplaybreaks
\begin{align}
Z_{\eta}^{(1,1)} &= +\frac{1}{6}(-10\CA+\NF)+\frac{1}{6}\NG C_A\,,\label{eq:wcn1}\\
Z_{\eta}^{(2,2)} &= +\frac{1}{24} \left(70 \CA^2-17 \CA \NF+\NFSquare\right)+\frac{1}{24} \NG \left(-17 \CA^2+2 \CA \NF\right)+\frac{1}{24} \NGSquare\CA^2\,,
  \\
\label{eq:Z21ren}Z_{\eta}^{(2,1)} &= +\frac{1}{24} \left(-46 \CA^2+10 \CA \NF+3 \CF \NF\right)+\frac{13}{24}\NG \CA^2\,,\\
Z_{\eta}^{(3,3)} &= +\frac{1}{432} \left(-2240 \CA^3 + 894 \CA^2 \NF - 117 \CA \NFSquare + 5 \CA^2 \NFCube - 10\CA\CF \NFCube \right)\nn\\
    &\phantom{\,=\,}+\frac{1}{144} \NG \left(298 \CA^3 - 78 \CA^2 \NF + 5 \CA \NFSquare \right)\nn\\
    &\phantom{\,=\,}+\frac{1}{144} \NGSquare \left(-39 \CA^3 + 5 \CA^2 \NF \right) +\frac{5}{432} \NGCube \CA^3  \,,\\
Z_{\eta}^{(3,2)} &= +\frac{1}{144} \left(1024 \CA^3 - 370 \CA^2\NF - 92 \CA\CF\NF + 30 \CA \NFSquare + 11 \CF \NFSquare \right)\nn\\
   &\phantom{\,=\,}+\frac{1}{144} \NG \left(-462 \CA^3 + 71 \CA^2 \NF + 11 \CA\CF\NF \right) + \frac{41}{144}\NGSquare\CA^3 \,,\\
Z_{\eta}^{(3,1)} &= +\frac{1}{2592}(-8335 \CA^3 + 2546 \CA^2 \NF + 1992 \CA\CF\NF - 54\CF^2\NF - 43\CA\NF^2 \nn\\
   &\phantom{\,=\,+\frac{1}{2592}(}- 66\CF\NF^2 +1296\CA^2\NF\zeta_{3} -1296\CA\CF\NF\zeta_{3})\nn\\
   &\phantom{\,=\,}+\frac{1}{1296}\NG \left(2242 \CA^3 - 76\CA^2\NF - 33 \CA\CF \NF \right) -\frac{109}{2592} \NGSquare\CA^3\,.
\end{align}
\endgroup
Finally the $R$-ratio for the decay of neutralino to gluinos and partons is 
\begin{align}\label{Rratio}
R &= \frac{\sigma(\tilde{\chi}\to \text{partons} + \text{gluinos})}{\sigma(\tilde{\chi}\to \glu g)} \nn\\
&= 1 + \left(\frac{\alpha_s}{2\pi}\right)\left[\frac{67}{6}N-\NF-\NG N\right]\nn\\
&+ \left(\frac{\alpha_s}{2\pi}\right)^2\Bigg[
  N^2 \left(\frac{11521}{81} - \frac{155}{54}\pi^2 - \frac{51}{2}\zeta_3 \right)
  + \NF N \left(  - \frac{39821}{1296} + \frac{71\pi^2}{108}+ 2\zeta_3 \right)\nn\\
&\phantom{+ \left(\frac{\alpha_s}{2\pi}\right)^2\Bigg[}
  + \NF N^{-1} \left( \frac{71}{48} - \zeta_3 \right)
  + \NFSquare \left( \frac{91}{81} - \frac{\pi^2}{27} \right)
  + \NG N^2 \left(  - \frac{20869}{648} + \frac{71\pi^2}{108} + 3\zeta_3 \right)\nn\\
&\phantom{+ \left(\frac{\alpha_s}{2\pi}\right)^2\Bigg[}    
  + \NG \NF N \left( \frac{182}{81} - \frac{2\pi^2}{27} \right)
  + \NGSquare N^2 \left( \frac{91}{81} - \frac{\pi^2}{27} \right)
  \Bigg]\nn\\
  &+ \left(\frac{\alpha_s}{2\pi}\right)^3\Bigg[
    + N^3 \left( \frac{45447757}{23328} - \frac{17731}{216}\pi^2 + \frac{365}{3}\zeta_5 - \frac{1407}{2}\zeta_3 \right)
    \nn\\&\phantom{+ \left(\frac{\alpha_s}{2\pi}\right)^3\Bigg[}
      + \NF N^2 \left(  - \frac{2702383}{3888} + \frac{1625}{54}\pi^2 - \frac{1}{120}\pi^4 - \frac{35}{3}\zeta_5 + \frac{3455}{24}\zeta_3 \right)
    \nn\\&\phantom{+ \left(\frac{\alpha_s}{2\pi}\right)^3\Bigg[}      
      + \NF \left( \frac{133685}{2592} - \frac{41}{72}\pi^2 - \frac{1}{120}\pi^4 + \frac{5}{2}\zeta_5 - \frac{439}{12}\zeta_3 \right)
    \nn\\&\phantom{+ \left(\frac{\alpha_s}{2\pi}\right)^3\Bigg[}            
       + \NF N^{-2} \left( \frac{155}{288} - \frac{5}{2}\zeta_5 + \frac{37}{24}\zeta_3 \right)
       + \NFSquare N \left( \frac{84127}{1296} - \frac{727}{216}\pi^2 - \frac{19}{3}\zeta_3 \right)
    \nn\\&\phantom{+ \left(\frac{\alpha_s}{2\pi}\right)^3\Bigg[}                   
       + \NFSquare N^{-1} \left(  - \frac{13745}{2592} + \frac{5}{72}\pi^2 + \frac{7}{2}\zeta_3 \right)
       + \NFCube \left(  - \frac{1055}{729} + \frac{1}{9}\pi^2 \right)
    \nn\\&\phantom{+ \left(\frac{\alpha_s}{2\pi}\right)^3\Bigg[}                       
      + \NG N^3 \left(  - \frac{1450409}{1944} + \frac{6623}{216}\pi^2 - \frac{50}{3}\zeta_5 + \frac{2185}{12}\zeta_3 \right)
    \nn\\&\phantom{+ \left(\frac{\alpha_s}{2\pi}\right)^3\Bigg[}                             
      + \NG \NF N^2 \left( \frac{38917}{288} - \frac{1469}{216}\pi^2 - \frac{97}{6}\zeta_3 \right)
    \nn\\&\phantom{+ \left(\frac{\alpha_s}{2\pi}\right)^3\Bigg[}                                   
       + \NG \NF \left(  - \frac{13745}{2592} + \frac{5}{72}\pi^2 + \frac{7}{2}\zeta_3 \right)
       + \NG \NFSquare N \left(  - \frac{1055}{243} + \frac{1}{3}\pi^2 \right)
    \nn\\&\phantom{+ \left(\frac{\alpha_s}{2\pi}\right)^3\Bigg[}                                          
       + \NGSquare N^3 \left( \frac{181999}{2592} - \frac{371}{108}\pi^2 - \frac{59}{6}\zeta_3 \right) 
       + \NGSquare \NF N^2 \left(  - \frac{1055}{243} + \frac{1}{3}\pi^2 \right)
    \nn\\&\phantom{+ \left(\frac{\alpha_s}{2\pi}\right)^3\Bigg[}                                                 
       + \NGCube N^3 \left(  - \frac{1055}{729} + \frac{1}{9}\pi^2 \right)
    \Bigg]\,.
\end{align}
Note that setting $\NG=0$ in~\eqref{eq:wcn1} reproduces the renormalization constant quoted in~(2.4) of~\cite{Gehrmann-DeRidder:2005svg}, where contributions due to additional gluino emissions were discarded. 

%%%%%%%%%%%%%%%%%%%%%%%%%%%%%%%%%%%%%%%%%%%%%%%%%%%%%%%%%%%%%%%%%%%%%%%%%%%%%%%%
\section{Results}\label{sec:results}

The notation we adopt is similar to the one in~\cite{Jakubcik:2022zdi,Chen:2023fba}. 
Given a set of $n$ final-state particles denoted by ${\cal I}$, we can
generically write the amplitude for the $\tilde{\chi} \to {\cal I}$ process as
\begin{equation}\label{eq:expansion}
	|{\cal M}\rangle_{{\cal I}} =
	|{\cal M}^{(0)}\rangle_{{\cal I}} 
	+ \left(\frac{\alpha_s}{2\pi}\right) |{\cal M}^{(1)}\rangle_{{\cal I}} 
	+ \left(\frac{\alpha_s}{2\pi}\right)^2 |{\cal M}^{(2)}\rangle_{{\cal I}}
	+ \left(\frac{\alpha_s}{2\pi}\right)^3 |{\cal M}^{(3)}\rangle_{{\cal I}}
	+ \ldots \,.
\end{equation}
In Appendix~\ref{app:ren} we describe how to renormalize of each term in~\eqref{eq:expansion}.
We denote the integration over the respective phase space of the matrix element
$\langle{\cal M}|{\cal M}\rangle _{{\cal I}}$ summed over spins,
colours and quark and gluino flavours as
\begin{equation}\label{eq:Tll}
	{\cal T}^{(k,\left[\ell \times \ell\right])}_{{\cal I}} = \int \d \Phi_n \,
	\langle{\cal M}^{(\ell)}|{\cal M}^{(\ell)}\rangle_{{\cal I}}\,,
\end{equation}
and for $\ell_1 \neq \ell_2$
\begin{equation}\label{eq:Tlm}
	{\cal T}^{(k,\left[\ell_1 \times \ell_2\right])}_{{\cal I}} = \int \d
	\Phi_n \, 2\,\Re\big[\langle{\cal M}^{(\ell_1)}|{\cal
		M}^{(\ell_2)}\rangle_{{\cal I}}\big]\,.
\end{equation}
The label $k$ denotes the perturbative order: contributions with
the same $k$ sum to the N$^k$LO correction to the total decay cross section.
The explicit expressions for ${\cal T}^{(3,\left[\ell_1 \times
	\ell_2\right])}_{{\cal I}}$ are provided in Appendix~\ref{app:results},
while in Appendix \ref{app:exprlower} we report the lower-order results
expanded up to transcendental weight six. We denote the coefficient of each colour factor
${\cal C}$ as ${\cal T}^{(k,\left[\ell_1 \times \ell_2\right])}_{{\cal
		I}}\big|_{\cal C}$, and we omit the superscript $\left[\ell_1 \times
\ell_2\right]$ in case of no ambiguity.
All results are also provided in the ancillary files in computer-readable
format, with the notation
\begin{equation}
	{\cal T}^{(k,\left[\ell_1 \times \ell_2\right])}_{{\cal I}}
	= \texttt{[X}\texttt{\_}\mathcal{I}\texttt{\_}k\texttt{\_}\ell_1\texttt{x}\ell_2\texttt{]}\,.
\end{equation}
All the expressions are renormalised and in time-like kinematics. In Appendix~\ref{app:coltables}, we tabulate all the colour factors $\mathcal{C}$ that appear for a given final state $\mathcal{I}$.
Higher-order results are normalised to
\begin{equation}\label{eq:Hgg0}
	{\cal T}^{(0)}_{\glu g} = 4(N^2-1)\eta^2(1-\e)(q^2)^2P_2\,,
\end{equation}
with $N$ the number of colours and $P_2$ is the volume of the two-particle phase space:
\begin{equation}
	P_2  = \int \d \Phi_2 =
	2^{-3+2\e}\, \pi^{-1+\e}\, \frac{\Gamma(1-\e)}{\Gamma(2-2\e)}\,
	(q^2)^{-\e} \,.
\end{equation}

Final states with multiple fermionic lines, either quark or gluino lines, are completely decomposed into same- or different-flavour contributions. Primed and double-primed fermion labels are used to explicitly indicate different-flavour lines. For the one-gluino four-quark final state we can differentiate the same- and different-flavour contributions simply by the powers of $N$ and $\NF$ they carry,
\begin{equation}
	{\cal T}^{(3)}_{\glu q\bar{q}q\bar{q}}=
	\dfrac{1}{\NF-1}{\cal T}^{(3)}_{\glu q\bar{q}q'\bar{q}'}
        +\Delta{\cal T}^{(3)}_{\glu q\bar{q}q\bar{q}}|_{\NF}
        +\Delta{\cal T}^{(3)}_{\glu q\bar{q}q\bar{q}}|_{\NF N^{-2}}\,,
\end{equation}
where $\Delta{\cal T}^{(3)}_{\glu q\bar{q}q\bar{q}}|_{\NF}$ and $\Delta{\cal T}^{(3)}_{\glu q\bar{q}q\bar{q}}|_{\NF N^{-2}}$ represent interference contributions only present in the same-flavour case. 

In multiple-gluino final states, same- and different-flavour contributions have the same colour factor, since gluinos transform in the adjoint representation of SU($3$). Let us first consider the emission of a gluino pair in addition to the single hard gluino present in any final state. We have
\begin{equation}\label{eq:gluFS1}
	{\cal T}^{(k)}_{\glu \glu \glu (ij)}
	= \dfrac{1}{\NG-1}{\cal T}^{(k)}_{\glu \glu' \glu' (ij)}+\Delta{\cal T}^{(k)}_{\glu \glu \glu (ij)}\,,
\end{equation}
where $\Delta{\cal T}^{(k)}_{\glu \glu \glu (ij)}$ indicates extra terms which can not be straightforwardly identified by a specific colour factor as in the same-flavour quark case above, and $i$ and $j$ are potential additional final-state particles which are not gluinos (quarks or gluons). At N$^3$LO, five-gluino final states are also allowed, with up to three different gluino flavours. The following relations hold:
\begin{eqnarray}
	\label{eq:gluFS2bis}{\cal T}^{(3)}_{\glu \glu' \glu' \glu' \glu'}
	&=& \dfrac{1}{\NG-2}\,{\cal T}^{(3)}_{\glu \glu' \glu' \glu'' \glu''}+\Delta{\cal T}^{(3)}_{\glu \glu' \glu' \glu' \glu'}\,,\\  
	\label{eq:gluFS2}{\cal T}^{(3)}_{\glu \glu \glu \glu' \glu'}
	&=& \dfrac{2}{\NG-2}\,{\cal T}^{(3)}_{\glu \glu' \glu' \glu'' \glu''}+\Delta{\cal T}^{(3)}_{\glu \glu \glu \glu' \glu'}\,,\\
	\label{eq:gluFS3}{\cal T}^{(3)}_{\glu \glu \glu \glu \glu}
	&=& \dfrac{1}{(\NG-1)(\NG-2)}\,{\cal T}^{(3)}_{\glu \glu' \glu' \glu'' \glu''}+\dfrac{1}{\NG-1}\Delta{\cal T}^{(3)}_{\glu \glu' \glu' \glu' \glu'}\nn\\&+&\dfrac{1}{\NG-1}\Delta{\cal T}^{(3)}_{\glu \glu \glu \glu' \glu'}+\Delta{\cal T}^{(3)}_{\glu \glu \glu \glu \glu}\,.
\end{eqnarray}
In the ancillary files, we specifically denote the $\Delta$-terms above as
\begin{equation}
	\Delta{\cal T}^{(k,\left[\ell_1 \times \ell_2\right])}_{{\cal I}}
	= \texttt{[Delta\_X}\texttt{\_}\mathcal{I}\texttt{\_}k\texttt{\_}\ell_1\texttt{x}\ell_2\texttt{]}\,.
\end{equation}

The natural check we perform on the obtained results is the complete cancellation of the infrared singularities in the sum over all possible final states at a given perturbative order, as well as the recovery of the fully inclusive decay rate in~\eqref{Rratio}. We find perfect agreement in all powers of $N$, $\NF$ and $\NG$.

It is possible to identify precise relations among the quark and gluino contributions to each term in~\eqref{eqn:matelems}. As observed also in~\cite{Clavelli:1996pz,Clavelli:1998jz,Bern:2003ck}, the perturbative corrections due to the emission of gluino pairs can be directly matched onto those coming from a quark-antiquark pair emission simply by adjusting certain colour factors in the final result. This can be easily seen for example in the $\beta$-function coefficients~\eqref{eq:beta0} and~\eqref{eq:beta1} for which we have 
\begin{eqnarray}
	\beta_0\big|_{C_A\NG}&=&\beta_0\big|_{\NF}\,,\\
	\beta_1\big|_{C_A^2\NG}&=&\beta_1\big|_{C_A\NF}+\beta_1\big|_{C_F\NF}\,.
\end{eqnarray}

For each final-state multiplicity and up to N$^3$LO, we can then consider the sum over all possible partonic configurations, denoted in the following by fixing the subscript $\mathcal{I}=n$, instead of indicating a specific set of partons. At NLO we have
\begin{eqnarray}
	\T{1}{n}\big|_{C_A\NG}&=&\T{1}{n}\big|_{\NF}
\end{eqnarray}
with $n=2,3$, at NNLO we have
\begin{eqnarray}
	\T{2}{n}\big|_{C_A^2\NG}&=&\T{2}{n}\big|_{C_A\NF}+\T{2}{n}\big|_{C_F\NF}\,,\\
	\T{2}{n}\big|_{C_A^2\NGSquare}&=&\T{2}{n}\big|_{\NFSquare}\,,\\
	\T{2}{n}\big|_{C_A\NG\NF}&=&2\T{2}{n}\big|_{\NFSquare}
\end{eqnarray}
with $n=2,3,4$ and at N$^3$LO we have
\begin{eqnarray}
	\label{eq:relstart}\T{3}{n}\big|_{C_A^3\NG}&=&\T{3}{n}\big|_{C_A^2\NF}+\T{3}{n}\big|_{C_A C_F\NF}+\T{3}{n}\big|_{C_F^2\NF}\,,\\ 
	\T{3}{n}\big|_{C_A^3\NGSquare}&=&\T{3}{n}\big|_{C_A\NFSquare}+\T{3}{n}\big|_{C_F\NFSquare}\,,\\
	\T{3}{n}\big|_{C_A^3\NGCube}&=&\T{3}{n}\big|_{\NFCube}\,,\\
	\T{3}{n}\big|_{C_A^2\NG\NF}+\T{3}{n}\big|_{C_A C_F\NG\NF}&=&2\T{3}{n}\big|_{C_A\NFSquare}+2\T{3}{n}\big|_{C_F\NFSquare}\,,\\
	\T{3}{n}\big|_{C_A^2\NGSquare\NF}&=&3\T{3}{n}\big|_{\NFCube}\,,\\
	\label{eq:relend}\T{3}{n}\big|_{C_A\NG\NFSquare}&=&3\T{3}{n}\big|_{\NFCube}
\end{eqnarray}
with $n=2,3,4,5$. Note that for a given choice of multiplicity and colour factor, some terms might be vanishing but the identities hold nonetheless. We verified that the relations above hold up to transcendental weight six, but it is perfectly reasonable to assume they are exactly satisfied. Such results indicate that the gluino contribution is straightforwardly obtained from the standard QCD matrix elements (i.e. with $\NG=0$) in two steps. Firstly, the kinematics of an emitted gluino pair and a quark-antiquark pair is identical, so the gluino contribution can be generated with the replacement $\NF\to\NF+C_A\NG$, or more generally $\TF\NF\to\TF\NF+C_A\NG/2$, as observed at lower orders in ref.~\cite{Clavelli:1998jz}. Secondly, one has to account for the fact that gluinos transform in the adjoint representation and quarks in the fundamental representation of SU($3$) and consequently the emission of a gluon from a gluino line comes with a factor $C_A$ rather than a factor of $C_F$ as for the quark. It follows that after the first mapping, one has to identify $C_F^n\NG\to C_A^n\NG$ in all the generated colour factors which do not contain a residual $\NF$. 
The observed correspondence and its simple interpretation provide another strong check on our results, particularly on the implementation of gluinos which is new in the current workflow.

%%%%%%%%%%%%%%%%%%%%%%%%%%%%%%%%%%%%%%%%%%%%%%%%%%%%%%%%%%%%%%%%%%%%%%%%%%%%%%%%
\section{Discussion}\label{sec:disc}

\subsection{Comments on the infrared singularity structure}

In the following, we comment on the structure of the infrared singularities of our results at N$^3$LO. 
For the two-particle final state $\glu g$, the infrared singularity structure of
the three-loop amplitude (trivially integrated over the two-particle phase
space) is predicted through universal IR factorisation
formulae~\cite{Catani:1998bh,Becher:2009qa}.
In order to avoid a proliferation of powers of $\pi^2$ originating from the analytic continuation, we perform this analysis for the space-like form factors ${\cal T}^{(k),SL}_{\glu g}$, denoted with the superscript $SL$. They differ from the results given in Appendix~\ref{app:results} only by the absence of the trivial analytic continuation of the prefactor $(-q^2)^{\ell\e}$, and we provide them for ease of reference in the ancillary files. We have
\begin{align}
	\mathrm{Poles}\left({\cal T}^{(1),\,SL}_{\glu g}\right) & = 2\,I^{(1)}\,, \label{eq:poles1} \\
	\mathrm{Poles}\left({\cal T}^{(2,\left[2\times 0\right]),\,SL}_{\glu g}\right) & = 2\,I^{(2)} + I^{(1)} \, {\cal T}^{(1),SL}_{\glu g}\,, \label{eq:poles2} \\
	\mathrm{Poles}\left({\cal T}^{(3,\left[3\times 0\right]),\,SL}_{\glu g}\right) & = 
	2\,I^{(3)} + I^{(2)} \, {\cal T}^{(1),SL}_{\glu g} + I^{(1)} \, {\cal T}^{(2,\left[2\times 0\right]),SL}_{\glu g} \,, \label{eq:poles3}
\end{align}
where we can consider the subtraction operators $I^{(\ell)}$ as scalars in colour space because there are only two external partons. According to our normalization, we have ${\cal T}^{(0)}_{\glu g} = {\cal T}^{(0),\,SL}_{\glu g} = 1$. The factors of $2$ in the equations above come from taking twice the real part of the interference of the $\ell$-loop amplitude with the Born-level amplitude. 
The subtraction operators are given by
\begin{align}
	{I}^{(1)} & = \mathcal{Z}^{(1)}\,,\label{eq:I1} \\
	{I}^{(2)} & = \mathcal{Z}^{(2)} - \big(\mathcal{Z}^{(1)}\big)^2\,,\label{eq:I2} \\
	{I}^{(3)} & = \mathcal{Z}^{(3)} - 2\mathcal{Z}^{(2)}\mathcal{Z}^{(1)} + \big(\mathcal{Z}^{(1)}\big)^3\,,
	\label{eq:I3}
\end{align}
where the coefficients $\mathcal{Z}^{(\ell)}$ can be directly extracted from ultraviolet renormalization in soft-collinear effective theory~\cite{Becher:2009qa}:
\begin{align}
	\mathcal{Z}^{(1)} &= \frac{\Gamma_{0}'}{4 \epsilon ^2} + \frac{\Gamma _0}{2 \epsilon }\,, \label{eq:Z1}\\
	\mathcal{Z}^{(2)} &= \frac{\Gamma_{0}^{\prime 2}}{32 \epsilon ^4}+\frac{\Gamma'_0}{8 \epsilon
		^3}\left(\Gamma _0-\frac{3 \beta _0}{2}\right)+\frac{1}{4 \epsilon ^2}\left(-\beta _0 \Gamma _0+\frac{\Gamma _0^2}{2}+\frac{\Gamma'
		_1}{4}\right)+\frac{\Gamma _1}{4 \epsilon }\,,\label{eq:Z2}\\
	\mathcal{Z}^{(3)} &= +\frac{\Gamma_0^{\prime 3}}{384 \epsilon ^6} +\frac{\Gamma_0^{\prime 2}}{64 \epsilon ^5}\left(\Gamma _0-3\beta _0\right) +\frac{\Gamma_{0}'}{ 9\epsilon ^4}\left(-\frac{5}{4} \beta _0 \Gamma _0 +\frac{11}{9}
	\beta _0^2 +\frac{1}{4} \Gamma _0^2 +\frac{
		\Gamma_{1}'}{8}\right)\nonumber\\
	&+ \frac{1}{\epsilon ^3}\bigg(\frac{1}{9} \beta_1\Gamma_{0}'
	+\Gamma _0
	\left(\frac{\beta _0^2}{6}+\frac{\Gamma_{1}'}{32}\right)-\frac{1}{8} \beta _0 \Gamma
	_0^2-\frac{5 \beta _0 \Gamma_{1}'}{72}+\frac{\Gamma _1 \Gamma_{0}'}{16}+\frac{\Gamma
		_0^3}{48}\bigg)\nonumber\\
	&+\frac{1}{\epsilon ^2}\left(-\dfrac{\beta_1 \Gamma_0}{6}-\frac{\beta _0 \Gamma _1}{6}+\frac{\Gamma _1\Gamma_0}{8}+\frac{\Gamma'
		_2}{36}\right)+\frac{\Gamma _2}{6 \epsilon }\label{eq:Z3}
\end{align}
with $\Gamma_{\ell}$ and $\Gamma_{\ell}'$ given by
\begin{align}
	\Gamma_{\ell} &= \gamma_{\ell}^{\glu} + \gamma_{\ell}^{g}\,, \label{eqn: GammaExpansion}\\
	\Gamma_{\ell}' &= - 2\,C_A\,\gamma_{\ell}^{K}\,,
	\label{eqn: gamma prime}
\end{align}
while $\gamma^i$ with $i=K,q,g,\glu$ represent the coefficients of the perturbative expansion
\begin{align}
\gamma^i=\sum_{i=0}^{\infty} \gamma_\ell^i\left(\frac{\alpha_s}{2\pi}\right)^{l+1}
\end{align}
for the cusp, quark, gluon and gluino anomalous dimensions. Note that the relevant $\beta_0$ and $\beta_1$ coefficients used in~\eqref{eq:Z1}-\eqref{eq:Z3} include the gluino contribution, as given in equations~\eqref{eq:beta0} and~\eqref{eq:beta1}. The same  holds for the anomalous dimension coefficients, for which however, to the best of our knowledge, the gluino contribution is not known. We can therefore impose the equations~\eqref{eq:poles1}-\eqref{eq:poles3} and seek a unique solution for the anomalous dimensions. The recovery of the usual anomalous dimension coefficients with $\NG=0$ is a strict test of our result for the two-particle final states. The coefficients of the powers of $\NG$ constitute a new result.

Since the cusp anomalous dimension alone describes the deepest poles, it can be easily isolated,
\begin{flalign}
	\gamma^{K}_0 &=2\,,&& \label{gammaK0} \\
	\gamma^{K}_1 &=\left(\frac{67}{9}-\frac{\pi ^2}{3}\right) C_A-\frac{10 \NF}{9}-\frac{10}{9}C_A\NG\,, &&\\
	\gamma^{K}_2 &=\left(\frac{11 \zeta _3}{3}+\frac{11 \pi ^4}{90}-\frac{67 \pi ^2}{27}+\frac{245}{12}\right) C_A^2+\left(-\frac{14 \zeta _3}{3}+\frac{10 \pi ^2}{27}-\frac{209}{54}\right) C_A \NF
	&&\nn \\ 
	& +\left(4 \zeta _3-\frac{55}{12}\right) C_F
	\NF-\frac{2}{27} \NFSquare+\left(-\frac{2 \zeta _3}{3}+\frac{10 \pi ^2}{27}-\frac{913}{108}\right) C_A^2 \NG
	&&\nn\\ &-\frac{2}{27}C_A^2\NGSquare -\frac{4 }{27}C_A\NF \NG\,.\label{gammaK2}
\end{flalign}
The standard QCD results computed to this order in~\cite{korchemksy:1987} are easily recovered setting ${\NG=0}$. It is possible to relate this result to the cusp anomalous dimension in the $\mathcal{N}=1$ supersymmetric Yang-Mills theory of gluons and gluinos by setting $\NF=0$ and $\NG=1$. Generally speaking, conventional dimensional regularization (CDR)~\cite{Collins:1984xc} used in this paper explicitly breaks supersymmetry~\cite{Bern:2002zk} and computations in supersymmetric theories often rely on other schemes such as the four-dimensional helicity (FDH)~\cite{Bern:2002zk} scheme or dimensional reduction (DR)~\cite{Siegel:1979wq}. For the cusp anomalous dimension up to this perturbative order, the conversion from the $\overline{\text{MS}}$ scheme with CDR to dimensional reduction is particularly straightforward~\cite{Grozin:2015kna}. We can simply redefine the strong coupling constant in the perturbative expansion of the cusp anomalous dimensions according to eq. (6.8) of~\cite{Grozin:2015kna}. Setting $\NF=0$ in our result and $n_s=0$, $n_f=\NG$ in eq. (6.13) of ref.~\cite{Grozin:2015kna}, we find perfect agreement.

The gluon and gluino anomalous dimensions both enter the two-particle final state matrix elements with the same colour factors and so they cannot be uniquely determined at three loops from this calculation alone. Hence we repeated the calculation of the gluon form factor up to three loops from the corrections to the vertex $H\to gg$ with the setup introduced in~\cite{Chen:2023fba}, this time including gluino-gluon interactions. We notice that for the renormalization of the strong coupling and the effective $Hgg$ coupling (which can be written entirely in terms of beta function coefficients) it is sufficient to consider the beta function coefficients given in~\eqref{eq:beta0} and~\eqref{eq:beta1} instead of the standard QCD ones, in order to include the gluino contribution. The low poles of this decay process are described in terms of the cusp and gluon collinear anomalous dimension, which reads
\begin{flalign}
	\gamma^{g}_0 &=-\frac{11}{6}C_A+\frac{1}{3}\NF+\frac{1}{3}\CA\NG \,,&&\\
	\gamma^{g}_1 &=\left(\frac{\zeta _3}{2}+\frac{11 \pi ^2}{72}-\frac{173}{27}\right) C_A^2+\left(\frac{32}{27}-\frac{\pi ^2}{36}\right) C_A \NF +\frac{C_F \NF}{2}\nn&&\\
	&-\left(\frac{91}{54}-\frac{1}{36}\pi^2\right)C_A^2\NG\, ,&&\\
	\gamma^{g}_2 &=\left(-\frac{5}{18} \pi ^2 \zeta _3+\frac{61 \zeta _3}{12}-2 \zeta _5-\frac{319 \pi ^4}{2160}+\frac{6109 \pi
		^2}{3888}-\frac{48593}{2916}\right) C_A^3
	\nn&&\\&
	+\left(\frac{89 \zeta_3}{54}+\frac{41 \pi ^4}{1080}-\frac{599 \pi ^2}{1944}+\frac{30715}{11664}\right) C_A^2 \NF
	\nn&&\\&	
	+\left(-\frac{19 \zeta _3}{9}-\frac{\pi ^4}{90}-\frac{\pi ^2}{24}+\frac{1217}{216}\right) C_A C_F \NF+\left(-\frac{7 \zeta _3}{27}+\frac{5 \pi ^2}{324}-\frac{269}{11664}\right) C_A \NFSquare&&\nn\\
	&
	-\frac{11}{72} C_F \NFSquare-\frac{1}{8} C_F^2 \NF  +\left(-\frac{25 \zeta_3}{54}+\frac{29 \pi ^4}{1080}-\frac{85 \pi ^2}{243}+\frac{94975}{11664}\right) C_A^3 \NG
	&&\nn\\&
	\left(-\frac{7\zeta_3}{27}+\frac{5\pi^2}{324}-\frac{2051}{1164}\right)C_A^3 \NGSquare +\left(-\frac{14\zeta_3}{27}+\frac{5\pi^2}{162}-\frac{145}{729}\right)C_A^2\NF\NG
	&&\nn\\&
	-\frac{11}{72}C_A C_F\NF\NG\,.
\end{flalign}
With this knowledge, one can use the result of this paper for the $\glu g$ dipole to extract also the gluino anomalous dimension coefficients
\begin{flalign}
	\gamma^{\glu}_0 &= - \frac{3}{2} C_A\,, &&\\
	\label{eq:anglu2loop}\gamma^{\glu}_1 &= \left(\frac{\zeta _3}{2}+\frac{\pi ^2}{24}-\frac{521}{108}\right) C_A^2+\left(\frac{65}{108}+\frac{\pi ^2}{12}\right) C_A \NF
	+\left(\frac{65}{108}+\frac{\pi ^2}{12}\right) C_A^2 \NG\, ,
	&&\\
	\gamma^{\glu}_2 &= \left( - \frac{145853}{11664} + \frac{2449\pi^2}{3888} + \frac{191\zeta_3}{36} - \frac{187\pi^4}{2160}
	- \frac{5\pi^2\zeta_3}{18} -2\zeta_5 \right) C_A^3
	\nn&&\\&
	+\left( -\frac{10757}{11664} + \frac{703\pi^2}{1944} + \frac{227\zeta_3}{54} -\frac{5\pi^4}{216}  \right) C_A^2 \NF
	\nn&&\\&
	+\left( \frac{1355}{216} + \frac{\pi^2}{8} - \frac{46\zeta_3}{9} - \frac{\pi^4}{90} \right) C_A C_F \NF
	+\left( \frac{2417}{5832} - \frac{5\pi^2}{108} - \frac{\zeta_3}{27} \right) C_A \NFSquare
	\nn&&\\&
	+\left(\frac{62413}{11664}+\frac{473}{972}\pi^2-\frac{49}{54}\zeta_3-\frac{37}{1080}\pi^4\right)C_A^3\NG
	+\left(\frac{2417}{5832} - \frac{5\pi^2}{108} - \frac{\zeta_3}{27}\right)C_A^3\NGSquare
	\nn&&\\&
	+\left(\frac{2417}{2916} - \frac{5\pi^2}{54} - \frac{2\zeta_3}{27}\right)C_A^2\NF\NG
	\,.&&
\end{flalign}
Again, the standard QCD results computed to this order in~\cite{matsuura:1988sm, vanNeerven:1985xr,Harlander:2000} are easily recovered setting $\NG=0$. As expected, we can observe in the anomalous dimensions the same correspondence between gluino and quark contributions discussed above:
\begin{eqnarray}
	\gamma_1^K\big|_{C_A\NG} &=& \gamma_1^K\big|_\NF,\\
	\gamma_2^K\big|_{C_A^2\NG} &=& \gamma_2^K\big|_{C_A\NF}+\gamma_2^K\big|_{C_F\NF},\\
	\gamma_2^K\big|_{C_A\NF\NG} &=& 2\gamma_2^K\big|_{C_A^2\NGSquare}=2\gamma_2^K\big|_{\NFSquare},\\
	\gamma_0^g\big|_\NG &=& \gamma_0^g\big|_\NF,\\
	\gamma_1^g\big|_{C_A^2\NG} &=& \gamma_1^g\big|_{C_A\NF}+\gamma_1^g\big|_{C_F\NF},\\
	\gamma_2^g\big|_{C_A^3\NG} &=& \gamma_2^g\big|_{C_A^2\NF}+\gamma_2^g\big|_{C_A C_F\NF}+\gamma_2^g\big|_{C_F^2\NF},\\
	\gamma_2^g\big|_{C_A^2\NF\NG}+\gamma_2^g\big|_{C_A C_F\NF\NG} &=& 2\gamma_2^g\big|_{C_A^3\NGSquare}=2\left(\gamma_2^g\big|_{C_A\NFSquare}+\gamma_2^g\big|_{C_F\NFSquare}\right),\\
	\gamma_1^\glu\big|_{C_A^2\NG} &=& \gamma_1^\glu\big|_{C_A\NF},\\
	\gamma_2^\glu\big|_{C_A^3\NG} &=& \gamma_2^\glu\big|_{C_A^2\NF}+\gamma_2^\glu\big|_{C_A C_F\NF},\\
	\gamma_2^\glu\big|_{C_A^2\NF\NG} &=& 2\gamma_2^\glu\big|_{C_A^3\NGSquare}=2\gamma_2^\glu\big|_{C_A\NFSquare}.
\end{eqnarray} 
We can use this correspondence to write down, without repeating an explicit calculation, the quark anomalous dimension with gluino contributions for completeness. The only caveat is that the quark anomalous dimension perturbative coefficients have an overall $\CF$ factor which should not undergo the conversion to $\CA$ described at the end of Section~\ref{sec:results}. We have
\begin{eqnarray}
	\gamma_1^q\big|_{C_F C_A\NG} &=& \gamma_1^q\big|_{C_F\NF},\\
	\gamma_2^q\big|_{C_F C_A^2\NG} &=& \gamma_2^q\big|_{C_F C_A\NF}+\gamma_2^q\big|_{C_F^2\NF},\\
	\gamma_2^q\big|_{C_F C_A\NF\NG} &=& 2\gamma_2^q\big|_{C_F C_A^2\NGSquare}=2\gamma_2^q\big|_{C_F\NFSquare},
\end{eqnarray}
hence the coefficients read
\begin{flalign}
    \gamma^{q}_0 &=-\frac{3 C_F}{2}\,, &&\\
    \gamma^{q}_1 &=\left(\frac{13 \zeta _3}{2}-\frac{11 \pi ^2}{24}-\frac{961}{216}\right) C_A C_F
    +\left(\frac{65}{108}+\frac{\pi ^2}{12}\right) C_F \NF\nonumber&&\\&+\left(-6 \zeta _3+\frac{\pi ^2}{2}-\frac{3}{8}\right) C_F^2+\left(\frac{65}{108}+\frac{\pi ^2}{12}\right) C_A C_F \NG\,,&&\\
    \label{eq:anq3loop}\gamma^{q}_2 &=\left(-\frac{241 \zeta _3}{54}+\frac{11 \pi ^4}{360}+\frac{1297 \pi ^2}{1944}-\frac{8659}{5832}\right) C_A C_F \NF\nonumber&&\\&+\left(-\frac{1}{3} \pi ^2 \zeta _3-\frac{211 \zeta _3}{6}-15 \zeta _5+\frac{247 \pi
   ^4}{1080}+\frac{205 \pi ^2}{72}-\frac{151}{32}\right) C_A C_F^2\nonumber&&\\
   &+\left(-\frac{11}{18} \pi ^2 \zeta _3+\frac{1763 \zeta _3}{36}-17 \zeta _5-\frac{83 \pi ^4}{720}-\frac{7163 \pi ^2}{3888}-\frac{139345}{23328}\right)
   C_A^2 C_F\nonumber&&\\&+\left(\frac{32 \zeta _3}{9}-\frac{7 \pi ^4}{108}-\frac{13 \pi ^2}{72}+\frac{2953}{432}\right) C_F^2 \NF\nonumber&&\\
   &+\left(-\frac{\zeta _3}{27}-\frac{5 \pi ^2}{108}+\frac{2417}{5832}\right) C_F \NFSquare+\left(\frac{2 \pi
   ^2 \zeta _3}{3}-\frac{17 \zeta _3}{2}+30 \zeta _5-\frac{\pi ^4}{5}-\frac{3 \pi ^2}{8}-\frac{29}{16}\right) C_F^3\nonumber&&\\& 
   +\left(\frac{62413}{11664}+\frac{473}{972}\pi^2-\frac{49}{54}\zeta_3-\frac{37}{1080}\pi^4\right)C_A^2 C_F\NG 
   \nonumber&&\\&
   +\left(\frac{2417}{5832}-\frac{5}{108}\pi^2-\frac{\zeta_3}{27}\right)C_A^2 C_F \NGSquare+\left(\frac{2417}{2916}-\frac{5}{54}\pi^2-\frac{2\zeta_3}{27}\right)C_A C_F \NF \NG
    \,.&&
\end{flalign}
We observe that to the loop order considered here, the quark and gluino anomalous dimensions coincide when $\CF$ is mapped to $\CA$ in both, effectively washing out the difference between the two fermions with different SU($3$) transformation properties,
\begin{align}\label{eq:qglurel}
\gamma_\ell^\glu\big|_{\CF\to\CA} = \gamma_\ell^q\big|_{\CF\to\CA}\,.
\end{align}

In the $\mathcal{N}=1$ supersymmetric Yang-Mills theory, the gluino and gluon anomalous dimensions are in fact equal. This was observed to two-loop order in~\cite{Bern:2003ck} in the FDH scheme. Since we use CDR, we can verify this relation only at first perturbative order: $\gamma_0^\glu=\gamma_0^g$ if $\NF=0$ and $\NG=1$.
	
In order to extend the observations about the infrared singularities to real emission layers along the lines of the analysis presented in~\cite{Chen:2023fba}, we summarize in Table~\ref{tab:polesNXg} the coefficients of the deepest infrared poles for each contribution to~\eqref{eqn:matelems} organized according to final-state particles. For the following discussion, it is more convenient to use the constants $C_A$ and $C_F$ instead of powers of $N$.
\begin{table}[t]
	\centering
	\begin{tabular}{c c c c c c c}
		\toprule
		& Final-state $\mathcal{I}$ &
		$C_A^3$ & $C_A^2 \NF$ & $C_A C_F \NF$ & $C_F^2 \NF$ & $C_A^3 \NG$ \\
		\midrule
		\midrule
		\multirow{1}{*}{{\footnotesize VVV}} & {\footnotesize $\glu g$}
		& $-\frac{4}{3}\frac{1}{\e^6} -\frac{73}{6}\frac{1}{\e^5}$
		& $+\frac{5}{3}\frac{1}{\e^5}$
		&
		&
		& $+\frac{5}{3}\frac{1}{\e^5}$
		\\
		\midrule
		\multirow{3}{*}[-0.2em]{{\footnotesize VVR}} & {\footnotesize $\glu g g$}
		& $+\frac{46}{9}\frac{1}{\e^6} +\frac{4333}{108}\frac{1}{\e^5}$
		& $-\frac{112}{27}\frac{1}{\e^5}$
		&
		&
		& $-\frac{112}{27}\frac{1}{\e^5}$
		\\
		\cmidrule{2-7}
		& {\footnotesize $\glu q \bar{q}$}
		&
		& $-\frac{7}{18}\frac{1}{\e^5}$
		& $-\frac{4}{9}\frac{1}{\e^5}$
		& $-\frac{2}{9}\frac{1}{\e^5}$
		&
		\\
		\cmidrule{2-7}
		& {\footnotesize $\glu \glu \glu + \glu \glu' \glu'$}
		&
		&
		&
		&
		& $-\frac{19}{18}\frac{1}{\e^5}$
		\\
		\midrule
		\multirow{3}{*}[-0.4em]{{\footnotesize VRR}} & {\footnotesize $\glu g g g$}
		& $-\frac{113}{18}\frac{1}{\e^6} -43\frac{1}{\e^5}$
		& $+\frac{5}{2}\frac{1}{\e^5}$
		&
		&
		& $+\frac{5}{2}\frac{1}{\e^5}$
		\\
		\cmidrule{2-7}
		& {\footnotesize $\glu q \bar{q} g$}
		&
		& $+\frac{139}{108}\frac{1}{\e^5}$
		& $+\frac{55}{54}\frac{1}{\e^5}$
		& $+\frac{4}{9}\frac{1}{\e^5}$
		&
		\\
		\cmidrule{2-7}
		& {\footnotesize $\glu \glu \glu g + \glu \glu' \glu' g$}
		&
		&
		&
		&
		& $+\frac{11}{4}\frac{1}{\e^5}$
		\\
		\midrule
		\multirow{6}{*}[-1em]{{\footnotesize RRR}} & {\footnotesize $\glu gggg$}
		& $+\frac{5}{2}\frac{1}{\e^6} + \frac{1625}{108}\frac{1}{\e^5}$
		&
		&
		&
		&
		\\
		\cmidrule{2-7}
		& {\footnotesize $\glu q\bar{q} gg$}
		&
		& $-\frac{11}{12}\frac{1}{\e^5}$
		& $-\frac{31}{54}\frac{1}{\e^5}$
		& $-\frac{2}{9}\frac{1}{\e^5}$
		&
		\\
		\cmidrule{2-7}
		& {\footnotesize $\glu q\bar{q} q'\bar{q}' + \glu q\bar{q}q\bar{q}$}
		&
		&
		&
		&
		&
		\\
		\cmidrule{2-7}
		& {\footnotesize $\glu \glu' \glu' \glu'' \glu'' + \glu \glu' \glu' \glu' \glu'$}
		& \multirow{2}{*}{}
		&
		&
		&
		&
		\\
		& {\footnotesize $+ \glu \glu \glu \glu' \glu' + \glu \glu \glu \glu \glu$}
		&
		&
		&
		&
		&
		\\
		\cmidrule{2-7}
		& {\footnotesize $\glu \glu' \glu' g g + \glu \glu \glu g g$}
		&
		&
		&
		&
		& $-\frac{185}{108}\frac{1}{\e^5}$
		\\
		\bottomrule
	\end{tabular}
	\caption{Coefficients of $\e^{-6}$ and $\e^{-5}$ poles for different colour
		factors of $\tilde{\chi} \to \glu g$ at N$^3$LO. The sum of the coefficients in each column vanishes. Blank cells indicate vanishing
		coefficients for these poles.}
	\label{tab:polesNXg}
\end{table}

First of all, it can be easily checked that for each layer the coefficients of the poles in the $C_A^3\NG$ colour factor can be obtained summing the coefficients of $C_A^2\NF$, $C_A C_F\NF$ and $C_F^2\NF$. This trivially follows from equations~\eqref{eq:relstart}-\eqref{eq:relend}. The poles in the $C_F^2\NF$ colour factor feature the usual
$1-2+1$ pattern of cancellation among the the VVR, VRR and RRR layers, typical of abelian emissions, described in detail in~\cite{Chen:2023fba}.

\begin{table}[t]
  \centering
  \begin{tabular}{c c c c c c}
    \toprule
    & Final-state $\mathcal{I}$ &
    $\CA^3$ & $\NF \CA^2$ & $\NF \CA \CF$ & $\NF \CF^2$ \\
    \midrule
    \midrule
    \multirow{1}{*}{{\footnotesize VVV}} & {\footnotesize $gg$}
    & $-\frac{4}{3}\frac{1}{\e^6} -\frac{77}{6}\frac{1}{\e^5}$
    & $+\frac{7}{3}\frac{1}{\e^5}$
    &
    &
    \\
    \midrule
    \multirow{2}{*}[-0.2em]{{\footnotesize VVR}} & {\footnotesize $ggg$}
    & $+\frac{46}{9}\frac{1}{\e^6} +\frac{4609}{108}\frac{1}{\e^5}$
    & $-\frac{305}{54}\frac{1}{\e^5}$
    &
    &
    \\
    \cmidrule{2-6}
    & {\footnotesize $q\bar{q}g$}
    &
    & $-\frac{7}{9}\frac{1}{\e^5}$
    & $-\frac{8}{9}\frac{1}{\e^5}$
    & $-\frac{4}{9}\frac{1}{\e^5}$
    \\
    \midrule
    \multirow{3}{*}[-0.4em]{{\footnotesize VRR}} & {\footnotesize $gggg$}
    & $-\frac{113}{18}\frac{1}{\e^6} -\frac{1661}{36}\frac{1}{\e^5}$
    & $+\frac{10}{3}\frac{1}{\e^5}$
    &
    &
    \\
    \cmidrule{2-6}
    & {\footnotesize $q\bar{q}gg$}
    &
    & $+\frac{119}{54}\frac{1}{\e^5}$
    & $+\frac{53}{27}\frac{1}{\e^5}$
    & $+\frac{8}{9}\frac{1}{\e^5}$
    \\
    \cmidrule{2-6}
    & {\footnotesize $q\bar{q}q'\bar{q}' + q\bar{q}q\bar{q}$}
    &
    &
    &
    &
    \\
    \midrule
    \multirow{3}{*}[-0.4em]{{\footnotesize RRR}} & {\footnotesize $ggggg$}
    & $+\frac{5}{2}\frac{1}{\e^6} +\frac{440}{27}\frac{1}{\e^5}$
    &
    &
    &
    \\
    \cmidrule{2-6}
    & {\footnotesize $q\bar{q}ggg$}
    &
    & $-\frac{13}{9}\frac{1}{\e^5}$
    & $-\frac{29}{27}\frac{1}{\e^5}$
    & $-\frac{4}{9}\frac{1}{\e^5}$
    \\
    \cmidrule{2-6}
    & {\footnotesize $q\bar{q}q'\bar{q}'g + q\bar{q}q\bar{q}g$}
    &
    &
    &
    &
    \\
    \bottomrule
  \end{tabular}
  \caption{Coefficients of $\e^{-6}$ and $\e^{-5}$ poles for different colour
    factors of $H \to gg$ at N$^3$LO. Blank cells indicate vanishing
    coefficients for these poles. The sum of the coefficients in each column vanishes. Adapted from Table~2 of~\cite{Chen:2023fba}.}
  \label{tab:polesHgg}
\end{table}
\begin{table}[t]
  \centering
  \begin{tabular}{c c c c c c c}
    \toprule
    & Final-state $\mathcal{I}$ &
    $\CA^2 \CF$ & $\CA \CF^2$ & $\CF^3$ & $\NF \CA \CF$ & $\NF \CF^2$ \\
    \midrule
    \midrule
    \multirow{1}{*}{{\footnotesize VVV}} & {\footnotesize $q\bar{q}$}
    &
    & $-\frac{11}{2}\frac{1}{\e^5}$
    & $-\frac{4}{3}\frac{1}{\e^6} - \frac{6}{\e^5}$
    &
    & $+\frac{1}{\e^5}$
    \\
    \midrule
    \multirow{1}{*}{{\footnotesize VVR}} & {\footnotesize $q\bar{q}g$}
    & $+\frac{1}{9}\frac{1}{\e^6} +\frac{241}{108}\frac{1}{\e^5}$
    & $+\frac{1}{\e^6} +\frac{52}{3}\frac{1}{\e^5}$
    & $+\frac{4}{\e^6} + \frac{18}{\e^5}$
    & $-\frac{17}{54}\frac{1}{\e^5}$
    & $-\frac{7}{3}\frac{1}{\e^5}$
    \\
    \midrule
    \multirow{2}{*}[-0.2em]{{\footnotesize VRR}} & {\footnotesize $q\bar{q}gg$}
    & $-\frac{5}{18}\frac{1}{\e^6} -\frac{133}{36}\frac{1}{\e^5}$
    & $-\frac{2}{\e^6} -\frac{109}{6}\frac{1}{\e^5}$
    & $-\frac{4}{\e^6} - \frac{18}{\e^5}$
    & $+\frac{1}{3}\frac{1}{\e^5}$
    & $+\frac{4}{3}\frac{1}{\e^5}$
    \\
    \cmidrule{2-7}
    & {\footnotesize $q\bar{q}q'\bar{q}' + q\bar{q}q\bar{q}$}
    &
    &
    &
    & $+\frac{1}{27}\frac{1}{\e^5}$
    & $+\frac{11}{27}\frac{1}{\e^5}$
    \\
    \midrule
    \multirow{2}{*}[-0.2em]{{\footnotesize RRR}} & {\footnotesize $q\bar{q}ggg$}
    & $+\frac{1}{6}\frac{1}{\e^6} +\frac{79}{54}\frac{1}{\e^5}$
    & $+\frac{1}{\e^6} +\frac{19}{3}\frac{1}{\e^5}$
    & $+\frac{4}{3}\frac{1}{\e^6} +\frac{6}{\e^5}$
    &
    &
    \\
    \cmidrule{2-7}
    & {\footnotesize $q\bar{q}q'\bar{q}'g + q\bar{q}q\bar{q}g$}
    &
    &
    &
    & $-\frac{1}{18}\frac{1}{\e^5}$
    & $-\frac{11}{27}\frac{1}{\e^5}$
    \\
    \bottomrule
  \end{tabular}
  \caption{Coefficients of $\e^{-6}$ and $\e^{-5}$ poles for different
    colour factors of $H \to b\bar{b}$ at N$^3$LO. Blank cells
    indicate vanishing coefficients for these poles. The sum of the coefficients in each column vanishes. Adapted from
    Table~1 of~\cite{Chen:2023fba}. Note that the overall factor of
    $2\CF$ that was factored out in~\cite{Chen:2023fba} is explicitly
    included here.}
  \label{tab:polesHbb}
\end{table}

It is instructive to compare the highest poles of $\tilde{\chi} \to \glu g$ with the highest
poles of $H \to gg$ and $H \to b\bar{b}$ (massless bottom quarks), presented in~\cite{Chen:2023fba}. Namely, we want to relate the infrared poles of a fermion-vector boson colour dipole with those of a vector-vector and a fermion-antifermion dipole.
For ease of reference, we re-print Table 1 and Table 2 from~\cite{Chen:2023fba} as Table~\ref{tab:polesHgg} and Table~\ref{tab:polesHbb} below. In Table~\ref{tab:polesHbb} we restore an overall factor $2C_F$ which was omitted in~\cite{Chen:2023fba} as part of the chosen normalization.

\newcommand{\Pol}{{\cal P}}

We introduce the notation $\mathcal{P}\left(\cdot\right)$ to extract the $\e^{-6}$ and $\e^{-5}$ poles. For any layer, summing over partonic final states, we find
\begin{eqnarray}
	\label{eq:poles_rel1}&&\hspace{-1cm}\Pol\left(\tilde{\chi}\to g\tilde{g}\right)\Big|_{C_A^3}=\dfrac{1}{2}\Bigg[\Pol\left(H\to gg\right)\Big|_{C_A^3}\nn\\
	&&\hspace{1.5cm}+\Pol\left(H\to b\bar{b} \right)\Big|_{C_F C_A^2}+\Pol\left(H\to b\bar{b} \right)\Big|_{C_F^2 C_A}+\Pol\left(H\to b\bar{b} \right)\Big|_{C_F^3}\Bigg],\\
	\label{eq:poles_rel2}&&\hspace{-1cm}\Pol\left(\tilde{\chi}\to g\tilde{g}\right)\Big|_{C_A^2\NF}+\Pol\left(\tilde{\chi}\to g\tilde{g}\right)\Big|_{C_A C_F\NF}+\Pol\left(\tilde{\chi}\to g\tilde{g}\right)\Big|_{C_F^2\NF}=\\
	&&\hspace{0.5cm}\dfrac{1}{2}\Bigg[\Pol\left(H\to gg\right)\Big|_{C_A^2 \NF}+\Pol\left(H\to gg\right)\Big|_{C_A C_F \NF}+\Pol\left(H\to gg\right)\Big|_{C_F^2\NF}\nn\\
	&&\hspace{3cm}+\Pol\left(H\to b\bar{b} \right)\Big|_{C_A C_F \NF}+\Pol\left(H\to b\bar{b} \right)\Big|_{C_F^2 \NF}\Bigg],
\end{eqnarray}
which indicates that after replacing $C_F$ with $C_A$ in the $H\to b\bar{b}$ result, the deepest infrared poles for radiation from a fermion-vector dipole are given by the average of the singularities in radiation from vector-vector and fermion-antifermion dipoles. We verified that equations analogous to~\eqref{eq:poles_rel1} and~\eqref{eq:poles_rel2} hold also for all layers at NLO and NNLO. This is explained by the fact that the deepest infrared singularities receive distinct contributions from the two hard partons.

\subsection{Applications in higher-order QCD calculations}\label{sec:QCD_applications}
In this section we discuss how the presented results can be adjusted for use in higher-order calculations in QCD. In~\cite{Gehrmann-DeRidder:2005svg}, it was shown that the matrix elements for the decay of a neutralino into a gluino and a gluon can be employed for the analysis of QCD radiation from a dipole formed by a hard quark and a hard gluon. This property was exploited to extract NLO and NNLO quark-gluon antenna functions~\cite{Kosower:1997zr,Campbell:1998nn,Gehrmann-DeRidder:2005btv} from physical matrix elements. 

In~\cite{Gehrmann-DeRidder:2005svg}, the matrix elements for neutralino decay were adapted to obtain quark-gluon antenna functions by discarding any contribution coming from final states with multiple gluinos, on top of neglecting any internal gluino loop. This strategy effectively retains the QCD (quarks and gluons) component of the radiation from the hard gluino.  We can recover the results in~\cite{Gehrmann-DeRidder:2005svg} and extend the same procedure to N$^3$LO by choosing
\begin{equation}\label{eq:restricted}
	\NG=0\quad\text{and}\quad\Delta{\cal T}^{(k)}_{\glu \glu \glu (ij)}=0\quad\text{and}\quad\Delta{\cal T}^{(3)}_{\glu \glu \glu \glu \glu}=0
\end{equation}
in the expressions provided in Appendix~\ref{app:results} and~\ref{app:exprlower}. With~\eqref{eq:restricted}, the results presented in this paper can be directly used for the definition of integrated quark-gluon antenna functions for final-state radiation at N$^3$LO.

\begin{figure}
	\centering
	\begin{subfigure}[t]{.99\linewidth}
		\centering
		\includegraphics[width=0.4\linewidth]{Figures/SE1-crop.pdf}
		\caption{}
	\end{subfigure}
	
	\begin{subfigure}[b]{.49\linewidth}
		\centering
		\includegraphics[width=0.8\linewidth]{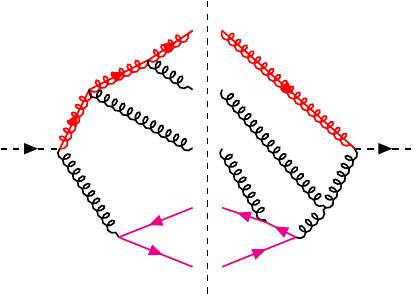}
		\caption{}
	\end{subfigure}
	\begin{subfigure}[b]{.49\linewidth}
		\centering
		\includegraphics[width=0.8\linewidth]{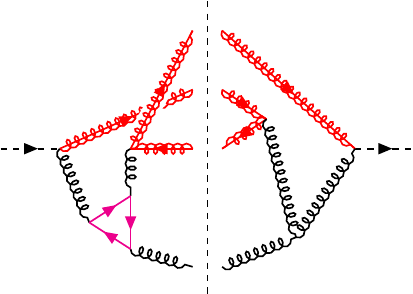}
		\caption{}
	\end{subfigure}
	\caption{The example self-energy diagram (a), considered in Section~\ref{sec:renorm} in Fig. 1(a) has both (b) a five-propagator cut with a single gluino on the cut and (c) a four-propagator cut with 3 gluinos originating from the hard gluino line in the final state. The former is considered as QCD radiation from the gluon-gluino dipole, the latter has to be discarded.}\label{fig:cuts_restricted}
\end{figure}

We can elaborate further on~\eqref{eq:restricted} and the interplay we observe between the layers of~\eqref{eqn:matelems} once this condition is imposed. Setting $\NG=0$ preserves any check and result discussed in the previous sections and is a consistent way to discard the contribution of cut or uncut closed gluino lines at the level of the unrenormalized matrix elements, renormalization coefficients, or anomalous dimensions. On the other hand, discarding specific final states, as prescribed by the second part of~\eqref{eq:restricted}, has implications for the validation against results obtained via the optical theorem in Section~\ref{sec:renorm}. As shown in Figure~\ref{fig:cuts_restricted}, whether a given cut of a self-energy diagram is kept or not depends on the particular choice of the cut propagators. Indeed, discarding multiple gluino final states affects the three-, four- and five-particles matrix elements at each perturbative order, while leaving the two-particle final-state intact.

In particular, one should not expect the full cancellation of infrared poles to be preserved in the full sum over different-multiplicities final states after the restrictions are imposed. At NLO, the removed term $\Delta{\cal T}^{(1)}_{\glu \glu \glu}$ happens to be infrared-finite, so the sum over the two- and three-particle final states is finite as well, but this is not the case at NNLO and N$^3$LO.

In~\cite{Gehrmann-DeRidder:2005svg} the authors argue that it is possible to define the infrared pole structure of the two-particle final-state at NNLO, namely the renormalized two-loop correction to $\tilde{\chi}\to \glu g$, in such a way that the full cancellation of infrared singularities is recovered in the sum with the integrated real emission layers. From our explicit calculation, we find that in order to obtain the quoted infrared structure of~\cite{Gehrmann-DeRidder:2005svg}, one would need to consider a different leading-colour term for the two-loop matrix element:
\begin{equation}\label{eq:ANTX22}
	{\cal T}^{(2)}_{\glu g}\Big|_{N^{2}}\to {\cal T}^{(2)}_{\glu g}\Big|_{N^{2}} + \Delta{\cal T}^{(2)}_{\glu \glu \glu}\Big|_{N^{2}} + \Delta{\cal T}^{(2)}_{\glu \glu \glu g}\Big|_{N^{2}}
\end{equation}
to effectively reabsorb the infrared poles which are removed in the three- and four-particle final states. Such a discrepancy is also the reason why the two-loop gluino anomalous dimension obtained in~\cite{Gehrmann-DeRidder:2005svg} differs from the one we give in~\eqref{eq:anglu2loop}, even after considering $\NG=0$. In~\cite{Gehrmann-DeRidder:2005svg}, the gluino hard collinear function~\cite{Catani:1998bh,Becher:2009cu} $H^{(2)}_\glu$ is extracted from the considered infrared singularity structure of~\eqref{eq:ANTX22} reads\footnote{We confirmed with the authors that there is a misprint in~(5.15) of~\cite{Gehrmann-DeRidder:2005svg}.}
\begin{equation}\label{H2gluino}
	H^{(2)}_\glu= \dfrac{e^{\e\gamma_E}}{4\Gamma(1-\e)\e}\left\lbrace  \left[-\dfrac{187}{216}+\dfrac{13}{48}\pi^2-\dfrac{1}{2}\zeta_3\right]C_A^2+C_A\NF\left[-\dfrac{25}{108}+\dfrac{\pi^2}{24}\right]\right\rbrace.
\end{equation}
Exploiting the explicit expression of the hard collinear function~\cite{Becher:2009cu}
\begin{equation}
	H^{(2)}_\glu=\frac{1}{4\e}\left(\gamma_1^\glu-\dfrac{\gamma_1^K}{\gamma_0^K}\gamma_0^\glu+\dfrac{\pi^2}{16}\beta_0\gamma_0^K C_A\right)\,,
\end{equation}
it is possible to extract the following gluino two-loop collinear anomalous dimension (${\NG=0}$):
\begin{flalign}
	\label{eq:anglu2loopANT}\gamma^{\glu}_{1,restricted} &= \left(-\frac{\zeta _3}{2}+\frac{7 \pi ^2}{24}-\frac{1393}{216}\right) C_A^2+\left(\frac{65}{108}+\frac{\pi ^2}{12}\right) C_A \NF\,,
\end{flalign}
where the subscript \textit{restricted} indicate that the quantity has been obtained after imposing~\eqref{eq:restricted}.
One can verify that
\begin{equation}
	\gamma^{\glu}_{1}-\gamma^{\glu}_{1,restricted}=-\frac{1}{4}\left(\Delta{\cal T}^{(2)}_{\glu \glu \glu}\Big|_{N^{2}} + \Delta{\cal T}^{(2)}_{\glu \glu \glu g}\Big|_{N^{2}}\right)+\mathcal{O}(\e)\,.
\end{equation}
The difference in the two anomalous dimensions is due to missing same-flavour triple-collinear limits $\glu\parallel\glu\parallel\glu$ in $\gamma^{\glu}_{1,restricted}$, which are the only type of infrared singularity present in $\Delta{\cal T}^{(2)}_{\glu \glu \glu}\Big|_{N^{2}}$ and $\Delta{\cal T}^{(2)}_{\glu \glu \glu g}\Big|_{N^{2}}$.

In the context of the antenna subtraction method at NNLO, the same-flavour $q\parallel\bar{q}\parallel q$ triple collinear limit (which is completely analogous to the $\glu\parallel\glu\parallel\glu$ limit) is removed in double-real emission corrections by means of the quark-antiquark antenna function $C_4^0$~\cite{Gehrmann-DeRidder:2005btv}. For this reason there has been no need to retain such a configuration in the definition of quark-gluon antenna functions. 
Due to the complicated infrared structure at N$^3$LO, only the first applications will indicate whether it is more advantageous to keep or remove the multiple gluino final states in the definition of the quark-gluon antenna functions.
Nevertheless, with the results presented in this paper, we expose clear relations between the infrared behaviour of the radiation from a gluino and off a quark after accounting for systematic differences in colour factors. 
Moreover, the full decomposition into partonic channels and colour factors of the integrated results gives us control over different infrared limits and therefore versatility in defining the precise transition from neutralino decay matrix elements and quark-gluon antenna functions in the future.

%%%%%%%%%%%%%%%%%%%%%%%%%%%%%%%%%%%%%%%%%%%%%%%%%%%%%%%%%%%%%%%%%%%%%%%%%%%%%%%%
\section{Conclusions}\label{sec:concl}

In this paper, we extended the computation presented in~\cite{Gehrmann-DeRidder:2005svg} to higher perturbative order by computing the QCD corrections to the decay of a neutralino to gluinos and partons at $\mathcal{O}(\alpha_s^3)$.
The main result of this work lies in the analytically integrated tree-level five-parton, one-loop four-parton, two-loop three-parton, and three-loop two-parton matrix elements for this process, separated by final-state particle species and colour factors.
The relation between the colour algebra of a gluino and a quark-antiquark pair with the same momenta~\cite{Gehrmann-DeRidder:2005svg} permits the use of the integrated matrix elements as integrated quark-gluon antenna functions in the antenna subtraction scheme.
Together with the quark-antiquark antenna functions from photon decay~\cite{Jakubcik:2022zdi} and the gluon-gluon antenna functions from Higgs decay~\cite{Chen:2023fba}, the results presented here complete the set of integrated antenna functions in final-final kinematics at $\mathcal{O}(\alpha_s^3)$.

In order to perform checks on our results, we extended the QCD Lagrangian by $\NG$ flavours of gluinos. In this theory, some of the renormalization coefficients of the couplings were not known and we derived them by considering higher-order corrections to the photon and neutrino propagators.
Having identified the UV and IR poles of the matrix elements, we were able to extract the gluino contributions to the cusp, quark and gluon anomalous dimensions and the gluino collinear anomalous dimension up to $\mathcal{O}(\alpha_s^3)$. Setting $\NF=0$ and $\NG=1$, we recovered a known result for the cusp anomalous dimension in $\mathcal{N}=1$ super Yang-Mills theory. Simple relations between the universal ingredients which we identified imply similarities in the IR structure of the amplitudes for the decay of a colour singlet to $\glu g$ and the decay of a colour singlet to $q\bar{q}$ and $gg$.

With all the analytical ingredients available, we are in the position to assemble a candidate extension of the antenna subtraction scheme for the calculation of the N$^3$LO correction for processes with final-state QCD radiation. %
Of course, the definition of a consistent set of subtraction terms at N$^3$LO is challenging. 
In the context of antenna subtraction, general observations can be made about the kind of structures we expect to appear in the subtraction term for each layer of the calculation~\cite{Marcoli:2023xtc}. Nonetheless, it is a priori not clear how to avoid over-subtraction due to the proliferation of possible combinations of NLO and NNLO antenna functions and matrix elements.
In practice (already at NNLO), antenna functions derived from neutralino decay contain an unphysical triple-collinear limit due to the cyclicity of the colour structure of gluino-gluon matrix element~\cite{Gehrmann-DeRidder:2005svg,Gehrmann-DeRidder:2007foh}. Such limits produce spurious infrared poles at the integrated level, which can however be systematically removed~\cite{MMTHESIS}.
Recently, a framework to compute \textit{idealised} antenna functions without spurious infrared limits has been proposed in~\cite{Braun-White:2023sgd,Braun-White:2023zwd,Fox:2023bma} and applied up to NNLO. 

The completion of the computation of antenna functions in final-final kinematics opens the door to the formulation of the first local subtraction scheme at N$^3$LO in the foreseeable future.
Jet production at lepton colliders is the ideal target for the implementation of such a subtraction scheme. In fact, all of the multiloop matrix elements for two-\cite{Gehrmann:2010ue, Garland:2001tf} and three-jet production~\cite{Gehrmann:2023jyv, Abreu:2021asb} (in the leading colour approximation) have recently become available. The calculation of jet observables at a new level of precision compared to present results~\cite{Gehrmann-DeRidder:2007foh,Weinzierl:2008iv,DelDuca:2016ily} is thus within reach. 

\section*{Acknowledgements}
We are indebted to Aude Gehrmann-De Ridder, Thomas Gehrmann and Nigel Glover for their feedback and encouragement
to pursue this work. We thank Oscar Braun-White, Kay Sch\"onwald, Tong-Zhi Yang and HanTian Zhang for enlightening discussions.
This work was supported
by the Swiss National Science Foundation (SNF) under contract 200020-204200,
by the UK Science and Technology
Facilities Council under contract ST/X000745/1
and by the European Research Council (ERC) under the
European Union's Horizon 2020 research and innovation programme grant agreement
101019620 (ERC Advanced Grant TOPUP). 

%%%%%%%%%%%%%%%%%%%%%%%%%%%%%%%%%%%%%%%%%%%%%%%%%%%%%%%%%%%%%%%%%%%%%%%%%%%%%%%%
\appendices

\section{The renormalization of amplitudes}\label{app:ren}

Using~\eqref{eq:alfaren} and~\eqref{eq:etarenorm} for the renormalization of $\alpha_s$ and the effective coupling $\eta$, the $\ell$-loop amplitude $|{\cal M}^{(\ell)}\rangle_{\cal I}$, where $\mathcal{I}$ indicates a generic final state, is renormalized as
\begingroup
\allowdisplaybreaks
\setlength{\jot}{12pt}
\begin{eqnarray}
	|{\cal M}^{(1)}\rangle_{ij} &=& |{\cal M}^{(1),U}\rangle_{ij}+\Zren{1}|{\cal M}^{(0)}\rangle_{ij}\,,\\
	|{\cal M}^{(2)}\rangle_{ij} &=& |{\cal M}^{(2),U}\rangle_{ij}+\left(\Zren{1} - \frac{\beta_0}{\e} \right)|{\cal M}^{(1),U}\rangle_{ij}+ \Zren{2}|{\cal M}^{(0)}\rangle_{ij}\,,\\
	|{\cal M}^{(3)}\rangle_{ij} &=& |{\cal M}^{(3),U}\rangle_{ij}+\left(\Zren{1} - \frac{2\beta_0}{\e} \right) |{\cal M}^{(2),U}\rangle_{ij}\nonumber \\
	&&+\left(\Zren{2} -\frac{\Zren{1}\beta_0}{\e}+\frac{\beta_0^2}{\e^2} - \frac{\beta_1}{2\e} \right) |{\cal M}^{(1),U}\rangle_{ij}+\Zren{3}|{\cal M}^{(0)}\rangle_{ij}\,,\\
	|{\cal M}^{(1)}\rangle_{ijk} &=& |{\cal M}^{(1),U}\rangle_{ijk}+\left(\Zren{1}-\frac{\beta_0}{2\e}\right)|{\cal M}^{(0)}\rangle_{ijk}\,,\\
	|{\cal M}^{(2)}\rangle_{ijk} &=& |{\cal M}^{(2),U}\rangle_{ijk}+\left(\Zren{1} - \frac{3\beta_0}{2\e} \right)|{\cal M}^{(1),U}\rangle_{ij}\nonumber \\
	&&+\left(\Zren{2}-\frac{\Zren{1}\beta_0}{2\e}+\frac{3\beta_0^2}{8\e^2} - \frac{\beta_1}{4\e}\right)|{\cal M}^{(0)}\rangle_{ijk}\,,\\
	|{\cal M}^{(1)}\rangle_{ijkl} &=& |{\cal M}^{(1),U}\rangle_{ijkl}+\left(\Zren{1}-\frac{\beta_0}{\e}\right)|{\cal M}^{(0)}\rangle_{ijkl}\,,
\end{eqnarray}
\endgroup
where the superscript $U$ denotes unrenormalised quantities.

%%%%%%%%%%%%%%%%%%%%%%%%%%%%%%%%%%%%%%%%%%%%%%%%%%%%%%%%%%%%%%%%%%%%%%%%%%%%%%%%

\section{N$^3$LO results}\label{app:results}

\subsection{Two-particle final states}

%NXg33x0

%N^{3}
\begin{flalign}
{\cal T}^{(3,\left[3\times 0\right])}_{\glu g}\Big|_{N^{3}} = &
+\frac{1}{\e^6}\left(-\frac{1}{3}\right)
+\frac{1}{\e^5}\left(-\frac{53}{12}\right)
+\frac{1}{\e^4}\left(-\frac{4103}{324}+\frac{3}{2}\pi^2\right)
\nonumber &\\&
+\frac{1}{\e^3}\left(\frac{7109}{1944}+\frac{8609}{1296}\pi^2+\frac{11}{6}\zeta_{3}\right)
\nonumber &\\&
+\frac{1}{\e^2}\left(\frac{1699}{3888}-\frac{23429}{3888}\pi^2+\frac{2011}{108}\zeta_{3}-\frac{14333}{12960}\pi^4\right)
\nonumber &\\&
+\frac{1}{\e}\left(-\frac{6608795}{69984}-\frac{331013}{23328}\pi^2+\frac{4195}{108}\zeta_{3}+\frac{21737}{51840}\pi^4-\frac{1867}{216}\pi^2\zeta_{3}-\frac{439}{30}\zeta_{5}\right)
\nonumber &\\&
-\frac{67273981}{419904}+\frac{8587951}{139968}\pi^2+\frac{58957}{1944}\zeta_{3}+\frac{319535}{31104}\pi^4-\frac{13469}{432}\pi^2\zeta_{3}
\nonumber &\\&
+\frac{2339}{36}\zeta_{5}+\frac{18101}{116640}\pi^6-\frac{883}{18}\zeta_{3}^2 + \mathcal{O}(\e), &
\end{flalign}

%NF N^{2}
\begin{flalign}
{\cal T}^{(3,\left[3\times 0\right])}_{\glu g}\Big|_{\NF N^{2}} = &
+\frac{1}{\e^5}\left(\frac{2}{3}\right)
+\frac{1}{\e^4}\left(\frac{2543}{648}\right)
+\frac{1}{\e^3}\left(\frac{251}{1944}-\frac{157}{162}\pi^2\right)
\nonumber &\\&
+\frac{1}{\e^2}\left(-\frac{6035}{1944}+\frac{199}{486}\pi^2-\frac{133}{108}\zeta_{3}\right)
\nonumber &\\&
+\frac{1}{\e}\left(\frac{1401337}{69984}+\frac{15611}{5832}\pi^2+\frac{61}{162}\zeta_{3}+\frac{977}{25920}\pi^4\right)
\nonumber &\\&
+\frac{1062379}{209952}-\frac{1817077}{139968}\pi^2+\frac{3707}{162}\zeta_{3}-\frac{150553}{155520}\pi^4
\nonumber &\\&
-\frac{161}{54}\pi^2\zeta_{3}-\frac{19}{60}\zeta_{5} + \mathcal{O}(\e), &
\end{flalign}

%NF N^{0}
\begin{flalign}
{\cal T}^{(3,\left[3\times 0\right])}_{\glu g}\Big|_{\NF} = &
+\frac{1}{\e^3}\left(\frac{25}{72}\right)
+\frac{1}{\e^2}\left(\frac{83}{216}+\frac{2}{9}\zeta_{3}\right)
+\frac{1}{\e}\left(-\frac{2545}{1296}-\frac{25}{288}\pi^2+\frac{65}{54}\zeta_{3}+\frac{\pi^4}{270}\right)
\nonumber &\\&
-\frac{6355}{1944}+\frac{1039}{864}\pi^2+\frac{529}{648}\zeta_{3}+\frac{19}{1620}\pi^4-\frac{19}{18}\pi^2\zeta_{3}+\frac{14}{9}\zeta_{5} + \mathcal{O}(\e), &
\end{flalign}

%NF N^{-2}
\begin{flalign}
{\cal T}^{(3,\left[3\times 0\right])}_{\glu g}\Big|_{\NF N^{-2}} = & +\frac{1}{\e}\left(-\frac{1}{96}\right), &
\end{flalign}

%NF^2 N^{1}
\begin{flalign}
{\cal T}^{(3,\left[3\times 0\right])}_{\glu g}\Big|_{\NFSquare N} = &
+\frac{1}{\e^4}\left(-\frac{49}{162}\right)
+\frac{1}{\e^3}\left(-\frac{763}{1944}\right)
\nonumber &\\&
+\frac{1}{\e^2}\left(\frac{85}{162}+\frac{77}{1296}\pi^2\right)
+\frac{1}{\e}\left(-\frac{22073}{69984}-\frac{215}{1944}\pi^2+\frac{19}{324}\zeta_{3}\right)
\nonumber &\\&
+\frac{1140755}{419904}-\frac{101}{7776}\pi^2-\frac{140}{243}\zeta_{3}-\frac{71}{10368}\pi^4 + \mathcal{O}(\e), &
\end{flalign}

%NF^2 N^{-1}
\begin{flalign}
{\cal T}^{(3,\left[3\times 0\right])}_{\glu g}\Big|_{\NFSquare N^{-1}} = &
+\frac{1}{\e^2}\left(-\frac{11}{144}\right)
+\frac{1}{\e}\left(\frac{11}{432}\right), &
\end{flalign}

%NF^3 N^{0}
\begin{flalign}
{\cal T}^{(3,\left[3\times 0\right])}_{\glu g}\Big|_{\NFCube} = &
+\frac{1}{\e^3}\left(\frac{5}{216}\right), &
\end{flalign}

%NG N^{3}
\begin{flalign}
{\cal T}^{(3,\left[3\times 0\right])}_{\glu g}\Big|_{\NG N^{3}} = &
+\frac{1}{\e^5}\left(\frac{2}{3}\right)
+\frac{1}{\e^4}\left(\frac{2543}{648}\right)
+\frac{1}{\e^3}\left(-\frac{53}{243}-\frac{157}{162}\pi^2\right)
\nonumber &\\&
+\frac{1}{\e^2}\left(-\frac{3391}{972}+\frac{199}{486}\pi^2-\frac{157}{108}\zeta_{3}\right)
\nonumber &\\&
+\frac{1}{\e}\left(\frac{769019}{34992}+\frac{64469}{23328}\pi^2-\frac{67}{81}\zeta_{3}+\frac{881}{25920}\pi^4\right)
\nonumber &\\&
+\frac{1748719}{209952}-\frac{1985395}{139968}\pi^2+\frac{14299}{648}\zeta_{3}-\frac{152377}{155520}\pi^4
\nonumber &\\&
-\frac{52}{27}\pi^2\zeta_{3}-\frac{337}{180}\zeta_{5} + \mathcal{O}(\e), &
\end{flalign}

%NG NF N^{2}
\begin{flalign}
{\cal T}^{(3,\left[3\times 0\right])}_{\glu g}\Big|_{\NG \NF N^{2}} = &
+\frac{1}{\e^4}\left(-\frac{49}{81}\right)
+\frac{1}{\e^3}\left(-\frac{763}{972}\right)
+\frac{1}{\e^2}\left(\frac{1459}{1296}+\frac{77}{648}\pi^2\right)
\nonumber &\\&
+\frac{1}{\e}\left(-\frac{5741}{8748}-\frac{215}{972}\pi^2+\frac{19}{162}\zeta_{3}\right)
\nonumber &\\&
+\frac{1140755}{209952}-\frac{101}{3888}\pi^2-\frac{280}{243}\zeta_{3}-\frac{71}{5184}\pi^4 + \mathcal{O}(\e), &
\end{flalign}

%NG NF N^{0}
\begin{flalign}
{\cal T}^{(3,\left[3\times 0\right])}_{\glu g}\Big|_{\NG \NF} = &
+\frac{1}{\e^2}\left(-\frac{11}{144}\right)
+\frac{1}{\e}\left(\frac{11}{432}\right), &
\end{flalign}

%NG NF^2 N^{1}
\begin{flalign}
{\cal T}^{(3,\left[3\times 0\right])}_{\glu g}\Big|_{\NG \NFSquare N} = &
+\frac{1}{\e^3}\left(\frac{5}{72}\right), &
\end{flalign}

%NG^2 N^{3}
\begin{flalign}
{\cal T}^{(3,\left[3\times 0\right])}_{\glu g}\Big|_{\NGSquare N^{3}} = &
+\frac{1}{\e^4}\left(-\frac{49}{162}\right)
+\frac{1}{\e^3}\left(-\frac{763}{1944}\right)
+\frac{1}{\e^2}\left(\frac{779}{1296}+\frac{77}{1296}\pi^2\right)
\nonumber &\\&
+\frac{1}{\e}\left(-\frac{23855}{69984}-\frac{215}{1944}\pi^2+\frac{19}{324}\zeta_{3}\right)
\nonumber &\\&
+\frac{1140755}{419904}-\frac{101}{7776}\pi^2-\frac{140}{243}\zeta_{3}-\frac{71}{10368}\pi^4 + \mathcal{O}(\e), &
\end{flalign}

%NG^2 NF N^{2}
\begin{flalign}
{\cal T}^{(3,\left[3\times 0\right])}_{\glu g}\Big|_{\NGSquare \NF N^{2}} = &
+\frac{1}{\e^3}\left(\frac{5}{72}\right), &
\end{flalign}

%NG^3 N^{3}
\begin{flalign}
{\cal T}^{(3,\left[3\times 0\right])}_{\glu g}\Big|_{\NGCube N^{3}} = &
+\frac{1}{\e^3}\left(\frac{5}{216}\right), &
\end{flalign}

%NXg32x1

%N^{3}
\begin{flalign}
{\cal T}^{(3,\left[2\times 1\right])}_{\glu g}\Big|_{N^{3}} = &
+\frac{1}{\e^6}\left(-1\right)
+\frac{1}{\e^5}\left(-\frac{31}{4}\right)
+\frac{1}{\e^4}\left(-\frac{127}{9}+\frac{2}{3}\pi^2\right)
\nonumber &\\&
+\frac{1}{\e^3}\left(-\frac{865}{216}+\frac{533}{144}\pi^2+\frac{13}{2}\zeta_{3}\right)
\nonumber &\\&
+\frac{1}{\e^2}\left(-\frac{25885}{1296}+\frac{1945}{432}\pi^2+\frac{1267}{36}\zeta_{3}-\frac{59}{1440}\pi^4\right)
\nonumber &\\&
+\frac{1}{\e}\left(-\frac{966289}{7776}-\frac{26681}{2592}\pi^2+\frac{5221}{108}\zeta_{3}-\frac{707}{1152}\pi^4-\frac{37}{8}\pi^2\zeta_{3}-\frac{9}{10}\zeta_{5}\right)
\nonumber &\\&
-\frac{18431953}{46656}+\frac{98035}{15552}\pi^2+\frac{99985}{648}\zeta_{3}+\frac{13937}{5760}\pi^4-\frac{6625}{432}\pi^2\zeta_{3}
\nonumber &\\&
+\frac{611}{12}\zeta_{5}-\frac{6647}{60480}\pi^6-\frac{125}{2}\zeta_{3}^2 + \mathcal{O}(\e), &
\end{flalign}

%NF N^{2}
\begin{flalign}
{\cal T}^{(3,\left[2\times 1\right])}_{\glu g}\Big|_{\NF N^{2}} = &
+\frac{1}{\e^5}\left(1\right)
+\frac{1}{\e^4}\left(\frac{85}{24}\right)
+\frac{1}{\e^3}\left(\frac{37}{24}-\frac{7}{18}\pi^2\right)
+\frac{1}{\e^2}\left(\frac{8}{9}-\frac{35}{27}\pi^2-\frac{115}{36}\zeta_{3}\right)
\nonumber &\\&
+\frac{1}{\e}\left(\frac{2965}{144}+\frac{779}{324}\pi^2-\frac{80}{27}\zeta_{3}+\frac{209}{2880}\pi^4\right)
\nonumber &\\&
+\frac{61645}{864}-\frac{10199}{15552}\pi^2+\frac{1787}{324}\zeta_{3}-\frac{4313}{17280}\pi^4
\nonumber &\\&
+\frac{35}{54}\pi^2\zeta_{3}-\frac{67}{60}\zeta_{5} + \mathcal{O}(\e), &
\end{flalign}

%NF N^{0}
\begin{flalign}
{\cal T}^{(3,\left[2\times 1\right])}_{\glu g}\Big|_{\NF} = &
+\frac{1}{\e^3}\left(\frac{1}{8}\right)
+\frac{1}{\e^2}\left(\frac{5}{24}\right)
+\frac{1}{\e}\left(-\frac{7}{96}\pi^2\right)
+\frac{1}{8}-\frac{7}{24}\zeta_{3} + \mathcal{O}(\e), &
\end{flalign}

%NF^2 N^{1}
\begin{flalign}
{\cal T}^{(3,\left[2\times 1\right])}_{\glu g}\Big|_{\NFSquare N} = &
+\frac{1}{\e^4}\left(-\frac{2}{9}\right)
+\frac{1}{\e^3}\left(-\frac{71}{216}\right)
+\frac{1}{\e^2}\left(\frac{55}{324}+\frac{37}{432}\pi^2\right)
\nonumber &\\&
+\frac{1}{\e}\left(-\frac{3701}{7776}-\frac{65}{648}\pi^2+\frac{31}{108}\zeta_{3}\right)
\nonumber &\\&
-\frac{91025}{46656}-\frac{205}{7776}\pi^2-\frac{40}{81}\zeta_{3}+\frac{37}{3456}\pi^4 + \mathcal{O}(\e), &
\end{flalign}

%NF^2 N^{-1}
\begin{flalign}
{\cal T}^{(3,\left[2\times 1\right])}_{\glu g}\Big|_{\NFSquare N^{-1}} = &
+\frac{1}{\e^2}\left(-\frac{1}{48}\right), &
\end{flalign}

%NF^3 N^{0}
\begin{flalign}
{\cal T}^{(3,\left[2\times 1\right])}_{\glu g}\Big|_{\NFCube} = &
+\frac{1}{\e^3}\left(\frac{1}{72}\right), &
\end{flalign}

%NG N^{3}
\begin{flalign}
{\cal T}^{(3,\left[2\times 1\right])}_{\glu g}\Big|_{\NG N^{3}} = &
+\frac{1}{\e^5}\left(1\right)
+\frac{1}{\e^4}\left(\frac{85}{24}\right)
+\frac{1}{\e^3}\left(\frac{17}{12}-\frac{7}{18}\pi^2\right)
+\frac{1}{\e^2}\left(\frac{49}{72}-\frac{35}{27}\pi^2-\frac{115}{36}\zeta_{3}\right)
\nonumber &\\&
+\frac{1}{\e}\left(\frac{2965}{144}+\frac{6421}{2592}\pi^2-\frac{80}{27}\zeta_{3}+\frac{209}{2880}\pi^4\right)
\nonumber &\\&
+\frac{61537}{864}-\frac{10199}{15552}\pi^2+\frac{3763}{648}\zeta_{3}-\frac{4313}{17280}\pi^4
\nonumber &\\&
+\frac{35}{54}\pi^2\zeta_{3}-\frac{67}{60}\zeta_{5} + \mathcal{O}(\e), &
\end{flalign}

%NG NF N^{2}
\begin{flalign}
{\cal T}^{(3,\left[2\times 1\right])}_{\glu g}\Big|_{\NG \NF N^{2}} = &
+\frac{1}{\e^4}\left(-\frac{4}{9}\right)
+\frac{1}{\e^3}\left(-\frac{71}{108}\right)
+\frac{1}{\e^2}\left(\frac{467}{1296}+\frac{37}{216}\pi^2\right)
\nonumber &\\&
+\frac{1}{\e}\left(-\frac{3701}{3888}-\frac{65}{324}\pi^2+\frac{31}{54}\zeta_{3}\right)
\nonumber &\\&
-\frac{91025}{23328}-\frac{205}{3888}\pi^2-\frac{80}{81}\zeta_{3}+\frac{37}{1728}\pi^4 + \mathcal{O}(\e), &
\end{flalign}

%NG NF N^{0}
\begin{flalign}
{\cal T}^{(3,\left[2\times 1\right])}_{\glu g}\Big|_{\NG \NF} = &
+\frac{1}{\e^2}\left(-\frac{1}{48}\right), &
\end{flalign}

%NG NF^2 N^{1}
\begin{flalign}
{\cal T}^{(3,\left[2\times 1\right])}_{\glu g}\Big|_{\NG \NFSquare N} = &
+\frac{1}{\e^3}\left(\frac{1}{24}\right), &
\end{flalign}

%NG^2 N^{3}
\begin{flalign}
{\cal T}^{(3,\left[2\times 1\right])}_{\glu g}\Big|_{\NGSquare N^{3}} = &
+\frac{1}{\e^4}\left(-\frac{2}{9}\right)
+\frac{1}{\e^3}\left(-\frac{71}{216}\right)
+\frac{1}{\e^2}\left(\frac{247}{1296}+\frac{37}{432}\pi^2\right)
\nonumber &\\&
+\frac{1}{\e}\left(-\frac{3701}{7776}-\frac{65}{648}\pi^2+\frac{31}{108}\zeta_{3}\right)
\nonumber &\\&
-\frac{91025}{46656}-\frac{205}{7776}\pi^2-\frac{40}{81}\zeta_{3}+\frac{37}{3456}\pi^4 + \mathcal{O}(\e), &
\end{flalign}

%NG^2 NF N^{2}
\begin{flalign}
{\cal T}^{(3,\left[2\times 1\right])}_{\glu g}\Big|_{\NGSquare \NF N^{2}} = &
+\frac{1}{\e^3}\left(\frac{1}{24}\right), &
\end{flalign}

%NG^3 N^{3}
\begin{flalign}
{\cal T}^{(3,\left[2\times 1\right])}_{\glu g}\Big|_{\NGCube N^{3}} = &
+\frac{1}{\e^3}\left(\frac{1}{72}\right). &
\end{flalign}

\subsection{Three-particle final states}

%NXgg32x0

%N^{3}
\begin{flalign}
{\cal T}^{(3,\left[2\times 0\right])}_{\glu gg}\Big|_{N^{3}} = &
+\frac{1}{\e^6}\left(\frac{23}{9}\right)
+\frac{1}{\e^5}\left(\frac{5015}{216}\right)
+\frac{1}{\e^4}\left(\frac{50089}{648}-\frac{139}{18}\pi^2\right)
\nonumber &\\&
+\frac{1}{\e^3}\left(\frac{1068031}{3888}-\frac{11867}{324}\pi^2-\frac{1195}{18}\zeta_{3}\right)
\nonumber &\\&
+\frac{1}{\e^2}\left(\frac{14295587}{11664}-\frac{952265}{7776}\pi^2-\frac{15881}{36}\zeta_{3}+\frac{15613}{2592}\pi^4\right)
\nonumber &\\&
+\frac{1}{\e}\left(\frac{376138991}{69984}-\frac{13078363}{23328}\pi^2-\frac{350261}{216}\zeta_{3}+\frac{712829}{51840}\pi^4\right.
\nonumber &\\&\phantom{+\frac{1}{\e}\bigg(}
\left.+\frac{48313}{216}\pi^2\zeta_{3}-\frac{77033}{90}\zeta_{5}\right)
\nonumber &\\&
+\frac{1263139087}{52488}-\frac{47111561}{17496}\pi^2-\frac{1636477}{216}\zeta_{3}+\frac{3385121}{62208}\pi^4
\nonumber &\\&
+\frac{381625}{432}\pi^2\zeta_{3}-\frac{831617}{180}\zeta_{5}-\frac{716087}{816480}\pi^6+\frac{24409}{18}\zeta_{3}^2 + \mathcal{O}(\e), &
\end{flalign}

%NF N^{2}
\begin{flalign}
{\cal T}^{(3,\left[2\times 0\right])}_{\glu gg}\Big|_{\NF N^{2}} = &
+\frac{1}{\e^5}\left(-\frac{143}{54}\right)
+\frac{1}{\e^4}\left(-\frac{1735}{162}\right)
+\frac{1}{\e^3}\left(-\frac{6370}{243}+\frac{1885}{648}\pi^2\right)
\nonumber &\\&
+\frac{1}{\e^2}\left(-\frac{1255655}{11664}+\frac{14765}{1944}\pi^2+\frac{229}{6}\zeta_{3}\right)
\nonumber &\\&
+\frac{1}{\e}\left(-\frac{27578543}{69984}+\frac{276335}{11664}\pi^2+\frac{6119}{54}\zeta_{3}-\frac{1277}{8640}\pi^4\right)
\nonumber &\\&
-\frac{593639723}{419904}+\frac{6905129}{69984}\pi^2+\frac{224831}{648}\zeta_{3}+\frac{113441}{77760}\pi^4
\nonumber &\\&
-\frac{8159}{216}\pi^2\zeta_{3}+\frac{14492}{45}\zeta_{5} + \mathcal{O}(\e), &
\end{flalign}

%NF N^{0}
\begin{flalign}
{\cal T}^{(3,\left[2\times 0\right])}_{\glu gg}\Big|_{\NF} = &
+\frac{1}{\e^3}\left(-\frac{1}{2}\right)
+\frac{1}{\e^2}\left(-\frac{5}{6}\right)
+\frac{1}{\e}\left(-\frac{67}{24}+\frac{7}{24}\pi^2\right)
\nonumber &\\&
-\frac{1189}{144}+\frac{17}{36}\pi^2+\frac{13}{3}\zeta_{3} + \mathcal{O}(\e), &
\end{flalign}

%NF^2 N^{1}
\begin{flalign}
{\cal T}^{(3,\left[2\times 0\right])}_{\glu gg}\Big|_{\NFSquare N} = &
+\frac{1}{\e^4}\left(\frac{4}{9}\right)
+\frac{1}{\e^3}\left(\frac{20}{27}\right)
+\frac{1}{\e^2}\left(\frac{23}{9}-\frac{7}{27}\pi^2\right)
\nonumber &\\&
+\frac{1}{\e}\left(8-\frac{35}{81}\pi^2-\frac{100}{27}\zeta_{3}\right)
\nonumber &\\&
+\frac{7945}{324}-\frac{965}{648}\pi^2-\frac{500}{81}\zeta_{3}-\frac{71}{3240}\pi^4 + \mathcal{O}(\e), &
\end{flalign}

%NG N^{3}
\begin{flalign}
{\cal T}^{(3,\left[2\times 0\right])}_{\glu gg}\Big|_{\NG N^{3}} = &
+\frac{1}{\e^5}\left(-\frac{143}{54}\right)
+\frac{1}{\e^4}\left(-\frac{1735}{162}\right)
+\frac{1}{\e^3}\left(-\frac{12497}{486}+\frac{1885}{648}\pi^2\right)
\nonumber &\\&
+\frac{1}{\e^2}\left(-\frac{1245935}{11664}+\frac{14765}{1944}\pi^2+\frac{229}{6}\zeta_{3}\right)
\nonumber &\\&
+\frac{1}{\e}\left(-\frac{27383171}{69984}+\frac{272933}{11664}\pi^2+\frac{6119}{54}\zeta_{3}-\frac{1277}{8640}\pi^4\right)
\nonumber &\\&
-\frac{590172599}{419904}+\frac{6872081}{69984}\pi^2+\frac{222023}{648}\zeta_{3}+\frac{113441}{77760}\pi^4
\nonumber &\\&
-\frac{8159}{216}\pi^2\zeta_{3}+\frac{14492}{45}\zeta_{5} + \mathcal{O}(\e), &
\end{flalign}

%NG NF N^{2}
\begin{flalign}
{\cal T}^{(3,\left[2\times 0\right])}_{\glu gg}\Big|_{\NG \NF N^{2}} = &
+\frac{1}{\e^4}\left(\frac{8}{9}\right)
+\frac{1}{\e^3}\left(\frac{40}{27}\right)
+\frac{1}{\e^2}\left(\frac{46}{9}-\frac{14}{27}\pi^2\right)
\nonumber &\\&
+\frac{1}{\e}\left(16-\frac{70}{81}\pi^2-\frac{200}{27}\zeta_{3}\right)
\nonumber &\\&
+\frac{7945}{162}-\frac{965}{324}\pi^2-\frac{1000}{81}\zeta_{3}-\frac{71}{1620}\pi^4 + \mathcal{O}(\e), &
\end{flalign}

%NG^2 N^{3}
\begin{flalign}
{\cal T}^{(3,\left[2\times 0\right])}_{\glu gg}\Big|_{\NGSquare N^{3}} = &
+\frac{1}{\e^4}\left(\frac{4}{9}\right)
+\frac{1}{\e^3}\left(\frac{20}{27}\right)
+\frac{1}{\e^2}\left(\frac{23}{9}-\frac{7}{27}\pi^2\right)
\nonumber &\\&
+\frac{1}{\e}\left(8-\frac{35}{81}\pi^2-\frac{100}{27}\zeta_{3}\right)
\nonumber &\\&
+\frac{7945}{324}-\frac{965}{648}\pi^2-\frac{500}{81}\zeta_{3}-\frac{71}{3240}\pi^4 + \mathcal{O}(\e), &
\end{flalign}

%NXgg31x1

%N^{3}
\begin{flalign}
{\cal T}^{(3,\left[1\times 1\right])}_{\glu gg}\Big|_{N^{3}} = &
+\frac{1}{\e^6}\left(\frac{23}{9}\right)
+\frac{1}{\e^5}\left(\frac{1217}{72}\right)
+\frac{1}{\e^4}\left(\frac{1441}{24}-\frac{301}{108}\pi^2\right)
\nonumber &\\&
+\frac{1}{\e^3}\left(\frac{303481}{1296}-\frac{3131}{144}\pi^2-\frac{607}{9}\zeta_{3}\right)
\nonumber &\\&
+\frac{1}{\e^2}\left(\frac{4057931}{3888}-\frac{185797}{2592}\pi^2-\frac{2101}{6}\zeta_{3}-\frac{3311}{4320}\pi^4\right)
\nonumber &\\&
+\frac{1}{\e}\left(\frac{36840731}{7776}-\frac{8614}{27}\pi^2-\frac{304165}{216}\zeta_{3}+\frac{3907}{2592}\pi^4\right.
\nonumber &\\&\phantom{+\frac{1}{\e}\bigg(}
\left.+\frac{10253}{108}\pi^2\zeta_{3}-\frac{13393}{15}\zeta_{5}\right)
\nonumber &\\&
+\frac{3089933381}{139968}-\frac{2898479}{1944}\pi^2-\frac{2218567}{324}\zeta_{3}-\frac{2141281}{311040}\pi^4
\nonumber &\\&
+\frac{112135}{216}\pi^2\zeta_{3}-\frac{122167}{30}\zeta_{5}-\frac{101317}{120960}\pi^6+\frac{23447}{18}\zeta_{3}^2 + \mathcal{O}(\e), &
\end{flalign}

%NF N^{2}
\begin{flalign}
{\cal T}^{(3,\left[1\times 1\right])}_{\glu gg}\Big|_{\NF N^{2}} = &
+\frac{1}{\e^5}\left(-\frac{3}{2}\right)
+\frac{1}{\e^4}\left(-\frac{56}{9}\right)
+\frac{1}{\e^3}\left(-\frac{1847}{108}+\frac{71}{36}\pi^2\right)
\nonumber &\\&
+\frac{1}{\e^2}\left(-\frac{85495}{1296}+\frac{209}{36}\pi^2+24 \zeta_{3}\right)
\nonumber &\\&
+\frac{1}{\e}\left(-\frac{322571}{1296}+\frac{4229}{216}\pi^2+\frac{695}{9}\zeta_{3}-\frac{199}{432}\pi^4\right)
\nonumber &\\&
-\frac{5558927}{5832}+\frac{617953}{7776}\pi^2+\frac{43687}{162}\zeta_{3}-\frac{2993}{4320}\pi^4
\nonumber &\\&
-\frac{940}{27}\pi^2\zeta_{3}+\frac{3224}{15}\zeta_{5} + \mathcal{O}(\e), &
\end{flalign}

%NF^2 N^{1}
\begin{flalign}
{\cal T}^{(3,\left[1\times 1\right])}_{\glu gg}\Big|_{\NFSquare N} = &
+\frac{1}{\e^4}\left(\frac{2}{9}\right)
+\frac{1}{\e^3}\left(\frac{10}{27}\right)
+\frac{1}{\e^2}\left(\frac{139}{108}-\frac{7}{54}\pi^2\right)
\nonumber &\\&
+\frac{1}{\e}\left(\frac{2657}{648}-\frac{35}{162}\pi^2-\frac{50}{27}\zeta_{3}\right)
\nonumber &\\&
+\frac{8407}{648}-\frac{503}{648}\pi^2-\frac{250}{81}\zeta_{3}-\frac{71}{6480}\pi^4 + \mathcal{O}(\e), &
\end{flalign}

%NG N^{3}
\begin{flalign}
{\cal T}^{(3,\left[1\times 1\right])}_{\glu gg}\Big|_{\NG N^{3}} = &
+\frac{1}{\e^5}\left(-\frac{3}{2}\right)
+\frac{1}{\e^4}\left(-\frac{56}{9}\right)
+\frac{1}{\e^3}\left(-\frac{1847}{108}+\frac{71}{36}\pi^2\right)
\nonumber &\\&
+\frac{1}{\e^2}\left(-\frac{85495}{1296}+\frac{209}{36}\pi^2+24 \zeta_{3}\right)
\nonumber &\\&
+\frac{1}{\e}\left(-\frac{322571}{1296}+\frac{4229}{216}\pi^2+\frac{695}{9}\zeta_{3}-\frac{199}{432}\pi^4\right)
\nonumber &\\&
-\frac{5558927}{5832}+\frac{617953}{7776}\pi^2+\frac{43687}{162}\zeta_{3}-\frac{2993}{4320}\pi^4
\nonumber &\\&
-\frac{940}{27}\pi^2\zeta_{3}+\frac{3224}{15}\zeta_{5} + \mathcal{O}(\e), &
\end{flalign}

%NG NF N^{2}
\begin{flalign}
{\cal T}^{(3,\left[1\times 1\right])}_{\glu gg}\Big|_{\NG \NF N^{2}} = &
+\frac{1}{\e^4}\left(\frac{4}{9}\right)
+\frac{1}{\e^3}\left(\frac{20}{27}\right)
+\frac{1}{\e^2}\left(\frac{139}{54}-\frac{7}{27}\pi^2\right)
\nonumber &\\&
+\frac{1}{\e}\left(\frac{2657}{324}-\frac{35}{81}\pi^2-\frac{100}{27}\zeta_{3}\right)
\nonumber &\\&
+\frac{8407}{324}-\frac{503}{324}\pi^2-\frac{500}{81}\zeta_{3}-\frac{71}{3240}\pi^4 + \mathcal{O}(\e), &
\end{flalign}

%NG^2 N^{3}
\begin{flalign}
{\cal T}^{(3,\left[1\times 1\right])}_{\glu gg}\Big|_{\NGSquare N^{3}} = &
+\frac{1}{\e^4}\left(\frac{2}{9}\right)
+\frac{1}{\e^3}\left(\frac{10}{27}\right)
+\frac{1}{\e^2}\left(\frac{139}{108}-\frac{7}{54}\pi^2\right)
\nonumber &\\&
+\frac{1}{\e}\left(\frac{2657}{648}-\frac{35}{162}\pi^2-\frac{50}{27}\zeta_{3}\right)
\nonumber &\\&
+\frac{8407}{648}-\frac{503}{648}\pi^2-\frac{250}{81}\zeta_{3}-\frac{71}{6480}\pi^4 + \mathcal{O}(\e), &
\end{flalign}

%NXqqb32x0

%N^{2}
\begin{flalign}
{\cal T}^{(3,\left[2\times 0\right])}_{\glu q\bar{q}}\Big|_{\NF N^{2}} = &
+\frac{1}{\e^5}\left(-\frac{1}{3}\right)
+\frac{1}{\e^4}\left(-\frac{131}{36}\right)
+\frac{1}{\e^3}\left(-\frac{1225}{81}+\frac{109}{108}\pi^2\right)
\nonumber &\\&
+\frac{1}{\e^2}\left(-\frac{141119}{2592}+\frac{7951}{1296}\pi^2+\frac{74}{9}\zeta_{3}\right)
\nonumber &\\&
+\frac{1}{\e}\left(-\frac{1197013}{5184}+\frac{179701}{7776}\pi^2+\frac{7819}{108}\zeta_{3}-\frac{3509}{4320}\pi^4\right)
\nonumber &\\&
-\frac{286872709}{279936}+\frac{9288529}{93312}\pi^2+\frac{23369}{72}\zeta_{3}-\frac{81611}{31104}\pi^4
\nonumber &\\&
-\frac{712}{27}\pi^2\zeta_{3}+\frac{857}{10}\zeta_{5} + \mathcal{O}(\e), &
\end{flalign}

%N^{0}
\begin{flalign}
{\cal T}^{(3,\left[2\times 0\right])}_{\glu q\bar{q}}\Big|_{\NF} = &
+\frac{1}{\e^5}\left(\frac{1}{6}\right)
+\frac{1}{\e^4}\left(\frac{179}{108}\right)
+\frac{1}{\e^3}\left(\frac{11537}{1296}-\frac{61}{108}\pi^2\right)
\nonumber &\\&
+\frac{1}{\e^2}\left(\frac{9353}{216}-\frac{11117}{2592}\pi^2-\frac{209}{36}\zeta_{3}\right)
\nonumber &\\&
+\frac{1}{\e}\left(\frac{2361235}{11664}-\frac{58801}{2592}\pi^2-\frac{10583}{216}\zeta_{3}+\frac{12941}{25920}\pi^4\right)
\nonumber &\\&
+\frac{33007549}{34992}-\frac{5207897}{46656}\pi^2-\frac{172183}{648}\zeta_{3}+\frac{960059}{311040}\pi^4
\nonumber &\\&
+\frac{8711}{432}\pi^2\zeta_{3}-\frac{5281}{60}\zeta_{5} + \mathcal{O}(\e), &
\end{flalign}

%N^{-2}
\begin{flalign}
{\cal T}^{(3,\left[2\times 0\right])}_{\glu q\bar{q}}\Big|_{\NF N^{-2}} = &
+\frac{1}{\e^5}\left(-\frac{1}{36}\right)
+\frac{1}{\e^4}\left(-\frac{13}{54}\right)
+\frac{1}{\e^3}\left(-\frac{2041}{1296}+\frac{41}{432}\pi^2\right)
\nonumber &\\&
+\frac{1}{\e^2}\left(-\frac{69793}{7776}+\frac{131}{162}\pi^2+\frac{43}{36}\zeta_{3}\right)
\nonumber &\\&
+\frac{1}{\e}\left(-\frac{2236849}{46656}+\frac{81413}{15552}\pi^2+\frac{1109}{108}\zeta_{3}-\frac{4033}{51840}\pi^4\right)
\nonumber &\\&
-\frac{69371737}{279936}+\frac{2764421}{93312}\pi^2+\frac{88249}{1296}\zeta_{3}-\frac{49063}{77760}\pi^4
\nonumber &\\&
-\frac{593}{144}\pi^2\zeta_{3}+\frac{991}{60}\zeta_{5} + \mathcal{O}(\e), &
\end{flalign}

%NF N^{1}
\begin{flalign}
{\cal T}^{(3,\left[2\times 0\right])}_{\glu q\bar{q}}\Big|_{\NFSquare N} = &
+\frac{1}{\e^4}\left(\frac{5}{18}\right)
+\frac{1}{\e^3}\left(\frac{85}{81}\right)
+\frac{1}{\e^2}\left(\frac{155}{216}-\frac{35}{648}\pi^2\right)
\nonumber &\\&
+\frac{1}{\e}\left(-\frac{14197}{3888}+\frac{1403}{1296}\pi^2-\frac{83}{54}\zeta_{3}\right)
\nonumber &\\&
-\frac{2643827}{69984}+\frac{66611}{7776}\pi^2+\frac{865}{108}\zeta_{3}-\frac{23633}{77760}\pi^4 + \mathcal{O}(\e), &
\end{flalign}

%NF N^{-1}
\begin{flalign}
{\cal T}^{(3,\left[2\times 0\right])}_{\glu q\bar{q}}\Big|_{\NFSquare N^{-1}} = &
+\frac{1}{\e^4}\left(-\frac{7}{108}\right)
+\frac{1}{\e^3}\left(-\frac{73}{324}\right)
+\frac{1}{\e^2}\left(-\frac{179}{648}+\frac{5}{432}\pi^2\right)
\nonumber &\\&
+\frac{1}{\e}\left(\frac{27457}{11664}-\frac{571}{1296}\pi^2+\frac{83}{108}\zeta_{3}\right)
\nonumber &\\&
+\frac{1932101}{69984}-\frac{11947}{2592}\pi^2-\frac{217}{324}\zeta_{3}+\frac{14077}{155520}\pi^4 + \mathcal{O}(\e), &
\end{flalign}

%NF^2 N^{0}
\begin{flalign}
{\cal T}^{(3,\left[2\times 0\right])}_{\glu q\bar{q}}\Big|_{\NFCube} = &
+\frac{1}{\e^3}\left(-\frac{2}{81}\right)
+\frac{1}{\e}\left(\frac{4}{243}\pi^2\right)
-\frac{2110}{2187}+\frac{2}{9}\pi^2 + \mathcal{O}(\e), &
\end{flalign}

%NG N^{2}
\begin{flalign}
{\cal T}^{(3,\left[2\times 0\right])}_{\glu q\bar{q}}\Big|_{\NG \NF N^{2}} = &
+\frac{1}{\e^4}\left(\frac{5}{18}\right)
+\frac{1}{\e^3}\left(\frac{85}{81}\right)
+\frac{1}{\e^2}\left(\frac{143}{216}-\frac{35}{648}\pi^2\right)
\nonumber &\\&
+\frac{1}{\e}\left(-\frac{13567}{3888}+\frac{1403}{1296}\pi^2-\frac{95}{54}\zeta_{3}\right)
\nonumber &\\&
-\frac{2445701}{69984}+\frac{66251}{7776}\pi^2+\frac{653}{108}\zeta_{3}-\frac{23921}{77760}\pi^4 + \mathcal{O}(\e), &
\end{flalign}

%NG N^{0}
\begin{flalign}
{\cal T}^{(3,\left[2\times 0\right])}_{\glu q\bar{q}}\Big|_{\NG \NF} = &
+\frac{1}{\e^4}\left(-\frac{7}{108}\right)
+\frac{1}{\e^3}\left(-\frac{73}{324}\right)
+\frac{1}{\e^2}\left(-\frac{215}{648}+\frac{5}{432}\pi^2\right)
\nonumber &\\&
+\frac{1}{\e}\left(\frac{29347}{11664}-\frac{571}{1296}\pi^2+\frac{59}{108}\zeta_{3}\right)
\nonumber &\\&
+\frac{2130227}{69984}-\frac{12067}{2592}\pi^2-\frac{853}{324}\zeta_{3}+\frac{13501}{155520}\pi^4 + \mathcal{O}(\e), &
\end{flalign}

%NG NF N^{1}
\begin{flalign}
{\cal T}^{(3,\left[2\times 0\right])}_{\glu q\bar{q}}\Big|_{\NG \NFSquare N} = &
+\frac{1}{\e^3}\left(-\frac{4}{81}\right)
+\frac{1}{\e}\left(\frac{8}{243}\pi^2\right)
-\frac{4220}{2187}+\frac{4}{9}\pi^2 + \mathcal{O}(\e), &
\end{flalign}

%NG^2 N^{2}
\begin{flalign}
{\cal T}^{(3,\left[2\times 0\right])}_{\glu q\bar{q}}\Big|_{\NGSquare \NF N^{2}} = &
+\frac{1}{\e^3}\left(-\frac{2}{81}\right)
+\frac{1}{\e}\left(\frac{4}{243}\pi^2\right)
-\frac{2110}{2187}+\frac{2}{9}\pi^2 + \mathcal{O}(\e), &
\end{flalign}

%NXqqb31x1

%N^{2}
\begin{flalign}
{\cal T}^{(3,\left[1\times 1\right])}_{\glu q\bar{q}}\Big|_{\NF N^{2}} = &
+\frac{1}{\e^5}\left(-\frac{1}{3}\right)
+\frac{1}{\e^4}\left(-\frac{49}{18}\right)
+\frac{1}{\e^3}\left(-\frac{7397}{648}+\frac{10}{27}\pi^2\right)
\nonumber &\\&
+\frac{1}{\e^2}\left(-\frac{57797}{1296}+\frac{290}{81}\pi^2+\frac{77}{9}\zeta_{3}\right)
\nonumber &\\&
+\frac{1}{\e}\left(-\frac{1511657}{7776}+\frac{2087}{144}\pi^2+\frac{1610}{27}\zeta_{3}+\frac{1087}{12960}\pi^4\right)
\nonumber &\\&
-\frac{124660217}{139968}+\frac{2860655}{46656}\pi^2+\frac{22090}{81}\zeta_{3}-\frac{212}{1215}\pi^4
\nonumber &\\&
-\frac{299}{27}\pi^2\zeta_{3}+\frac{3091}{30}\zeta_{5} + \mathcal{O}(\e), &
\end{flalign}

%N^{0}
\begin{flalign}
{\cal T}^{(3,\left[1\times 1\right])}_{\glu q\bar{q}}\Big|_{\NF} = &
+\frac{1}{\e^5}\left(\frac{1}{6}\right)
+\frac{1}{\e^4}\left(\frac{101}{72}\right)
+\frac{1}{\e^3}\left(\frac{3311}{432}-\frac{17}{72}\pi^2\right)
\nonumber &\\&
+\frac{1}{\e^2}\left(\frac{1393}{36}-\frac{151}{72}\pi^2-\frac{35}{6}\zeta_{3}\right)
\nonumber &\\&
+\frac{1}{\e}\left(\frac{27325}{144}-\frac{15317}{1296}\pi^2-\frac{1231}{27}\zeta_{3}-\frac{173}{2880}\pi^4\right)
\nonumber &\\&
+\frac{1350823}{1458}-\frac{946739}{15552}\pi^2-\frac{42035}{162}\zeta_{3}-\frac{12971}{51840}\pi^4
\nonumber &\\&
+\frac{619}{72}\pi^2\zeta_{3}-\frac{811}{10}\zeta_{5} + \mathcal{O}(\e), &
\end{flalign}

%N^{-2}
\begin{flalign}
{\cal T}^{(3,\left[1\times 1\right])}_{\glu q\bar{q}}\Big|_{\NF N^{-2}} = &
+\frac{1}{\e^5}\left(-\frac{1}{36}\right)
+\frac{1}{\e^4}\left(-\frac{13}{54}\right)
+\frac{1}{\e^3}\left(-\frac{2041}{1296}+\frac{17}{432}\pi^2\right)
\nonumber &\\&
+\frac{1}{\e^2}\left(-\frac{34937}{3888}+\frac{221}{648}\pi^2+\frac{37}{36}\zeta_{3}\right)
\nonumber &\\&
+\frac{1}{\e}\left(-\frac{559921}{11664}+\frac{34697}{15552}\pi^2+\frac{481}{54}\zeta_{3}+\frac{277}{17280}\pi^4\right)
\nonumber &\\&
-\frac{8679233}{34992}+\frac{593929}{46656}\pi^2+\frac{75517}{1296}\zeta_{3}+\frac{3601}{25920}\pi^4
\nonumber &\\&
-\frac{629}{432}\pi^2\zeta_{3}+\frac{327}{20}\zeta_{5} + \mathcal{O}(\e), &
\end{flalign}

%NF N^{1}
\begin{flalign}
{\cal T}^{(3,\left[1\times 1\right])}_{\glu q\bar{q}}\Big|_{\NFSquare N} = &
+\frac{1}{\e^4}\left(\frac{1}{9}\right)
+\frac{1}{\e^3}\left(\frac{31}{108}\right)
+\frac{1}{\e^2}\left(-\frac{361}{324}-\frac{5}{36}\pi^2\right)
\nonumber &\\&
+\frac{1}{\e}\left(-\frac{811}{72}+\frac{49}{324}\pi^2+\frac{11}{27}\zeta_{3}\right)
\nonumber &\\&
-\frac{836173}{11664}+\frac{13633}{3888}\pi^2+\frac{43}{3}\zeta_{3}+\frac{167}{1440}\pi^4 + \mathcal{O}(\e), &
\end{flalign}

%NF N^{-1}
\begin{flalign}
{\cal T}^{(3,\left[1\times 1\right])}_{\glu q\bar{q}}\Big|_{\NFSquare N^{-1}} = &
+\frac{1}{\e^4}\left(-\frac{1}{54}\right)
+\frac{1}{\e^3}\left(\frac{1}{324}\right)
+\frac{1}{\e^2}\left(\frac{175}{324}+\frac{5}{162}\pi^2\right)
\nonumber &\\&
+\frac{1}{\e}\left(\frac{3689}{729}+\frac{11}{486}\pi^2-\frac{2}{9}\zeta_{3}\right)
\nonumber &\\&
+\frac{75737}{2187}-\frac{137}{216}\pi^2-\frac{146}{27}\zeta_{3}-\frac{523}{12960}\pi^4 + \mathcal{O}(\e), &
\end{flalign}

%NF^2 N^{0}
\begin{flalign}
{\cal T}^{(3,\left[1\times 1\right])}_{\glu q\bar{q}}\Big|_{\NFCube} = &
+\frac{1}{\e^3}\left(-\frac{1}{81}\right)
+\frac{1}{\e}\left(-\frac{4}{243}\pi^2\right)
-\frac{1055}{2187}-\frac{\pi^2}{9} + \mathcal{O}(\e), &
\end{flalign}

%NG N^{2}
\begin{flalign}
{\cal T}^{(3,\left[1\times 1\right])}_{\glu q\bar{q}}\Big|_{\NG \NF N^{2}} = &
+\frac{1}{\e^4}\left(\frac{1}{9}\right)
+\frac{1}{\e^3}\left(\frac{31}{108}\right)
+\frac{1}{\e^2}\left(-\frac{361}{324}-\frac{5}{36}\pi^2\right)
\nonumber &\\&
+\frac{1}{\e}\left(-\frac{811}{72}+\frac{49}{324}\pi^2+\frac{11}{27}\zeta_{3}\right)
\nonumber &\\&
-\frac{836173}{11664}+\frac{13633}{3888}\pi^2+\frac{43}{3}\zeta_{3}+\frac{167}{1440}\pi^4 + \mathcal{O}(\e), &
\end{flalign}

%NG N^{0}
\begin{flalign}
{\cal T}^{(3,\left[1\times 1\right])}_{\glu q\bar{q}}\Big|_{\NG \NF} = &
+\frac{1}{\e^4}\left(-\frac{1}{54}\right)
+\frac{1}{\e^3}\left(\frac{1}{324}\right)
+\frac{1}{\e^2}\left(\frac{175}{324}+\frac{5}{162}\pi^2\right)
\nonumber &\\&
+\frac{1}{\e}\left(\frac{3689}{729}+\frac{11}{486}\pi^2-\frac{2}{9}\zeta_{3}\right)
\nonumber &\\&
+\frac{75737}{2187}-\frac{137}{216}\pi^2-\frac{146}{27}\zeta_{3}-\frac{523}{12960}\pi^4 + \mathcal{O}(\e), &
\end{flalign}

%NG NF N^{1}
\begin{flalign}
{\cal T}^{(3,\left[1\times 1\right])}_{\glu q\bar{q}}\Big|_{\NG \NFSquare N} = &
+\frac{1}{\e^3}\left(-\frac{2}{81}\right)
+\frac{1}{\e}\left(-\frac{8}{243}\pi^2\right)
-\frac{2110}{2187}-\frac{2}{9}\pi^2 + \mathcal{O}(\e), &
\end{flalign}

%NG^2 N^{2}
\begin{flalign}
{\cal T}^{(3,\left[1\times 1\right])}_{\glu q\bar{q}}\Big|_{\NGSquare \NF N^{2}} = &
+\frac{1}{\e^3}\left(-\frac{1}{81}\right)
+\frac{1}{\e}\left(-\frac{4}{243}\pi^2\right)
-\frac{1055}{2187}-\frac{\pi^2}{9} + \mathcal{O}(\e), &
\end{flalign}

%NXXX32x0

%N^{3}
\begin{flalign}
{\cal T}^{(3,\left[2\times 0\right])}_{\glu \glu' \glu'}\Big|_{(\NG-1)N^{3}} = &
+\frac{1}{\e^5}\left(-\frac{19}{36}\right)
+\frac{1}{\e^4}\left(-\frac{299}{54}\right)
+\frac{1}{\e^3}\left(-\frac{16589}{648}+\frac{721}{432}\pi^2\right)
\nonumber &\\&
+\frac{1}{\e^2}\left(-\frac{416837}{3888}+\frac{3235}{288}\pi^2+\frac{137}{9}\zeta_{3}\right)
\nonumber &\\&
+\frac{1}{\e}\left(-\frac{11296141}{23328}+\frac{793621}{15552}\pi^2+\frac{28439}{216}\zeta_{3}-\frac{72023}{51840}\pi^4\right)
\nonumber &\\&
-\frac{312220691}{139968}+\frac{2817179}{11664}\pi^2+\frac{284419}{432}\zeta_{3}-\frac{1972421}{311040}\pi^4
\nonumber &\\&
-\frac{3647}{72}\pi^2\zeta_{3}+\frac{5707}{30}\zeta_{5} + \mathcal{O}(\e), &
\end{flalign}

%NF N^{2}
\begin{flalign}
{\cal T}^{(3,\left[2\times 0\right])}_{\glu \glu' \glu'}\Big|_{(\NG-1)\NF N^{2}} = &
+\frac{1}{\e^4}\left(\frac{37}{108}\right)
+\frac{1}{\e^3}\left(\frac{413}{324}\right)
+\frac{1}{\e^2}\left(\frac{91}{81}-\frac{85}{1296}\pi^2\right)
\nonumber &\\&
+\frac{1}{\e}\left(-\frac{33377}{5832}+\frac{329}{216}\pi^2-\frac{25}{12}\zeta_{3}\right)
\nonumber &\\&
-\frac{2308979}{34992}+\frac{25487}{1944}\pi^2+\frac{862}{81}\zeta_{3}-\frac{60767}{155520}\pi^4 + \mathcal{O}(\e), &
\end{flalign}

%NF N^{0}
\begin{flalign}
{\cal T}^{(3,\left[2\times 0\right])}_{\glu \glu' \glu'}\Big|_{(\NG-1)\NF} = &
+\frac{1}{\e^2}\left(\frac{1}{18}\right)
+\frac{1}{\e}\left(-\frac{35}{216}+\frac{2}{9}\zeta_{3}\right)
\nonumber &\\&
-\frac{1223}{432}+\frac{5}{108}\pi^2+\frac{53}{27}\zeta_{3}+\frac{\pi^4}{270} + \mathcal{O}(\e), &
\end{flalign}

%NF^2 N^{1}
\begin{flalign}
{\cal T}^{(3,\left[2\times 0\right])}_{\glu \glu' \glu'}\Big|_{(\NG-1)\NFSquare N} = &
+\frac{1}{\e^3}\left(-\frac{2}{81}\right)
+\frac{1}{\e}\left(\frac{4}{243}\pi^2\right)
-\frac{2110}{2187}+\frac{2}{9}\pi^2 + \mathcal{O}(\e), &
\end{flalign}

%NG N^{3}
\begin{flalign}
{\cal T}^{(3,\left[2\times 0\right])}_{\glu \glu' \glu'}\Big|_{(\NG-1)\NG N^{3}} = &
+\frac{1}{\e^4}\left(\frac{37}{108}\right)
+\frac{1}{\e^3}\left(\frac{413}{324}\right)
+\frac{1}{\e^2}\left(\frac{173}{162}-\frac{85}{1296}\pi^2\right)
\nonumber &\\&
+\frac{1}{\e}\left(-\frac{4054}{729}+\frac{329}{216}\pi^2-\frac{83}{36}\zeta_{3}\right)
\nonumber &\\&
-\frac{552479}{8748}+\frac{25397}{1944}\pi^2+\frac{703}{81}\zeta_{3}-\frac{61343}{155520}\pi^4 + \mathcal{O}(\e), &
\end{flalign}

%NG NF N^{2}
\begin{flalign}
{\cal T}^{(3,\left[2\times 0\right])}_{\glu \glu' \glu'}\Big|_{(\NG-1)\NG \NF N^{2}} = &
+\frac{1}{\e^3}\left(-\frac{4}{81}\right)
+\frac{1}{\e}\left(\frac{8}{243}\pi^2\right)
-\frac{4220}{2187}+\frac{4}{9}\pi^2 + \mathcal{O}(\e), &
\end{flalign}

%NG^2 N^{3}
\begin{flalign}
{\cal T}^{(3,\left[2\times 0\right])}_{\glu \glu' \glu'}\Big|_{(\NG-1)\NGSquare N^{3}} = &
+\frac{1}{\e^3}\left(-\frac{2}{81}\right)
+\frac{1}{\e}\left(\frac{4}{243}\pi^2\right)
-\frac{2110}{2187}+\frac{2}{9}\pi^2 + \mathcal{O}(\e), &
\end{flalign}

%NXXX31x1

%N^{3}
\begin{flalign}
{\cal T}^{(3,\left[1\times 1\right])}_{\glu \glu' \glu'}\Big|_{(\NG-1)N^{3}} = &
+\frac{1}{\e^5}\left(-\frac{19}{36}\right)
+\frac{1}{\e^4}\left(-\frac{943}{216}\right)
+\frac{1}{\e^3}\left(-\frac{1673}{81}+\frac{31}{48}\pi^2\right)
\nonumber &\\&
+\frac{1}{\e^2}\left(-\frac{22493}{243}+\frac{325}{54}\pi^2+\frac{185}{12}\zeta_{3}\right)
\nonumber &\\&
+\frac{1}{\e}\left(-\frac{10121639}{23328}+\frac{443897}{15552}\pi^2+\frac{6163}{54}\zeta_{3}+\frac{8293}{51840}\pi^4\right)
\nonumber &\\&
-\frac{290265901}{139968}+\frac{2104963}{15552}\pi^2+\frac{255079}{432}\zeta_{3}+\frac{33383}{155520}\pi^4
\nonumber &\\&
-\frac{9127}{432}\pi^2\zeta_{3}+\frac{12029}{60}\zeta_{5} + \mathcal{O}(\e), &
\end{flalign}

%NF N^{2}
\begin{flalign}
{\cal T}^{(3,\left[1\times 1\right])}_{\glu \glu' \glu'}\Big|_{(\NG-1)\NF N^{2}} = &
+\frac{1}{\e^4}\left(\frac{7}{54}\right)
+\frac{1}{\e^3}\left(\frac{23}{81}\right)
+\frac{1}{\e^2}\left(-\frac{131}{81}-\frac{55}{324}\pi^2\right)
\nonumber &\\&
+\frac{1}{\e}\left(-\frac{93907}{5832}+\frac{125}{972}\pi^2+\frac{17}{27}\zeta_{3}\right)
\nonumber &\\&
-\frac{3681287}{34992}+\frac{15883}{3888}\pi^2+\frac{533}{27}\zeta_{3}+\frac{1013}{6480}\pi^4 + \mathcal{O}(\e), &
\end{flalign}

%NF^2 N^{1}
\begin{flalign}
{\cal T}^{(3,\left[1\times 1\right])}_{\glu \glu' \glu'}\Big|_{(\NG-1)\NFSquare N} = &
+\frac{1}{\e^3}\left(-\frac{1}{81}\right)
+\frac{1}{\e}\left(-\frac{4}{243}\pi^2\right)
-\frac{1055}{2187}-\frac{\pi^2}{9} + \mathcal{O}(\e), &
\end{flalign}

%NG N^{3}
\begin{flalign}
{\cal T}^{(3,\left[1\times 1\right])}_{\glu \glu' \glu'}\Big|_{(\NG-1)\NG N^{3}} = &
+\frac{1}{\e^4}\left(\frac{7}{54}\right)
+\frac{1}{\e^3}\left(\frac{23}{81}\right)
+\frac{1}{\e^2}\left(-\frac{131}{81}-\frac{55}{324}\pi^2\right)
\nonumber &\\&
+\frac{1}{\e}\left(-\frac{93907}{5832}+\frac{125}{972}\pi^2+\frac{17}{27}\zeta_{3}\right)
\nonumber &\\&
-\frac{3681287}{34992}+\frac{15883}{3888}\pi^2+\frac{533}{27}\zeta_{3}+\frac{1013}{6480}\pi^4 + \mathcal{O}(\e), &
\end{flalign}

%NG NF N^{2}
\begin{flalign}
{\cal T}^{(3,\left[1\times 1\right])}_{\glu \glu' \glu'}\Big|_{(\NG-1)\NG \NF N^{2}} = &
+\frac{1}{\e^3}\left(-\frac{2}{81}\right)
+\frac{1}{\e}\left(-\frac{8}{243}\pi^2\right)
-\frac{2110}{2187}-\frac{2}{9}\pi^2 + \mathcal{O}(\e), &
\end{flalign}

%NG^2 N^{3}
\begin{flalign}
{\cal T}^{(3,\left[1\times 1\right])}_{\glu \glu' \glu'}\Big|_{(\NG-1)\NGSquare N^{3}} = &
+\frac{1}{\e^3}\left(-\frac{1}{81}\right)
+\frac{1}{\e}\left(-\frac{4}{243}\pi^2\right)
-\frac{1055}{2187}-\frac{\pi^2}{9} + \mathcal{O}(\e), &
\end{flalign}

%NXXXhard32x0

%N^{3}
\begin{flalign}
\Delta{\cal T}^{(3,\left[2\times 0\right])}_{\glu \glu \glu}\Big|_{N^{3}} = &
+\frac{1}{\e^4}\left(-\frac{3}{8}\right)
+\frac{1}{\e^3}\left(-\frac{31}{8}\right)
+\frac{1}{\e^2}\left(-\frac{5839}{288}+\frac{125}{96}\pi^2+\zeta_{3}\right)
\nonumber &\\&
+\frac{1}{\e}\left(-\frac{20011}{192}+\frac{8351}{864}\pi^2+\frac{247}{12}\zeta_{3}+\frac{5}{108}\pi^4\right)
\nonumber &\\&
-\frac{17252495}{31104}+\frac{555101}{10368}\pi^2+\frac{22177}{144}\zeta_{3}-\frac{92543}{103680}\pi^4
\nonumber &\\&
-\frac{211}{36}\pi^2\zeta_{3}+\frac{277}{6}\zeta_{5} + \mathcal{O}(\e), &
\end{flalign}

%NF N^{2}
\begin{flalign}
\Delta{\cal T}^{(3,\left[2\times 0\right])}_{\glu \glu \glu}\Big|_{\NF N^{2}} = &
+\frac{1}{\e^3}\left(\frac{1}{8}\right)
+\frac{1}{\e^2}\left(-\frac{13}{144}\right)
+\frac{1}{\e}\left(-\frac{1247}{288}+\frac{187}{864}\pi^2\right)
\nonumber &\\&
-\frac{504463}{15552}+\frac{17195}{5184}\pi^2+\frac{199}{72}\zeta_{3}+\frac{\pi^4}{1620} + \mathcal{O}(\e), &
\end{flalign}

%NF N^{0}
\begin{flalign}
\Delta{\cal T}^{(3,\left[2\times 0\right])}_{\glu \glu \glu}\Big|_{\NF} = &
-\frac{77}{144}+\frac{\zeta_{3}}{3} + \mathcal{O}(\e), &
\end{flalign}

%NF^2 N^{1}
\begin{flalign}
\Delta{\cal T}^{(3,\left[2\times 0\right])}_{\glu \glu \glu}\Big|_{\NFSquare N} = &
-\frac{245}{486}+\frac{\pi^2}{27} + \mathcal{O}(\e), &
\end{flalign}

%NG N^{3}
\begin{flalign}
\Delta{\cal T}^{(3,\left[2\times 0\right])}_{\glu \glu \glu}\Big|_{\NG N^{3}} = &
+\frac{1}{\e^3}\left(\frac{1}{8}\right)
+\frac{1}{\e^2}\left(\frac{173}{432}\right)
+\frac{1}{\e}\left(-\frac{133}{96}+\frac{187}{864}\pi^2\right)
\nonumber &\\&
-\frac{266339}{15552}+\frac{13379}{5184}\pi^2+\frac{175}{72}\zeta_{3}+\frac{\pi^4}{1620} + \mathcal{O}(\e), &
\end{flalign}

%NG NF N^{2}
\begin{flalign}
\Delta{\cal T}^{(3,\left[2\times 0\right])}_{\glu \glu \glu}\Big|_{\NG \NF N^{2}} = &
+\frac{1}{\e^2}\left(-\frac{2}{27}\right)
+\frac{1}{\e}\left(-\frac{4}{9}\right)
-\frac{787}{243}+\frac{5}{27}\pi^2 + \mathcal{O}(\e), &
\end{flalign}

%NG^2 N^{3}
\begin{flalign}
\Delta{\cal T}^{(3,\left[2\times 0\right])}_{\glu \glu \glu}\Big|_{\NGSquare N^{3}} = &
+\frac{1}{\e^2}\left(-\frac{2}{27}\right)
+\frac{1}{\e}\left(-\frac{4}{9}\right)
-\frac{443}{162}+\frac{4}{27}\pi^2 + \mathcal{O}(\e), &
\end{flalign}

%NXXXhard31x1

%N^{3}
\begin{flalign}
\Delta{\cal T}^{(3,\left[1\times 1\right])}_{\glu \glu \glu}\Big|_{N^{3}} = &
+\frac{1}{\e^4}\left(-\frac{3}{8}\right)
+\frac{1}{\e^3}\left(-\frac{51}{16}\right)
+\frac{1}{\e^2}\left(-\frac{847}{48}+\frac{55}{96}\pi^2+\zeta_{3}\right)
\nonumber &\\&
+\frac{1}{\e}\left(-\frac{21061}{216}+\frac{9409}{1728}\pi^2+\frac{497}{24}\zeta_{3}+\frac{5}{108}\pi^4\right)
\nonumber &\\&
-\frac{8485661}{15552}+\frac{40319}{1296}\pi^2+\frac{7319}{48}\zeta_{3}+\frac{15767}{34560}\pi^4
\nonumber &\\&
-\frac{137}{36}\pi^2\zeta_{3}+\frac{241}{6}\zeta_{5} + \mathcal{O}(\e), &
\end{flalign}

%NF N^{2}
\begin{flalign}
\Delta{\cal T}^{(3,\left[1\times 1\right])}_{\glu \glu \glu}\Big|_{\NF N^{2}} = &
+\frac{1}{\e^2}\left(-\frac{7}{12}\right)
+\frac{1}{\e}\left(-\frac{154}{27}+\frac{\pi^2}{216}\right)
\nonumber &\\&
-\frac{74095}{1944}+\frac{65}{54}\pi^2+\frac{71}{18}\zeta_{3}+\frac{\pi^4}{1620} + \mathcal{O}(\e), &
\end{flalign}

%NF^2 N^{1}
\begin{flalign}
\Delta{\cal T}^{(3,\left[1\times 1\right])}_{\glu \glu \glu}\Big|_{\NFSquare N} = &
-\frac{245}{972}-\frac{5}{324}\pi^2 + \mathcal{O}(\e), &
\end{flalign}

%NG N^{3}
\begin{flalign}
\Delta{\cal T}^{(3,\left[1\times 1\right])}_{\glu \glu \glu}\Big|_{\NG N^{3}} = &
+\frac{1}{\e^2}\left(-\frac{8}{27}\right)
+\frac{1}{\e}\left(-\frac{215}{54}+\frac{\pi^2}{216}\right)
\nonumber &\\&
-\frac{57293}{1944}+\frac{167}{216}\pi^2+\frac{71}{18}\zeta_{3}+\frac{\pi^4}{1620} + \mathcal{O}(\e), &
\end{flalign}

%NG NF N^{2}
\begin{flalign}
\Delta{\cal T}^{(3,\left[1\times 1\right])}_{\glu \glu \glu}\Big|_{\NG \NF N^{2}} = &
+\frac{1}{\e^2}\left(-\frac{1}{27}\right)
+\frac{1}{\e}\left(-\frac{2}{9}\right)
-\frac{787}{486}+\frac{2}{81}\pi^2 + \mathcal{O}(\e), &
\end{flalign}

%NG^2 N^{3}
\begin{flalign}
\Delta{\cal T}^{(3,\left[1\times 1\right])}_{\glu \glu \glu}\Big|_{\NGSquare N^{3}} = &
+\frac{1}{\e^2}\left(-\frac{1}{27}\right)
+\frac{1}{\e}\left(-\frac{2}{9}\right)
-\frac{443}{324}+\frac{13}{324}\pi^2 + \mathcal{O}(\e). &
\end{flalign}

\subsection{Four-particle final states}

%NXggg3

%N^{3}
\begin{flalign}
{\cal T}^{(3)}_{\glu ggg}\Big|_{N^{3}} = &
+\frac{1}{\e^6}\left(-\frac{113}{18}\right)
+\frac{1}{\e^5}\left(-43\right)
+\frac{1}{\e^4}\left(-\frac{33217}{162}+\frac{3233}{216}\pi^2\right)
\nonumber &\\&
+\frac{1}{\e^3}\left(-\frac{2057893}{1944}+\frac{57299}{648}\pi^2+\frac{4655}{18}\zeta_{3}\right)
\nonumber &\\&
+\frac{1}{\e^2}\left(-\frac{10366801}{1944}+\frac{1757471}{3888}\pi^2+\frac{84745}{54}\zeta_{3}-\frac{205097}{25920}\pi^4\right)
\nonumber &\\&
+\frac{1}{\e}\left(-\frac{3739171981}{139968}+\frac{28013821}{11664}\pi^2+\frac{2704571}{324}\zeta_{3}-\frac{962707}{25920}\pi^4\right.
\nonumber &\\&\phantom{+\frac{1}{\e}\bigg(}
\left.-\frac{5337}{8}\pi^2\zeta_{3}+\frac{307039}{90}\zeta_{5}\right)
\nonumber &\\&
-\frac{56105413465}{419904}+\frac{435252581}{34992}\pi^2+\frac{9803681}{216}\zeta_{3}-\frac{9924221}{51840}\pi^4
\nonumber &\\&
-\frac{28985}{8}\pi^2\zeta_{3}+\frac{3504919}{180}\zeta_{5}+\frac{15608711}{6531840}\pi^6-\frac{230473}{36}\zeta_{3}^2 + \mathcal{O}(\e), &
\end{flalign}

%NF N^{2}
\begin{flalign}
{\cal T}^{(3)}_{\glu ggg}\Big|_{\NF N^{2}} = &
+\frac{1}{\e^5}\left(\frac{5}{2}\right)
+\frac{1}{\e^4}\left(\frac{37}{4}\right)
+\frac{1}{\e^3}\left(\frac{2401}{54}-\frac{11}{3}\pi^2\right)
\nonumber &\\&
+\frac{1}{\e^2}\left(\frac{258299}{1296}-\frac{55}{4}\pi^2-\frac{188}{3}\zeta_{3}\right)
\nonumber &\\&
+\frac{1}{\e}\left(\frac{563651}{648}-\frac{259907}{3888}\pi^2-\frac{6520}{27}\zeta_{3}+\frac{1559}{2160}\pi^4\right)
\nonumber &\\&
+\frac{43566037}{11664}-\frac{294227}{972}\pi^2-\frac{257987}{216}\zeta_{3}+\frac{92927}{38880}\pi^4
\nonumber &\\&
+\frac{375}{4}\pi^2\zeta_{3}-\frac{9569}{18}\zeta_{5} + \mathcal{O}(\e), &
\end{flalign}

%NG N^{3}
\begin{flalign}
{\cal T}^{(3)}_{\glu ggg}\Big|_{\NG N^{3}} = &
+\frac{1}{\e^5}\left(\frac{5}{2}\right)
+\frac{1}{\e^4}\left(\frac{37}{4}\right)
+\frac{1}{\e^3}\left(\frac{2401}{54}-\frac{11}{3}\pi^2\right)
\nonumber &\\&
+\frac{1}{\e^2}\left(\frac{258299}{1296}-\frac{55}{4}\pi^2-\frac{188}{3}\zeta_{3}\right)
\nonumber &\\&
+\frac{1}{\e}\left(\frac{563651}{648}-\frac{259907}{3888}\pi^2-\frac{6520}{27}\zeta_{3}+\frac{1559}{2160}\pi^4\right)
\nonumber &\\&
+\frac{43566037}{11664}-\frac{294227}{972}\pi^2-\frac{257987}{216}\zeta_{3}+\frac{92927}{38880}\pi^4
\nonumber &\\&
+\frac{375}{4}\pi^2\zeta_{3}-\frac{9569}{18}\zeta_{5} + \mathcal{O}(\e), &
\end{flalign}

%NXqqbg3

%N^{2}
\begin{flalign}
{\cal T}^{(3)}_{\glu q\bar{q}g}\Big|_{\NF N^{2}} = &
+\frac{1}{\e^5}\left(\frac{103}{54}\right)
+\frac{1}{\e^4}\left(\frac{5179}{324}\right)
+\frac{1}{\e^3}\left(\frac{41695}{486}-\frac{2999}{648}\pi^2\right)
\nonumber &\\&
+\frac{1}{\e^2}\left(\frac{5275099}{11664}-\frac{132557}{3888}\pi^2-\frac{1469}{18}\zeta_{3}\right)
\nonumber &\\&
+\frac{1}{\e}\left(\frac{81435089}{34992}-\frac{1099223}{5832}\pi^2-\frac{66833}{108}\zeta_{3}+\frac{62333}{25920}\pi^4\right)
\nonumber &\\&
+\frac{2483622149}{209952}-\frac{142941515}{139968}\pi^2-\frac{2271103}{648}\zeta_{3}+\frac{9081}{640}\pi^4
\nonumber &\\&
+\frac{45271}{216}\pi^2\zeta_{3}-\frac{10837}{10}\zeta_{5} + \mathcal{O}(\e), &
\end{flalign}

%N^{0}
\begin{flalign}
{\cal T}^{(3)}_{\glu q\bar{q}g}\Big|_{\NF} = &
+\frac{1}{\e^5}\left(-\frac{79}{108}\right)
+\frac{1}{\e^4}\left(-\frac{515}{81}\right)
+\frac{1}{\e^3}\left(-\frac{37897}{972}+\frac{817}{432}\pi^2\right)
\nonumber &\\&
+\frac{1}{\e^2}\left(-\frac{1278703}{5832}+\frac{9991}{648}\pi^2+\frac{3773}{108}\zeta_{3}\right)
\nonumber &\\&
+\frac{1}{\e}\left(-\frac{20662283}{17496}+\frac{93157}{972}\pi^2+\frac{23317}{81}\zeta_{3}-\frac{50093}{51840}\pi^4\right)
\nonumber &\\&
-\frac{40789202}{6561}+\frac{6377351}{11664}\pi^2+\frac{1753931}{972}\zeta_{3}-\frac{562619}{77760}\pi^4
\nonumber &\\&
-\frac{39811}{432}\pi^2\zeta_{3}+\frac{92029}{180}\zeta_{5} + \mathcal{O}(\e), &
\end{flalign}

%N^{-2}
\begin{flalign}
{\cal T}^{(3)}_{\glu q\bar{q}g}\Big|_{\NF N^{-2}} = &
+\frac{1}{\e^5}\left(\frac{1}{9}\right)
+\frac{1}{\e^4}\left(\frac{26}{27}\right)
+\frac{1}{\e^3}\left(\frac{4181}{648}-\frac{31}{108}\pi^2\right)
\nonumber &\\&
+\frac{1}{\e^2}\left(\frac{74041}{1944}-\frac{797}{324}\pi^2-\frac{50}{9}\zeta_{3}\right)
\nonumber &\\&
+\frac{1}{\e}\left(\frac{614119}{2916}-\frac{127541}{7776}\pi^2-\frac{2555}{54}\zeta_{3}+\frac{1733}{12960}\pi^4\right)
\nonumber &\\&
+\frac{19639597}{17496}-\frac{2254897}{23328}\pi^2-\frac{204217}{648}\zeta_{3}+\frac{10927}{9720}\pi^4
\nonumber &\\&
+\frac{133}{9}\pi^2\zeta_{3}-\frac{3778}{45}\zeta_{5} + \mathcal{O}(\e), &
\end{flalign}

%NF N^{1}
\begin{flalign}
{\cal T}^{(3)}_{\glu q\bar{q}g}\Big|_{\NFSquare N} = &
+\frac{1}{\e^4}\left(-\frac{95}{162}\right)
+\frac{1}{\e^3}\left(-\frac{2143}{972}\right)
\nonumber &\\&
+\frac{1}{\e^2}\left(-\frac{809}{108}+\frac{193}{324}\pi^2\right)
+\frac{1}{\e}\left(-\frac{309853}{17496}+\frac{215}{216}\pi^2+\frac{800}{81}\zeta_{3}\right)
\nonumber &\\&
+\frac{983413}{104976}-\frac{3497}{972}\pi^2+\frac{2984}{243}\zeta_{3}+\frac{527}{3888}\pi^4 + \mathcal{O}(\e), &
\end{flalign}

%NF N^{-1}
\begin{flalign}
{\cal T}^{(3)}_{\glu q\bar{q}g}\Big|_{\NF N^{-1}} = &
+\frac{1}{\e^4}\left(\frac{5}{54}\right)
+\frac{1}{\e^3}\left(\frac{103}{324}\right)
\nonumber &\\&
+\frac{1}{\e^2}\left(\frac{121}{216}-\frac{23}{324}\pi^2\right)
+\frac{1}{\e}\left(-\frac{30247}{11664}+\frac{239}{1944}\pi^2-\frac{34}{27}\zeta_{3}\right)
\nonumber &\\&
-\frac{2619503}{69984}+\frac{1331}{432}\pi^2+\frac{179}{81}\zeta_{3}-\frac{127}{3240}\pi^4 + \mathcal{O}(\e), &
\end{flalign}

%NG N^{2}
\begin{flalign}
{\cal T}^{(3)}_{\glu q\bar{q}g}\Big|_{\NG \NF N^{2}} = &
+\frac{1}{\e^4}\left(-\frac{95}{162}\right)
+\frac{1}{\e^3}\left(-\frac{2143}{972}\right)
+\frac{1}{\e^2}\left(-\frac{809}{108}+\frac{193}{324}\pi^2\right)
\nonumber &\\&
+\frac{1}{\e}\left(-\frac{309853}{17496}+\frac{215}{216}\pi^2+\frac{800}{81}\zeta_{3}\right)
\nonumber &\\&
+\frac{983413}{104976}-\frac{3497}{972}\pi^2+\frac{2984}{243}\zeta_{3}+\frac{527}{3888}\pi^4 + \mathcal{O}(\e), &
\end{flalign}

%NG N^{0}
\begin{flalign}
{\cal T}^{(3)}_{\glu q\bar{q}g}\Big|_{\NG \NF} = &
+\frac{1}{\e^4}\left(\frac{5}{54}\right)
+\frac{1}{\e^3}\left(\frac{103}{324}\right)
\nonumber &\\&
+\frac{1}{\e^2}\left(\frac{121}{216}-\frac{23}{324}\pi^2\right)
+\frac{1}{\e}\left(-\frac{30247}{11664}+\frac{239}{1944}\pi^2-\frac{34}{27}\zeta_{3}\right)
\nonumber &\\&
-\frac{2619503}{69984}+\frac{1331}{432}\pi^2+\frac{179}{81}\zeta_{3}-\frac{127}{3240}\pi^4 + \mathcal{O}(\e), &
\end{flalign}

%NXXXg3

%N^{3}
\begin{flalign}
{\cal T}^{(3)}_{\glu \glu' \glu' g}\Big|_{(\NG-1)N^{3}} = &
+\frac{1}{\e^5}\left(\frac{11}{4}\right)
+\frac{1}{\e^4}\left(\frac{839}{36}\right)
+\frac{1}{\e^3}\left(\frac{85039}{648}-\frac{8821}{1296}\pi^2\right)
\nonumber &\\&
+\frac{1}{\e^2}\left(\frac{919639}{1296}-\frac{202067}{3888}\pi^2-\frac{13187}{108}\zeta_{3}\right)
\nonumber &\\&
+\frac{1}{\e}\left(\frac{14458787}{3888}-\frac{7015283}{23328}\pi^2-\frac{309097}{324}\zeta_{3}+\frac{181691}{51840}\pi^4\right)
\nonumber &\\&
+\frac{1341517259}{69984}-\frac{232999109}{139968}\pi^2-\frac{5466911}{972}\zeta_{3}+\frac{3506753}{155520}\pi^4
\nonumber &\\&
+\frac{15193}{48}\pi^2\zeta_{3}-\frac{302207}{180}\zeta_{5} + \mathcal{O}(\e), &
\end{flalign}

%NF N^{2}
\begin{flalign}
{\cal T}^{(3)}_{\glu \glu' \glu' g}\Big|_{(\NG-1)\NF N^{2}} = &
+\frac{1}{\e^4}\left(-\frac{55}{81}\right)
+\frac{1}{\e^3}\left(-\frac{613}{243}\right)
+\frac{1}{\e^2}\left(-\frac{1739}{216}+\frac{2}{3}\pi^2\right)
\nonumber &\\&
+\frac{1}{\e}\left(-\frac{528965}{34992}+\frac{212}{243}\pi^2+\frac{902}{81}\zeta_{3}\right)
\nonumber &\\&
+\frac{9825335}{209952}-\frac{25967}{3888}\pi^2+\frac{2447}{243}\zeta_{3}+\frac{3397}{19440}\pi^4 + \mathcal{O}(\e), &
\end{flalign}

%NG N^{3}
\begin{flalign}
{\cal T}^{(3)}_{\glu \glu' \glu' g}\Big|_{(\NG-1)\NG N^{3}} = &
+\frac{1}{\e^4}\left(-\frac{55}{81}\right)
+\frac{1}{\e^3}\left(-\frac{613}{243}\right)
+\frac{1}{\e^2}\left(-\frac{1739}{216}+\frac{2}{3}\pi^2\right)
\nonumber &\\&
+\frac{1}{\e}\left(-\frac{528965}{34992}+\frac{212}{243}\pi^2+\frac{902}{81}\zeta_{3}\right)
\nonumber &\\&
+\frac{9825335}{209952}-\frac{25967}{3888}\pi^2+\frac{2447}{243}\zeta_{3}+\frac{3397}{19440}\pi^4 + \mathcal{O}(\e), &
\end{flalign}

%NXXXghard3

%N^{3}
\begin{flalign}
\Delta{\cal T}^{(3)}_{\glu \glu \glu g}\Big|_{N^{3}} = &
+\frac{1}{\e^4}\left(\frac{13}{8}\right)
+\frac{1}{\e^3}\left(\frac{279}{16}-\frac{5}{12}\pi^2+\frac{5}{3}\zeta_{3}\right)
\nonumber &\\&
+\frac{1}{\e^2}\left(\frac{37777}{288}-\frac{1879}{288}\pi^2-\frac{409}{18}\zeta_{3}+\frac{37}{216}\pi^4\right)
\nonumber &\\&
+\frac{1}{\e}\left(\frac{376769}{432}-\frac{86713}{1728}\pi^2-\frac{44717}{216}\zeta_{3}-\frac{3269}{6480}\pi^4-\frac{199}{36}\pi^2\zeta_{3}+\frac{314}{3}\zeta_{5}\right)
\nonumber &\\&
+\frac{18427153}{3456}-\frac{1227661}{3456}\pi^2-\frac{1550117}{1296}\zeta_{3}-\frac{398077}{311040}\pi^4
\nonumber &\\&
+\frac{8021}{108}\pi^2\zeta_{3}-\frac{26983}{36}\zeta_{5}+\frac{743}{12960}\pi^6-\frac{338}{3}\zeta_{3}^2 + \mathcal{O}(\e), &
\end{flalign}

%NF N^{2}
\begin{flalign}
\Delta{\cal T}^{(3)}_{\glu \glu \glu g}\Big|_{\NF N^{2}} = &
+\frac{1}{\e^3}\left(-\frac{1}{6}\right)
+\frac{1}{\e^2}\left(-\frac{29}{144}+\frac{5}{72}\pi^2-\frac{5}{18}\zeta_{3}\right)
\nonumber &\\&
+\frac{1}{\e}\left(\frac{3275}{864}-\frac{\pi^2}{54}+\frac{82}{27}\zeta_{3}-\frac{13}{648}\pi^4\right)
\nonumber &\\&
+\frac{97051}{1728}-\frac{715}{216}\pi^2-\frac{1081}{324}\zeta_{3}+\frac{547}{4860}\pi^4+\frac{25}{108}\pi^2\zeta_{3}-\frac{68}{9}\zeta_{5} + \mathcal{O}(\e), &
\end{flalign}

%NG N^{3}
\begin{flalign}
\Delta{\cal T}^{(3)}_{\glu \glu \glu g}\Big|_{\NG N^{3}} = &
+\frac{1}{\e^3}\left(-\frac{1}{6}\right)
+\frac{1}{\e^2}\left(-\frac{29}{144}+\frac{5}{72}\pi^2-\frac{5}{18}\zeta_{3}\right)
\nonumber &\\&
+\frac{1}{\e}\left(\frac{3275}{864}-\frac{\pi^2}{54}+\frac{82}{27}\zeta_{3}-\frac{13}{648}\pi^4\right)
\nonumber &\\&
+\frac{97051}{1728}-\frac{715}{216}\pi^2-\frac{1081}{324}\zeta_{3}+\frac{547}{4860}\pi^4+\frac{25}{108}\pi^2\zeta_{3}-\frac{68}{9}\zeta_{5} + \mathcal{O}(\e). &
\end{flalign}

\subsection{Five-particle final states}

%NXgggg3

%N^{3}
\begin{flalign}
{\cal T}^{(3)}_{\glu gggg}\Big|_{N^{3}} = &
+\frac{1}{\e^6}\left(\frac{5}{2}\right)
+\frac{1}{\e^5}\left(\frac{1625}{108}\right)
+\frac{1}{\e^4}\left(\frac{30611}{324}-\frac{53}{8}\pi^2\right)
\nonumber &\\&
+\frac{1}{\e^3}\left(\frac{89111}{162}-\frac{52357}{1296}\pi^2-\frac{1198}{9}\zeta_{3}\right)
\nonumber &\\&
+\frac{1}{\e^2}\left(\frac{35947711}{11664}-\frac{996125}{3888}\pi^2-\frac{29969}{36}\zeta_{3}+\frac{98561}{25920}\pi^4\right)
\nonumber &\\&
+\frac{1}{\e}\left(\frac{1177775537}{69984}-\frac{17491783}{11664}\pi^2-\frac{3495541}{648}\zeta_{3}+\frac{8501}{384}\pi^4\right.
\nonumber &\\&\phantom{+\frac{1}{\e}\bigg(}
\left.+\frac{39049}{108}\pi^2\zeta_{3}-\frac{14843}{9}\zeta_{5}\right)
\nonumber &\\&
+\frac{467664317}{5184}-\frac{294776213}{34992}\pi^2-\frac{124216589}{3888}\zeta_{3}+\frac{4131497}{31104}\pi^4
\nonumber &\\&
+\frac{980143}{432}\pi^2\zeta_{3}-\frac{772663}{72}\zeta_{5}-\frac{142039}{186624}\pi^6+\frac{15355}{4}\zeta_{3}^2 + \mathcal{O}(\e), &
\end{flalign}

%NXqqbgg3

%N^{2}
\begin{flalign}
{\cal T}^{(3)}_{\glu q \bar{q} gg}\Big|_{\NF N^{2}} = &
+\frac{1}{\e^5}\left(-\frac{34}{27}\right)
+\frac{1}{\e^4}\left(-\frac{254}{27}\right)
+\frac{1}{\e^3}\left(-\frac{120667}{1944}+\frac{1099}{324}\pi^2\right)
\nonumber &\\&
+\frac{1}{\e^2}\left(-\frac{8783047}{23328}+\frac{49741}{1944}\pi^2+\frac{3773}{54}\zeta_{3}\right)
\nonumber &\\&
+\frac{1}{\e}\left(-\frac{11249585}{5184}+\frac{3955307}{23328}\pi^2+\frac{43642}{81}\zeta_{3}-\frac{6139}{3240}\pi^4\right)
\nonumber &\\&
-\frac{10152741389}{839808}+\frac{288511147}{279936}\pi^2+\frac{2338273}{648}\zeta_{3}-\frac{104899}{7776}\pi^4
\nonumber &\\&
-\frac{41171}{216}\pi^2\zeta_{3}+\frac{79103}{90}\zeta_{5} + \mathcal{O}(\e), &
\end{flalign}

%N^{0}
\begin{flalign}
{\cal T}^{(3)}_{\glu q \bar{q} gg}\Big|_{\NF} = &
+\frac{1}{\e^5}\left(\frac{43}{108}\right)
+\frac{1}{\e^4}\left(\frac{2137}{648}\right)
+\frac{1}{\e^3}\left(\frac{43643}{1944}-\frac{157}{144}\pi^2\right)
\nonumber &\\&
+\frac{1}{\e^2}\left(\frac{1604873}{11664}-\frac{23447}{2592}\pi^2-\frac{1267}{54}\zeta_{3}\right)
\nonumber &\\&
+\frac{1}{\e}\left(\frac{55654211}{69984}-\frac{479309}{7776}\pi^2-\frac{126305}{648}\zeta_{3}+\frac{9167}{17280}\pi^4\right)
\nonumber &\\&
+\frac{1860006521}{419904}-\frac{5879509}{15552}\pi^2-\frac{322961}{243}\zeta_{3}+\frac{1372859}{311040}\pi^4
\nonumber &\\&
+\frac{3473}{54}\pi^2\zeta_{3}-\frac{30319}{90}\zeta_{5} + \mathcal{O}(\e), &
\end{flalign}

%N^{-2}
\begin{flalign}
{\cal T}^{(3)}_{\glu q \bar{q} gg}\Big|_{\NF N^{-2}} = &
+\frac{1}{\e^5}\left(-\frac{1}{18}\right)
+\frac{1}{\e^4}\left(-\frac{13}{27}\right)
+\frac{1}{\e^3}\left(-\frac{535}{162}+\frac{11}{72}\pi^2\right)
\nonumber &\\&
+\frac{1}{\e^2}\left(-\frac{157199}{7776}+\frac{143}{108}\pi^2+\frac{59}{18}\zeta_{3}\right)
\nonumber &\\&
+\frac{1}{\e}\left(-\frac{5427671}{46656}+\frac{11785}{1296}\pi^2+\frac{767}{27}\zeta_{3}-\frac{1993}{25920}\pi^4\right)
\nonumber &\\&
-\frac{180508649}{279936}+\frac{1733569}{31104}\pi^2+\frac{126665}{648}\zeta_{3}-\frac{25909}{38880}\pi^4
\nonumber &\\&
-\frac{1951}{216}\pi^2\zeta_{3}+\frac{4153}{90}\zeta_{5} + \mathcal{O}(\e), &
\end{flalign}

%NXqqbQQb3

%N^{1}
\begin{flalign}
{\cal T}^{(3)}_{\glu q \bar{q} q' \bar{q}'}\Big|_{(\NF-1)\NF N} = &
+\frac{1}{\e^4}\left(\frac{1}{18}\right)
+\frac{1}{\e^3}\left(\frac{155}{324}\right)
+\frac{1}{\e^2}\left(\frac{2171}{648}-\frac{103}{648}\pi^2\right)
\nonumber &\\&
+\frac{1}{\e}\left(\frac{13813}{648}-\frac{5327}{3888}\pi^2-\frac{191}{54}\zeta_{3}\right)
\nonumber &\\&
+\frac{4461685}{34992}-\frac{24773}{2592}\pi^2-\frac{9923}{324}\zeta_{3}+\frac{6331}{77760}\pi^4 + \mathcal{O}(\e), &
\end{flalign}

%N^{-1}
\begin{flalign}
{\cal T}^{(3)}_{\glu q \bar{q} q' \bar{q}'}\Big|_{(\NF-1)\NF N^{-1}} = &
+\frac{1}{\e^4}\left(-\frac{1}{108}\right)
+\frac{1}{\e^3}\left(-\frac{31}{324}\right)
+\frac{1}{\e^2}\left(-\frac{157}{216}+\frac{37}{1296}\pi^2\right)
\nonumber &\\&
+\frac{1}{\e}\left(-\frac{56531}{11664}+\frac{1147}{3888}\pi^2+\frac{77}{108}\zeta_{3}\right)
\nonumber &\\&
-\frac{2107297}{69984}+\frac{5785}{2592}\pi^2+\frac{2387}{324}\zeta_{3}-\frac{341}{31104}\pi^4 + \mathcal{O}(\e), &
\end{flalign}

%NXqqbqqb3

%N^{0}
\begin{flalign}
\Delta{\cal T}^{(3)}_{\glu q \bar{q} q \bar{q}}\Big|_{\NF} = &
+\frac{1}{\e^2}\left(-\frac{13}{144}+\frac{\pi^2}{72}-\frac{\zeta_{3}}{18}\right)
+\frac{1}{\e}\left(-\frac{1477}{864}+\frac{37}{216}\pi^2+\frac{47}{108}\zeta_{3}-\frac{23}{3240}\pi^4\right)
\nonumber &\\&
-\frac{98227}{5184}+\frac{8059}{5184}\pi^2+\frac{5195}{648}\zeta_{3}-\frac{1357}{38880}\pi^4+\frac{29}{216}\pi^2\zeta_{3}-\frac{13}{3}\zeta_{5} + \mathcal{O}(\e), &
\end{flalign}

%N^{-2}
\begin{flalign}
\Delta{\cal T}^{(3)}_{\glu q \bar{q} q \bar{q}}\Big|_{\NF N^{-2}} = &
+\frac{1}{\e^2}\left(\frac{13}{144}-\frac{\pi^2}{72}+\frac{\zeta_{3}}{18}\right)
+\frac{1}{\e}\left(\frac{1459}{864}-\frac{17}{108}\pi^2-\frac{29}{108}\zeta_{3}+\frac{2}{405}\pi^4\right)
\nonumber &\\&
+\frac{96877}{5184}-\frac{7411}{5184}\pi^2-\frac{833}{162}\zeta_{3}+\frac{1331}{38880}\pi^4-\frac{37}{216}\pi^2\zeta_{3}+\frac{22}{9}\zeta_{5} + \mathcal{O}(\e), &
\end{flalign}

%NXXXgg3

%N^{3}
\begin{flalign}
{\cal T}^{(3)}_{\glu \glu' \glu' gg}\Big|_{(\NG-1)N^{3}} = &
+\frac{1}{\e^5}\left(-\frac{185}{108}\right)
+\frac{1}{\e^4}\left(-\frac{8545}{648}\right)
+\frac{1}{\e^3}\left(-\frac{9485}{108}+\frac{6007}{1296}\pi^2\right)
\nonumber &\\&
+\frac{1}{\e^2}\left(-\frac{6232195}{11664}+\frac{279601}{7776}\pi^2+\frac{1739}{18}\zeta_{3}\right)
\nonumber &\\&
+\frac{1}{\e}\left(-\frac{215665115}{69984}+\frac{1401341}{5832}\pi^2+\frac{493849}{648}\zeta_{3}-\frac{43237}{17280}\pi^4\right)
\nonumber &\\&
-\frac{2402380063}{139968}+\frac{204972215}{139968}\pi^2+\frac{4989251}{972}\zeta_{3}-\frac{5776091}{311040}\pi^4
\nonumber &\\&
-\frac{28507}{108}\pi^2\zeta_{3}+\frac{22715}{18}\zeta_{5} + \mathcal{O}(\e), &
\end{flalign}

%NXXXgghard3

%N^{3}
\begin{flalign}
\Delta{\cal T}^{(3)}_{\glu \glu \glu gg}\Big|_{N^{3}} = &
+\frac{1}{\e^4}\left(-\frac{7}{8}\right)
+\frac{1}{\e^3}\left(-\frac{83}{8}+\frac{5}{12}\pi^2-\frac{5}{3}\zeta_{3}\right)
\nonumber &\\&
+\frac{1}{\e^2}\left(-\frac{13285}{144}+\frac{1295}{288}\pi^2+\frac{64}{3}\zeta_{3}-\frac{37}{216}\pi^4\right)
\nonumber &\\&
+\frac{1}{\e}\left(-\frac{146719}{216}+\frac{595}{16}\pi^2+\frac{3719}{24}\zeta_{3}+\frac{19}{54}\pi^4+\frac{23}{4}\pi^2\zeta_{3}-\frac{308}{3}\zeta_{5}\right)
\nonumber &\\&
-\frac{23181667}{5184}+\frac{20929}{72}\pi^2+\frac{212881}{216}\zeta_{3}+\frac{407}{3840}\pi^4-\frac{797}{12}\pi^2\zeta_{3}
\nonumber &\\&
+\frac{2477}{4}\zeta_{5}-\frac{499}{30240}\pi^6+\frac{773}{6}\zeta_{3}^2 + \mathcal{O}(\e), &
\end{flalign}

%NXXXqqb3

%N^{2}
\begin{flalign}
{\cal T}^{(3)}_{\glu \glu' \glu' q \bar{q}}\Big|_{(\NG-1)\NF N^{2}} = &
+\frac{1}{\e^4}\left(\frac{13}{108}\right)
+\frac{1}{\e^3}\left(\frac{341}{324}\right)
+\frac{1}{\e^2}\left(\frac{4813}{648}-\frac{449}{1296}\pi^2\right)
\nonumber &\\&
+\frac{1}{\e}\left(\frac{553799}{11664}-\frac{11801}{3888}\pi^2-\frac{841}{108}\zeta_{3}\right)
\nonumber &\\&
+\frac{19954037}{69984}-\frac{55331}{2592}\pi^2-\frac{7411}{108}\zeta_{3}+\frac{27029}{155520}\pi^4 + \mathcal{O}(\e), &
\end{flalign}

%N^{0}
\begin{flalign}
{\cal T}^{(3)}_{\glu \glu' \glu' q \bar{q}}\Big|_{(\NG-1)\NF} = &
+\frac{1}{\e^4}\left(-\frac{1}{108}\right)
+\frac{1}{\e^3}\left(-\frac{31}{324}\right)
+\frac{1}{\e^2}\left(-\frac{157}{216}+\frac{37}{1296}\pi^2\right)
\nonumber &\\&
+\frac{1}{\e}\left(-\frac{56531}{11664}+\frac{1147}{3888}\pi^2+\frac{77}{108}\zeta_{3}\right)
\nonumber &\\&
-\frac{2107297}{69984}+\frac{5785}{2592}\pi^2+\frac{2387}{324}\zeta_{3}-\frac{341}{31104}\pi^4 + \mathcal{O}(\e), &
\end{flalign}

%NXXXqqbhard3

%N^{2}
\begin{flalign}
\Delta{\cal T}^{(3)}_{\glu \glu \glu q \bar{q}}\Big|_{\NF N^{2}} = &
+\frac{1}{\e^3}\left(\frac{1}{24}\right)
+\frac{1}{\e^2}\left(\frac{25}{36}-\frac{\pi^2}{24}+\frac{\zeta_{3}}{6}\right)
\nonumber &\\&
+\frac{1}{\e}\left(\frac{743}{108}-\frac{109}{288}\pi^2-\frac{3}{2}\zeta_{3}+\frac{2}{135}\pi^4\right)
\nonumber &\\&
+\frac{129761}{2592}-\frac{2729}{864}\pi^2-\frac{2377}{216}\zeta_{3}+\frac{251}{6480}\pi^4-\frac{43}{72}\pi^2\zeta_{3}+8 \zeta_{5} + \mathcal{O}(\e), &
\end{flalign}

%NXXXXX3

%N^{3}
\begin{flalign}
{\cal T}^{(3)}_{\glu \glu' \glu' \glu'' \glu''}\Big|_{(\NG-2)(\NG-1)N^{3}} = &
+\frac{1}{\e^4}\left(\frac{7}{108}\right)
+\frac{1}{\e^3}\left(\frac{31}{54}\right)
+\frac{1}{\e^2}\left(\frac{1321}{324}-\frac{3}{16}\pi^2\right)
\nonumber &\\&
+\frac{1}{\e}\left(\frac{305165}{11664}-\frac{1079}{648}\pi^2-\frac{17}{4}\zeta_{3}\right)
\nonumber &\\&
+\frac{3676889}{23328}-\frac{5093}{432}\pi^2-\frac{6155}{162}\zeta_{3}+\frac{4789}{51840}\pi^4 + \mathcal{O}(\e), &
\end{flalign}

%NXXXXXSAMEFLAV3

%N^{3}
\begin{flalign}
\Delta{\cal T}^{(3)}_{\glu \glu' \glu' \glu' \glu'}\Big|_{(\NG-1)N^{3}} = &
+\frac{1}{\e^2}\left(\frac{13}{72}-\frac{\pi^2}{36}+\frac{\zeta_{3}}{9}\right)
+\frac{1}{\e}\left(\frac{367}{108}-\frac{71}{216}\pi^2-\frac{19}{27}\zeta_{3}+\frac{13}{1080}\pi^4\right)
\nonumber &\\&
+\frac{6097}{162}-\frac{7735}{2592}\pi^2-\frac{8527}{648}\zeta_{3}+\frac{28}{405}\pi^4-\frac{11}{36}\pi^2\zeta_{3}+\frac{61}{9}\zeta_{5} + \mathcal{O}(\e), &
\end{flalign}

%NXXXXXhard3

%N^{3}
\begin{flalign}
\Delta{\cal T}^{(3)}_{\glu \glu \glu \glu' \glu'}\Big|_{(\NG-1)N^{3}} = &
+\frac{1}{\e^3}\left(\frac{1}{24}\right)
+\frac{1}{\e^2}\left(\frac{25}{36}-\frac{\pi^2}{24}+\frac{\zeta_{3}}{6}\right)
\nonumber &\\&
+\frac{1}{\e}\left(\frac{743}{108}-\frac{109}{288}\pi^2-\frac{3}{2}\zeta_{3}+\frac{2}{135}\pi^4\right)
\nonumber &\\&
+\frac{129761}{2592}-\frac{2729}{864}\pi^2-\frac{2377}{216}\zeta_{3}+\frac{251}{6480}\pi^4-\frac{43}{72}\pi^2\zeta_{3}+8 \zeta_{5} + \mathcal{O}(\e), &
\end{flalign}

%NXXXXXhardhard3

%N^{3}
\begin{flalign}
\Delta{\cal T}^{(3)}_{\glu \glu \glu \glu \glu}\Big|_{N^{3}} = &
+\frac{1}{\e}\left(\frac{13}{32}-\frac{\pi^2}{16}+\frac{\zeta_{3}}{4}\right)
\nonumber &\\&
+\frac{25}{6}-\frac{397}{576}\pi^2+\frac{11}{12}\zeta_{3}-\frac{\pi^4}{288}+\frac{5}{72}\pi^2\zeta_{3}+\frac{41}{24}\zeta_{5} + \mathcal{O}(\e). &
\end{flalign}

%%%%%%%%%%%%%%%%%%%%%%%%%%%%%%%%%%%%%%%%%%%%%%%%%%%%%%%%%%%%%%%%%%%%%%%%%%%%%%%%

\section{Lower order results}\label{app:exprlower}

\subsection{NLO}
%NXg1

%N^{1}
\begin{flalign}
{\cal T}^{(1)}_{\glu g}\Big|_{N} = &
+\frac{1}{\e^2}\left(-2\right)
+\frac{1}{\e}\left(-\frac{10}{3}\right)
+\left(\frac{7}{6}\pi^2\right)
+{\e}\left(-2+\frac{14}{3}\zeta_{3}\right)
\nonumber &\\&
+{\e^2}\left(-6-\frac{73}{720}\pi^4\right)
+{\e^3}\left(-14+\frac{7}{6}\pi^2-\frac{49}{18}\pi^2\zeta_{3}+\frac{62}{5}\zeta_{5}\right)
\nonumber &\\&
+{\e^4}\left(-30+\frac{7}{2}\pi^2+\frac{14}{3}\zeta_{3}-\frac{437}{60480}\pi^6-\frac{49}{9}\zeta_{3}^2\right) + \mathcal{O}({\e^5}), &
\end{flalign}

%NF N^{0}
\begin{flalign}
{\cal T}^{(1)}_{\glu g}\Big|_{\NF} = & +\frac{1}{\e}\left(\frac{1}{3}\right), &
\end{flalign}

%NG N^{1}
\begin{flalign}
{\cal T}^{(1)}_{\glu g}\Big|_{\NG N} = & +\frac{1}{\e}\left(\frac{1}{3}\right), &
\end{flalign}

%NXgg1

%N^{1}
\begin{flalign}
{\cal T}^{(1)}_{\glu gg}\Big|_{N} = &
+\frac{1}{\e^2}\left(2\right)
+\frac{1}{\e}\left(\frac{10}{3}\right)
+\left(\frac{34}{3}-\frac{7}{6}\pi^2\right)
+{\e}\left(\frac{209}{6}-\frac{35}{18}\pi^2-\frac{50}{3}\zeta_{3}\right)
\nonumber &\\&
+{\e^2}\left(\frac{421}{4}-\frac{119}{18}\pi^2-\frac{250}{9}\zeta_{3}-\frac{71}{720}\pi^4\right)
\nonumber &\\&
+{\e^3}\left(\frac{2531}{8}-\frac{1463}{72}\pi^2-\frac{850}{9}\zeta_{3}-\frac{71}{432}\pi^4+\frac{175}{18}\pi^2\zeta_{3}-\frac{482}{5}\zeta_{5}\right)
\nonumber &\\&
+{\e^4}\bigg(\frac{15193}{16}-\frac{2947}{48}\pi^2-\frac{5225}{18}\zeta_{3}-\frac{1207}{2160}\pi^4
\nonumber &\\& \phantom{+{\e^4}\bigg(}
+\frac{875}{54}\pi^2\zeta_{3}-\frac{482}{3}\zeta_{5}-\frac{4027}{60480}\pi^6+\frac{625}{9}\zeta_{3}^2\bigg) + \mathcal{O}({\e^5}), &
\end{flalign}

%NXqqb1

%N^{0}
\begin{flalign}
{\cal T}^{(1)}_{\glu q\bar{q}}\Big|_{\NF} = &
+\frac{1}{\e}\left(-\frac{1}{3}\right)
+\left(-1\right)
+{\e}\left(-3+\frac{7}{36}\pi^2\right)
+{\e^2}\left(-9+\frac{7}{12}\pi^2+\frac{25}{9}\zeta_{3}\right)
\nonumber &\\&
+{\e^3}\left(-27+\frac{7}{4}\pi^2+\frac{25}{3}\zeta_{3}+\frac{71}{4320}\pi^4\right)
\nonumber &\\&
+{\e^4}\left(-81+\frac{21}{4}\pi^2+25 \zeta_{3}+\frac{71}{1440}\pi^4-\frac{175}{108}\pi^2\zeta_{3}+\frac{241}{15}\zeta_{5}\right) + \mathcal{O}({\e^5}), &
\end{flalign}

%NXXX1

%N^{1}
\begin{flalign}
{\cal T}^{(1)}_{\glu \glu' \glu'}\Big|_{(\NG-1)N} = &
+\frac{1}{\e}\left(-\frac{1}{3}\right)
+\left(-1\right)
+{\e}\left(-3+\frac{7}{36}\pi^2\right)
+{\e^2}\left(-9+\frac{7}{12}\pi^2+\frac{25}{9}\zeta_{3}\right)
\nonumber &\\&
+{\e^3}\left(-27+\frac{7}{4}\pi^2+\frac{25}{3}\zeta_{3}+\frac{71}{4320}\pi^4\right)
\nonumber &\\&
+{\e^4}\left(-81+\frac{21}{4}\pi^2+25 \zeta_{3}+\frac{71}{1440}\pi^4-\frac{175}{108}\pi^2\zeta_{3}+\frac{241}{15}\zeta_{5}\right) + \mathcal{O}({\e^5}), &
\end{flalign}

%NXXXhard1

%N^{1}
\begin{flalign}
\Delta{\cal T}^{(1)}_{\glu \glu \glu}\Big|_{N} = &
+\left(-\frac{1}{6}\right)
+{\e}\left(-\frac{5}{12}\right)
+{\e^2}\left(-\frac{31}{24}+\frac{7}{72}\pi^2\right)
+{\e^3}\left(-\frac{197}{48}+\frac{35}{144}\pi^2+\frac{25}{18}\zeta_{3}\right)
\nonumber &\\&
+{\e^4}\left(-\frac{1231}{96}+\frac{217}{288}\pi^2+\frac{125}{36}\zeta_{3}+\frac{71}{8640}\pi^4\right) + \mathcal{O}({\e^5}). &
\end{flalign}

\subsection{NNLO}

%NXg22x0

%N^{2}
\begin{flalign}
{\cal T}^{(2,\left[2\times 0\right])}_{\glu g}\Big|_{N^{2}} = &
+\frac{1}{\e^4}\left(1\right)
+\frac{1}{\e^3}\left(\frac{73}{12}\right)
+\frac{1}{\e^2}\left(\frac{143}{36}-\frac{25}{12}\pi^2\right)
+\frac{1}{\e}\left(-\frac{781}{216}-\frac{133}{72}\pi^2-\frac{25}{6}\zeta_{3}\right)
\nonumber &\\&
+\left(\frac{21923}{1296}+\frac{997}{216}\pi^2-\frac{253}{18}\zeta_{3}+\frac{31}{40}\pi^4\right)
\nonumber &\\&
+{\e}\left(\frac{519827}{7776}+\frac{937}{1296}\pi^2-\frac{845}{54}\zeta_{3}-\frac{21}{20}\pi^4+\frac{323}{36}\pi^2\zeta_{3}+\frac{71}{10}\zeta_{5}\right)
\nonumber &\\&
+{\e^2}\bigg(\frac{10159787}{46656}-\frac{172967}{7776}\pi^2-\frac{21085}{162}\zeta_{3}-\frac{14591}{4320}\pi^4
\nonumber &\\&\phantom{+{\e^2}\bigg(}
+\frac{185}{54}\pi^2\zeta_{3}-\frac{751}{30}\zeta_{5}+\frac{491}{10080}\pi^6+\frac{901}{18}\zeta_{3}^2\bigg) + \mathcal{O}({\e^3}), &
\end{flalign}

%NF N^{1}
\begin{flalign}
{\cal T}^{(2,\left[2\times 0\right])}_{\glu g}\Big|_{\NF N} = &
+\frac{1}{\e^3}\left(-\frac{5}{6}\right)
+\frac{1}{\e^2}\left(-\frac{41}{36}\right)
+\frac{1}{\e}\left(\frac{55}{54}+\frac{2}{9}\pi^2\right)
\nonumber &\\&
+\left(-\frac{3053}{1296}-\frac{65}{108}\pi^2+\frac{5}{9}\zeta_{3}\right)
+{\e}\left(-\frac{79361}{7776}-\frac{205}{1296}\pi^2-\frac{80}{27}\zeta_{3}+\frac{43}{480}\pi^4\right)
\nonumber &\\&
+{\e^2}\bigg(-\frac{1631345}{46656}+\frac{25937}{7776}\pi^2-\frac{131}{162}\zeta_{3}+\frac{101}{432}\pi^4
\nonumber &\\&\phantom{+{\e^2}\bigg(}
+\frac{269}{108}\pi^2\zeta_{3}-\frac{43}{15}\zeta_{5}\bigg) + \mathcal{O}({\e^3}), &
\end{flalign}

%NF N^{-1}
\begin{flalign}
{\cal T}^{(2,\left[2\times 0\right])}_{\glu g}\Big|_{\NF N^{-1}} = & +\frac{1}{\e}\left(-\frac{1}{8}\right), &
\end{flalign}

%NF^2 N^{0}
\begin{flalign}
{\cal T}^{(2,\left[2\times 0\right])}_{\glu g}\Big|_{\NFSquare} = & +\frac{1}{\e^2}\left(\frac{1}{12}\right), &
\end{flalign}

%NG N^{2}
\begin{flalign}
{\cal T}^{(2,\left[2\times 0\right])}_{\glu g}\Big|_{\NG N^{2}} = &
+\frac{1}{\e^3}\left(-\frac{5}{6}\right)
+\frac{1}{\e^2}\left(-\frac{41}{36}\right)
+\frac{1}{\e}\left(\frac{247}{216}+\frac{2}{9}\pi^2\right)
\nonumber &\\&
+\left(-\frac{3053}{1296}-\frac{65}{108}\pi^2+\frac{5}{9}\zeta_{3}\right)
+{\e}\left(-\frac{79361}{7776}-\frac{205}{1296}\pi^2-\frac{80}{27}\zeta_{3}+\frac{43}{480}\pi^4\right)
\nonumber &\\&
+{\e^2}\bigg(-\frac{1631345}{46656}+\frac{25937}{7776}\pi^2-\frac{131}{162}\zeta_{3}+\frac{101}{432}\pi^4
\nonumber &\\&\phantom{+{\e^2}\bigg(}
+\frac{269}{108}\pi^2\zeta_{3}-\frac{43}{15}\zeta_{5}\bigg) + \mathcal{O}({\e^3}), &
\end{flalign}

%NG NF N^{1}
\begin{flalign}
{\cal T}^{(2,\left[2\times 0\right])}_{\glu g}\Big|_{\NG \NF N} = & +\frac{1}{\e^2}\left(\frac{1}{6}\right), &
\end{flalign}

%NG^2 N^{2}
\begin{flalign}
{\cal T}^{(2,\left[2\times 0\right])}_{\glu g}\Big|_{\NGSquare N^{2}} = & +\frac{1}{\e^2}\left(\frac{1}{12}\right), &
\end{flalign}

%NXg21x1

%N^{2}
\begin{flalign}
{\cal T}^{(2,\left[1\times 1\right])}_{\glu g}\Big|_{N^{2}} = &
+\frac{1}{\e^4}\left(1\right)
+\frac{1}{\e^3}\left(\frac{10}{3}\right)
+\frac{1}{\e^2}\left(\frac{25}{9}-\frac{\pi^2}{6}\right)
+\frac{1}{\e}\left(2-\frac{35}{18}\pi^2-\frac{14}{3}\zeta_{3}\right)
\nonumber &\\&
+\left(\frac{28}{3}-\frac{70}{9}\zeta_{3}-\frac{7}{120}\pi^4\right)
+{\e}\left(24-\frac{\pi^2}{3}+\frac{73}{432}\pi^4+\frac{7}{9}\pi^2\zeta_{3}-\frac{62}{5}\zeta_{5}\right)
\nonumber &\\&
+{\e^2}\left(\frac{163}{3}-\frac{53}{18}\pi^2-\frac{28}{3}\zeta_{3}+\frac{245}{54}\pi^2\zeta_{3}-\frac{62}{3}\zeta_{5}-\frac{31}{3024}\pi^6+\frac{98}{9}\zeta_{3}^2\right) + \mathcal{O}({\e^3}), &
\end{flalign}

%NF N^{1}
\begin{flalign}
{\cal T}^{(2,\left[1\times 1\right])}_{\glu g}\Big|_{\NF N} = &
+\frac{1}{\e^3}\left(-\frac{1}{3}\right)
+\frac{1}{\e^2}\left(-\frac{5}{9}\right)
+\frac{1}{\e}\left(\frac{7}{36}\pi^2\right)
+\left(-\frac{1}{3}+\frac{7}{9}\zeta_{3}\right)
\nonumber &\\&
+{\e}\left(-1-\frac{73}{4320}\pi^4\right)
+{\e^2}\left(-\frac{7}{3}+\frac{7}{36}\pi^2-\frac{49}{108}\pi^2\zeta_{3}+\frac{31}{15}\zeta_{5}\right) + \mathcal{O}({\e^3}), &
\end{flalign}

%NF^2 N^{0}
\begin{flalign}
{\cal T}^{(2,\left[1\times 1\right])}_{\glu g}\Big|_{\NFSquare} = & +\frac{1}{\e^2}\left(\frac{1}{36}\right), &
\end{flalign}

%NG N^{2}
\begin{flalign}
{\cal T}^{(2,\left[1\times 1\right])}_{\glu g}\Big|_{\NG N^{2}} = &
+\frac{1}{\e^3}\left(-\frac{1}{3}\right)
+\frac{1}{\e^2}\left(-\frac{5}{9}\right)
+\frac{1}{\e}\left(\frac{7}{36}\pi^2\right)
+\left(-\frac{1}{3}+\frac{7}{9}\zeta_{3}\right)
\nonumber &\\&
+{\e}\left(-1-\frac{73}{4320}\pi^4\right)
+{\e^2}\left(-\frac{7}{3}+\frac{7}{36}\pi^2-\frac{49}{108}\pi^2\zeta_{3}+\frac{31}{15}\zeta_{5}\right) + \mathcal{O}({\e^3}), &
\end{flalign}

%NG NF N^{1}
\begin{flalign}
{\cal T}^{(2,\left[1\times 1\right])}_{\glu g}\Big|_{\NG \NF N} = & +\frac{1}{\e^2}\left(\frac{1}{18}\right), &
\end{flalign}

%NG^2 N^{2}
\begin{flalign}
{\cal T}^{(2,\left[1\times 1\right])}_{\glu g}\Big|_{\NGSquare N^{2}} = & +\frac{1}{\e^2}\left(\frac{1}{36}\right), &
\end{flalign}

%NXgg2

%N^{2}
\begin{flalign}
{\cal T}^{(2)}_{\glu gg}\Big|_{N^{2}} = &
+\frac{1}{\e^4}\left(-\frac{9}{2}\right)
+\frac{1}{\e^3}\left(-\frac{56}{3}\right)
+\frac{1}{\e^2}\left(-\frac{1835}{36}+\frac{71}{12}\pi^2\right)
\nonumber &\\&
+\frac{1}{\e}\left(-\frac{20977}{108}+\frac{209}{12}\pi^2+72 \zeta_{3}\right)
+\left(-\frac{19499}{27}+\frac{4195}{72}\pi^2+\frac{695}{3}\zeta_{3}-\frac{199}{144}\pi^4\right)
\nonumber &\\&
+{\e}\left(-\frac{2646667}{972}+\frac{151027}{648}\pi^2+\frac{43021}{54}\zeta_{3}-\frac{2993}{1440}\pi^4-\frac{940}{9}\pi^2\zeta_{3}+\frac{3224}{5}\zeta_{5}\right)
\nonumber &\\&
+{\e^2}\bigg(-\frac{20245145}{1944}+\frac{3545503}{3888}\pi^2+\frac{529357}{162}\zeta_{3}-\frac{3887}{405}\pi^4
\nonumber &\\&\phantom{+{\e^2}\bigg(}
-\frac{28685}{108}\pi^2\zeta_{3}+\frac{28021}{15}\zeta_{5}-\frac{2591}{7560}\pi^6+\frac{4616}{9}\zeta_{3}^2\bigg) + \mathcal{O}({\e^3}), &
\end{flalign}

%NF N^{1}
\begin{flalign}
{\cal T}^{(2)}_{\glu gg}\Big|_{\NF N} = &
+\frac{1}{\e^3}\left(\frac{4}{3}\right)
+\frac{1}{\e^2}\left(\frac{20}{9}\right)
+\frac{1}{\e}\left(\frac{275}{36}-\frac{7}{9}\pi^2\right)
+\left(\frac{287}{12}-\frac{35}{27}\pi^2-\frac{100}{9}\zeta_{3}\right)
\nonumber &\\&
+{\e}\left(\frac{7999}{108}-\frac{979}{216}\pi^2-\frac{500}{27}\zeta_{3}-\frac{71}{1080}\pi^4\right)
\nonumber &\\&
+{\e^2}\left(\frac{18595}{81}-\frac{3151}{216}\pi^2-\frac{3493}{54}\zeta_{3}-\frac{71}{648}\pi^4+\frac{175}{27}\pi^2\zeta_{3}-\frac{964}{15}\zeta_{5}\right) + \mathcal{O}({\e^3}), &
\end{flalign}

%NG N^{2}
\begin{flalign}
{\cal T}^{(2)}_{\glu gg}\Big|_{\NG N^{2}} = &
+\frac{1}{\e^3}\left(\frac{4}{3}\right)
+\frac{1}{\e^2}\left(\frac{20}{9}\right)
+\frac{1}{\e}\left(\frac{275}{36}-\frac{7}{9}\pi^2\right)
+\left(\frac{287}{12}-\frac{35}{27}\pi^2-\frac{100}{9}\zeta_{3}\right)
\nonumber &\\&
+{\e}\left(\frac{7999}{108}-\frac{979}{216}\pi^2-\frac{500}{27}\zeta_{3}-\frac{71}{1080}\pi^4\right)
\nonumber &\\&
+{\e^2}\left(\frac{18595}{81}-\frac{3151}{216}\pi^2-\frac{3493}{54}\zeta_{3}-\frac{71}{648}\pi^4+\frac{175}{27}\pi^2\zeta_{3}-\frac{964}{15}\zeta_{5}\right) + \mathcal{O}({\e^3}), &
\end{flalign}

%NXqqb2

%N^{1}
\begin{flalign}
{\cal T}^{(2)}_{\glu q\bar{q}}\Big|_{\NF N} = &
+\frac{1}{\e^3}\left(\frac{2}{3}\right)
+\frac{1}{\e^2}\left(\frac{67}{18}\right)
+\frac{1}{\e}\left(\frac{326}{27}-\frac{8}{9}\pi^2\right)
+\left(\frac{9215}{216}-\frac{275}{72}\pi^2-\frac{94}{9}\zeta_{3}\right)
\nonumber &\\&
+{\e}\left(\frac{612779}{3888}-\frac{8605}{648}\pi^2-\frac{2707}{54}\zeta_{3}+\frac{41}{180}\pi^4\right)
\nonumber &\\&
+{\e^2}\bigg(\frac{4644205}{7776}-\frac{195637}{3888}\pi^2-\frac{14818}{81}\zeta_{3}+\frac{15511}{25920}\pi^4
\nonumber &\\&\phantom{+{\e^2}\bigg(}
+\frac{391}{27}\pi^2\zeta_{3}-\frac{1294}{15}\zeta_{5}\bigg) + \mathcal{O}({\e^3}), &
\end{flalign}

%N^{-1}
\begin{flalign}
{\cal T}^{(2)}_{\glu q\bar{q}}\Big|_{\NF N^{-1}} = &
+\frac{1}{\e^3}\left(-\frac{1}{6}\right)
+\frac{1}{\e^2}\left(-\frac{35}{36}\right)
+\frac{1}{\e}\left(-\frac{509}{108}+\frac{\pi^2}{4}\right)
\nonumber &\\&
+\left(-\frac{1670}{81}+\frac{35}{24}\pi^2+\frac{31}{9}\zeta_{3}\right)
+{\e}\left(-\frac{20936}{243}+\frac{509}{72}\pi^2+\frac{1085}{54}\zeta_{3}-\frac{41}{720}\pi^4\right)
\nonumber &\\&
+{\e^2}\left(-\frac{256760}{729}+\frac{835}{27}\pi^2+\frac{15779}{162}\zeta_{3}-\frac{287}{864}\pi^4-\frac{31}{6}\pi^2\zeta_{3}+\frac{511}{15}\zeta_{5}\right) + \mathcal{O}({\e^3}), &
\end{flalign}

%NF N^{0}
\begin{flalign}
{\cal T}^{(2)}_{\glu q\bar{q}}\Big|_{\NFSquare} = &
+\frac{1}{\e^2}\left(-\frac{1}{9}\right)
+\left(\frac{91}{81}-\frac{\pi^2}{27}\right)
+{\e}\left(\frac{602}{81}-\frac{11}{18}\pi^2-\frac{4}{9}\zeta_{3}\right)
\nonumber &\\&
+{\e^2}\left(\frac{27442}{729}-\frac{95}{27}\pi^2-\frac{74}{9}\zeta_{3}+\frac{317}{6480}\pi^4\right) + \mathcal{O}({\e^3}), &
\end{flalign}

%NG N^{1}
\begin{flalign}
{\cal T}^{(2)}_{\glu q\bar{q}}\Big|_{\NG \NF N} = &
+\frac{1}{\e^2}\left(-\frac{1}{9}\right)
+\left(\frac{91}{81}-\frac{\pi^2}{27}\right)
+{\e}\left(\frac{602}{81}-\frac{11}{18}\pi^2-\frac{4}{9}\zeta_{3}\right)
\nonumber &\\&
+{\e^2}\left(\frac{27442}{729}-\frac{95}{27}\pi^2-\frac{74}{9}\zeta_{3}+\frac{317}{6480}\pi^4\right) + \mathcal{O}({\e^3}), &
\end{flalign}

%NXXX2

%N^{2}
\begin{flalign}
{\cal T}^{(2)}_{\glu \glu' \glu'}\Big|_{(\NG-1)N^{2}} = &
+\frac{1}{\e^3}\left(\frac{5}{6}\right)
+\frac{1}{\e^2}\left(\frac{169}{36}\right)
+\frac{1}{\e}\left(\frac{1825}{108}-\frac{41}{36}\pi^2\right)
\nonumber &\\&
+\left(\frac{41437}{648}-\frac{95}{18}\pi^2-\frac{125}{9}\zeta_{3}\right)
+{\e}\left(\frac{960763}{3888}-\frac{6647}{324}\pi^2-\frac{632}{9}\zeta_{3}+\frac{41}{144}\pi^4\right)
\nonumber &\\&
+{\e^2}\bigg(\frac{22508935}{23328}-\frac{319765}{3888}\pi^2-\frac{45787}{162}\zeta_{3}+\frac{24121}{25920}\pi^4
\nonumber &\\&\phantom{+{\e^2}\bigg(}
+\frac{1061}{54}\pi^2\zeta_{3}-\frac{361}{3}\zeta_{5}\bigg) + \mathcal{O}({\e^3}), &
\end{flalign}

%NF N^{1}
\begin{flalign}
{\cal T}^{(2)}_{\glu \glu' \glu'}\Big|_{(\NG-1) \NF N} = &
+\frac{1}{\e^2}\left(-\frac{1}{9}\right)
+\left(\frac{91}{81}-\frac{\pi^2}{27}\right)
+{\e}\left(\frac{602}{81}-\frac{11}{18}\pi^2-\frac{4}{9}\zeta_{3}\right)
\nonumber &\\&
+{\e^2}\left(\frac{27442}{729}-\frac{95}{27}\pi^2-\frac{74}{9}\zeta_{3}+\frac{317}{6480}\pi^4\right) + \mathcal{O}({\e^3}), &
\end{flalign}

%NG N^{2}
\begin{flalign}
{\cal T}^{(2)}_{\glu \glu' \glu'}\Big|_{(\NG-1) \NG N^{2}} = &
+\frac{1}{\e^2}\left(-\frac{1}{9}\right)
+\left(\frac{91}{81}-\frac{\pi^2}{27}\right)
+{\e}\left(\frac{602}{81}-\frac{11}{18}\pi^2-\frac{4}{9}\zeta_{3}\right)
\nonumber &\\&
+{\e^2}\left(\frac{27442}{729}-\frac{95}{27}\pi^2-\frac{74}{9}\zeta_{3}+\frac{317}{6480}\pi^4\right) + \mathcal{O}({\e^3}), &
\end{flalign}

%NXXXhard2

%N^{2}
\begin{flalign}
\Delta{\cal T}^{(2)}_{\glu \glu \glu}\Big|_{N^{2}} = &
+\frac{1}{\e^2}\left(\frac{1}{2}\right)
+\frac{1}{\e}\left(\frac{11}{4}\right)
+\left(\frac{401}{36}-\frac{3}{4}\pi^2-\frac{2}{3}\zeta_{3}\right)
\nonumber &\\&
+{\e}\left(\frac{16061}{324}-\frac{1573}{432}\pi^2-\frac{110}{9}\zeta_{3}-\frac{7}{270}\pi^4\right)
\nonumber &\\&
+{\e^2}\left(\frac{141481}{648}-\frac{4717}{288}\pi^2-\frac{6829}{108}\zeta_{3}+\frac{631}{6480}\pi^4+\frac{17}{9}\pi^2\zeta_{3}-\frac{50}{3}\zeta_{5}\right) + \mathcal{O}({\e^3}), &
\end{flalign}

%NF N^{1}
\begin{flalign}
\Delta{\cal T}^{(2)}_{\glu \glu \glu}\Big|_{\NF N} = &
+\left(\frac{7}{18}\right)
+{\e}\left(\frac{805}{324}-\frac{11}{108}\pi^2\right)
+{\e^2}\left(\frac{8227}{648}-\frac{181}{216}\pi^2-\frac{37}{27}\zeta_{3}\right) + \mathcal{O}({\e^3}), &
\end{flalign}

%NG N^{2}
\begin{flalign}
\Delta{\cal T}^{(2)}_{\glu \glu \glu}\Big|_{\NG N^{2}} = &
+\frac{1}{\e}\left(-\frac{1}{9}\right)
+\left(-\frac{5}{18}\right)
+{\e}\left(-\frac{31}{36}+\frac{7}{108}\pi^2\right)
\nonumber &\\&
+{\e^2}\left(-\frac{197}{72}+\frac{35}{216}\pi^2+\frac{25}{27}\zeta_{3}\right) + \mathcal{O}({\e^3}), &
\end{flalign}

%NXggg2

%N^{2}
\begin{flalign}
{\cal T}^{(2)}_{\glu ggg}\Big|_{N^{2}} = &
+\frac{1}{\e^4}\left(\frac{5}{2}\right)
+\frac{1}{\e^3}\left(\frac{37}{4}\right)
+\frac{1}{\e^2}\left(\frac{398}{9}-\frac{11}{3}\pi^2\right)
+\frac{1}{\e}\left(\frac{28319}{144}-\frac{55}{4}\pi^2-\frac{188}{3}\zeta_{3}\right)
\nonumber &\\&
+\left(\frac{2201527}{2592}-\frac{529}{8}\pi^2-\frac{722}{3}\zeta_{3}+\frac{511}{720}\pi^4\right)
\nonumber &\\&
+{\e}\left(\frac{6214571}{1728}-\frac{28295}{96}\pi^2-\frac{31624}{27}\zeta_{3}+\frac{10333}{4320}\pi^4+\frac{844}{9}\pi^2\zeta_{3}-\frac{1085}{2}\zeta_{5}\right)
\nonumber &\\&
+{\e^2}\bigg(\frac{2070937579}{93312}-\frac{947713}{576}\pi^2-\frac{1592867}{216}\zeta_{3}-\frac{58583}{12960}\pi^4
\nonumber &\\&\phantom{+{\e^2}\bigg(}
+\frac{4672}{9}\pi^2\zeta_{3}-\frac{555569}{120}\zeta_{5}-\frac{1459}{1890}\pi^6+\frac{86227}{72}\zeta_{3}^2\bigg) + \mathcal{O}({\e^3}), &
\end{flalign}

%NXqqbg2

%N^{1}
\begin{flalign}
{\cal T}^{(2)}_{\glu q\bar{q}g}\Big|_{\NF N} = &
+\frac{1}{\e^3}\left(-\frac{5}{6}\right)
+\frac{1}{\e^2}\left(-\frac{17}{4}\right)
+\frac{1}{\e}\left(-\frac{2239}{108}+\frac{5}{4}\pi^2\right)
\nonumber &\\&
+\left(-\frac{20521}{216}+\frac{51}{8}\pi^2+\frac{200}{9}\zeta_{3}\right)
+{\e}\left(-\frac{1624069}{3888}+\frac{2237}{72}\pi^2+\frac{340}{3}\zeta_{3}-\frac{29}{144}\pi^4\right)
\nonumber &\\&
+{\e^2}\bigg(-\frac{13887251}{7776}+\frac{61483}{432}\pi^2+\frac{89317}{162}\zeta_{3}-\frac{493}{480}\pi^4
\nonumber &\\&\phantom{+{\e^2}\bigg(}
-\frac{100}{3}\pi^2\zeta_{3}+\frac{616}{3}\zeta_{5}\bigg) + \mathcal{O}({\e^3}), &
\end{flalign}

%N^{-1}
\begin{flalign}
{\cal T}^{(2)}_{\glu q\bar{q}g}\Big|_{\NF N^{-1}} = &
+\frac{1}{\e^3}\left(\frac{1}{6}\right)
+\frac{1}{\e^2}\left(\frac{35}{36}\right)
+\frac{1}{\e}\left(\frac{1045}{216}-\frac{\pi^2}{4}\right)
+\left(\frac{28637}{1296}-\frac{35}{24}\pi^2-\frac{40}{9}\zeta_{3}\right)
\nonumber &\\&
+{\e}\left(\frac{749845}{7776}-\frac{1045}{144}\pi^2-\frac{700}{27}\zeta_{3}+\frac{29}{720}\pi^4\right)
\nonumber &\\&
+{\e^2}\bigg(\frac{19106909}{46656}-\frac{28637}{864}\pi^2-\frac{10450}{81}\zeta_{3}+\frac{203}{864}\pi^4
\nonumber &\\&\phantom{+{\e^2}\bigg(}
+\frac{20}{3}\pi^2\zeta_{3}-\frac{616}{15}\zeta_{5}\bigg) + \mathcal{O}({\e^3}), &
\end{flalign}

%NXXXg2

%N^{2}
\begin{flalign}
{\cal T}^{(2)}_{\glu \glu' \glu' g}\Big|_{(\NG-1)N^{2}} = &
+\frac{1}{\e^3}\left(-1\right)
+\frac{1}{\e^2}\left(-\frac{47}{9}\right)
+\frac{1}{\e}\left(-\frac{1841}{72}+\frac{3}{2}\pi^2\right)
\nonumber &\\&
+\left(-\frac{151763}{1296}+\frac{47}{6}\pi^2+\frac{80}{3}\zeta_{3}\right)
\nonumber &\\&
+{\e}\left(-\frac{1332661}{2592}+\frac{5519}{144}\pi^2+\frac{3760}{27}\zeta_{3}-\frac{29}{120}\pi^4\right)
\nonumber &\\&
+{\e^2}\bigg(-\frac{102430415}{46656}+\frac{151603}{864}\pi^2+\frac{36739}{54}\zeta_{3}-\frac{1363}{1080}\pi^4
\nonumber &\\&\phantom{+{\e^2}\bigg(}
-40 \pi^2\zeta_{3}+\frac{1232}{5}\zeta_{5}\bigg) + \mathcal{O}({\e^3}), &
\end{flalign}

%NXXXghard2

%N^{2}
\begin{flalign}
\Delta{\cal T}^{(2)}_{\glu \glu \glu g}\Big|_{N^{2}} = &
+\frac{1}{\e^2}\left(-\frac{1}{2}\right)
+\frac{1}{\e}\left(-\frac{57}{16}+\frac{\pi^2}{8}-\frac{\zeta_{3}}{2}\right)
+\left(-\frac{2143}{96}+\frac{9}{8}\pi^2+6 \zeta_{3}-\frac{2}{45}\pi^4\right)
\nonumber &\\&
+{\e}\left(-\frac{69323}{576}+\frac{1811}{288}\pi^2+\frac{91}{3}\zeta_{3}+\frac{91}{540}\pi^4+\frac{11}{12}\pi^2\zeta_{3}-22 \zeta_{5}\right)
\nonumber &\\&
+{\e^2}\bigg(\frac{4475885}{3456}-\frac{106079}{1728}\pi^2-\frac{3423}{8}\zeta_{3}-\frac{9647}{2592}\pi^4
\nonumber &\\&\phantom{+{\e^2}\bigg(}
+\frac{571}{18}\pi^2\zeta_{3}-\frac{12527}{24}\zeta_{5}-\frac{2167}{11340}\pi^6+\frac{2675}{24}\zeta_{3}^2\bigg) + \mathcal{O}({\e^3}). &
\end{flalign}

%%%%%%%%%%%%%%%%%%%%%%%%%%%%%%%%%%%%%%%%%%%%%%%%%%%%%%%%%%%%%%%%%%%%%%%%%%%%%%%%

\section{Colour factors up to N$^3$LO}\label{app:coltables}
\renewcommand{\arraystretch}{1.2}

\begin{table}[p]
  \centering
  \begin{tabular}{c c c}
    \cmidrule{1-3}\morecmidrules\cmidrule{1-3}  
    $\mathcal{I}$ & N$^k$LO, $\ell_1 \times \ell_2$ & Colour factors \\
    \cmidrule{1-3}\morecmidrules\cmidrule{1-3}  
    \multirow{10}{*}{$\glu g$}
    & 1 & $N$, $\NF$, $\NG N$ \\
    \cmidrule{2-3}
    & 2, $2 \times 0$ & $N^2$, $\NF N$, $\NF N^{-1}$, $\NFSquare$, $\NG N^2$, $\NG \NF N$, $\NGSquare N^2$ \\
    \cmidrule{2-3}
    & 2, $1 \times 1$ & $N^2$, $\NF N$, $\NFSquare$, $\NG N^2$, $\NG \NF N$, $\NGSquare N^2$ \\
    \cmidrule{2-3}
    & \multirow{2}{*}{3, $3 \times 0$} & $N^3$, $\NF N^2$, $\NF$, $\NF N^{-2}$, $\NFSquare N$, $\NFSquare N^{-1}$, $\NFCube$,
     \\ &&  $\NG N^3$,$\NG \NF N^2$, $\NG \NF$, $\NG \NFSquare N$,
    $\NGSquare N^3$, $\NGSquare \NF N^2$, $\NGCube N^3$ \\
    \cmidrule{2-3}  
    & \multirow{2}{*}{3, $2 \times 1$} & $N^3$, $\NF N^2$, $\NF$, $\NFSquare N$, $\NFSquare N^{-1}$, $\NFCube$
     \\ && $\NG N^3$, $\NG \NF N^2$, $\NG \NF$, $\NG \NFSquare N$,
    $\NGSquare N^3$, $\NGSquare \NF N^2$, $\NGCube N^3$ \\
    \specialrule{1pt}{0.2em}{0.2em}
    \multirow{4}{*}{$\glu g g$}
    & 1 & $N$ \\
    \cmidrule{2-3}  
    & 2 & $N^2$, $\NF N$, $\NG N^2$ \\
    \cmidrule{2-3}  
    & 3, $2 \times 0$ & $N^3$, $\NF N^2$, $\NF$, $\NFSquare N$, $\NG N^3$, $\NG \NF N^2$, $\NGSquare N^3$ \\
    \cmidrule{2-3}  
    & 3, $1 \times 1$ & $N^3$, $\NF N^2$, $\NFSquare N$, $\NG N^3$, $\NG \NF N^2$, $\NGSquare N^3$ \\
    \specialrule{1pt}{0.2em}{0.2em}  
    \multirow{7}{*}{$\glu q \bar{q}$}
    & 1 & $\NF$ \\
    \cmidrule{2-3}  
    & 2 & $\NF N$, $\NF N^{-1}$, $\NFSquare$, $\NG \NF N$ \\
    \cmidrule{2-3}  
    & \multirow{2}{*}{3, $2 \times 0$} & $\NF N^2$, $\NF$, $\NF N^{-2}$, $\NFSquare N$, $\NFSquare N^{-1}$, $\NFCube$, \\
    && $\NG \NF N^2$, $\NG \NF$, $\NG \NFSquare N$, $\NGSquare \NF N^2$ \\
    \cmidrule{2-3}  
    & \multirow{2}{*}{3, $1 \times 1$} & $\NF N^2$, $\NF$, $\NF N^{-2}$, $\NFSquare N$, $\NFSquare N^{-1}$, $\NFCube$, \\
    && $\NG \NF N^2$, $\NG \NF$, $\NG \NFSquare N$, $\NGSquare \NF N^2$ \\
    \specialrule{1pt}{0.2em}{0.2em}  
    \multirow{7}{*}{$\glu \glu' \glu'$}
    & 1 & $(\NG-1)N$ \\
    \cmidrule{2-3}  
    & 2 & $(\NG-1)N^2$, $(\NG-1)\NF N$, $(\NG-1)\NG N^2$ \\
    \cmidrule{2-3}  
    & \multirow{3}{*}{3, $2 \times 0$} & $(\NG-1)N^3$, $(\NG-1)\NF N^2$, $(\NG-1)\NF$, $(\NG-1)\NFSquare N$, \\
    && $(\NG-1)\NG N^3$, $(\NG-1)\NG \NF N^2$, $(\NG-1)\NGSquare N^3$ \\
    \cmidrule{2-3}  
    & \multirow{2}{*}{3, $1 \times 1$} & $(\NG-1)N^3$, $(\NG-1)\NF N^2$, $(\NG-1)\NFSquare N$, \\
    && $(\NG-1)\NG N^3$, $(\NG-1)\NG \NF N^2$, $(\NG-1)\NGSquare N^3$ \\
    \specialrule{1pt}{0.2em}{0.2em}      
    \multirow{5}{*}{$\glu \glu \glu$}
    & 1 & $N$ \\
    \cmidrule{2-3}  
    & 2 & $N^2$, $\NF N$, $\NG N^2$ \\
    \cmidrule{2-3}  
    & 3, $2 \times 0$ & $N^3$, $\NF N^2$, $\NF$, $\NFSquare N$, $\NG N^3$, $\NG \NF N^2$, $\NGSquare N^3$ \\
    \cmidrule{2-3}  
    & 3, $1 \times 1$ & $N^3$, $\NF N^2$, $\NFSquare N$, $\NG N^3$, $\NG \NF N^2$, $\NGSquare N^3$ \\    
    \cmidrule{1-3}\morecmidrules\cmidrule{1-3}  
  \end{tabular}
  \caption{%
    Colour factors appearing in the neutralino decay into two- and
    three-particle final states, organised by final-state particles
    $\mathcal{I}$, perturbative order $k$ and loop configuration
    $\ell_1 \times \ell_2$, in case of ambiguity.
  }
  \label{tab:CF23}
\end{table}

\begin{table}[p]
  \centering
  \begin{tabular}{c c c}
    \cmidrule{1-3}\morecmidrules\cmidrule{1-3}  
    $\mathcal{I}$ & N$^k$LO & Colour factors \\
    \cmidrule{1-3}\morecmidrules\cmidrule{1-3}  
    \multirow{2}{*}{$\glu g g g$}
    & 2 & $N^2$ \\
    \cmidrule{2-3}  
    & 3 & $N^3$, $\NF N^2$, $\NG N^3$ \\
    \specialrule{1pt}{0.2em}{0.2em}  
    \multirow{2}{*}{$\glu q \bar{q} g$}
    & 2 & $\NF N$, $\NF N^{-1}$ \\
    \cmidrule{2-3}  
    & 3 & $\NF N^2$, $\NF$, $\NF N^{-2}$, $\NFSquare N$, $\NFSquare N^{-1}$, $\NG \NF N^2$, $\NG \NF$ \\
    \specialrule{1pt}{0.2em}{0.2em}  
    \multirow{2}{*}{$\glu \glu' \glu' g$}
    & 2 & $(\NG-1)N^2$ \\
    \cmidrule{2-3}  
    & 3 & $(\NG-1)N^3$, $(\NG-1)\NF N^2$, $(\NG-1)\NG N^3$ \\
    \specialrule{1pt}{0.2em}{0.2em}  
    \multirow{2}{*}{$\glu \glu \glu g$}
    & 2 & $N^2$ \\
    \cmidrule{2-3}  
    & 3 & $N^3$, $\NF N^2$, $\NG N^3$ \\
    \specialrule{1pt}{0.2em}{0.2em}  
    \multirow{1}{*}{$\glu g g g g$}
    & 3 & $N^3$ \\
    \specialrule{1pt}{0.2em}{0.2em}  
    \multirow{1}{*}{$\glu q \bar{q} g g$}
    & 3 & $\NF N^2$, $\NF$, $\NF N^{-2}$ \\
    \specialrule{1pt}{0.2em}{0.2em}  
    \multirow{1}{*}{$\glu q \bar{q} q' \bar{q}'$}
    & 3 & $(\NF-1)\NF N$, $(\NF-1)\NF N^{-1}$ \\
    \specialrule{1pt}{0.2em}{0.2em}  
    \multirow{1}{*}{$\glu q \bar{q} q \bar{q}$}
    & 3 & $\NF$, $\NF N^{-2}$ \\
    \specialrule{1pt}{0.2em}{0.2em}  
    \multirow{1}{*}{$\glu \glu' \glu' g g$}
    & 3 & $(\NG-1)N^3$ \\
    \specialrule{1pt}{0.2em}{0.2em}  
    \multirow{1}{*}{$\glu \glu \glu g g$}
    & 3 & $N^3$ \\
    \specialrule{1pt}{0.2em}{0.2em}  
    \multirow{1}{*}{$\glu \glu' \glu' q \bar{q}$}
    & 3 & $(\NG-1)\NF N^2$, $(\NG-1)\NF$ \\
    \specialrule{1pt}{0.2em}{0.2em}  
    \multirow{1}{*}{$\glu \glu \glu q \bar{q}$}
    & 3 & $\NF N^2$ \\
    \specialrule{1pt}{0.2em}{0.2em}  
    \multirow{1}{*}{$\glu \glu' \glu' \glu'' \glu''$}
    & 3 & $(\NG-2)(\NG-1)N^3$ \\
    \specialrule{1pt}{0.2em}{0.2em}  
    \multirow{1}{*}{$\glu \glu' \glu' \glu' \glu'$}
    & 3 & $(\NG-1)N^3$ \\
    \specialrule{1pt}{0.2em}{0.2em}  
    \multirow{1}{*}{$\glu \glu \glu \glu' \glu'$}
    & 3 & $(\NG-1)N^3$ \\
    \specialrule{1pt}{0.2em}{0.2em}  
    \multirow{1}{*}{$\glu \glu \glu \glu \glu$}
    & 3 & $N^3$ \\            
    \cmidrule{1-3}\morecmidrules\cmidrule{1-3}  
  \end{tabular}
  \caption{%
    Colour factors appearing in the neutralino decay into four- and
    five-particle final states, organised by final-state particles
    $\mathcal{I}$ and perturbative order $k$.
  }
  \label{tab:CF45}
\end{table}
    
\newpage

\bibliographystyle{JHEP}
\bibliography{main}

\providecommand{\href}[2]{#2}\begingroup\raggedright\begin{thebibliography}{10}

\bibitem{Caola:2022ayt}
F.~Caola, W.~Chen, C.~Duhr, X.~Liu, B.~Mistlberger, F.~Petriello et~al.,
  \emph{{The Path forward to N$^3$LO}},  in \emph{{Snowmass 2021}}, 3, 2022
  [\href{https://arxiv.org/abs/2203.06730}{{\ttfamily 2203.06730}}].

\bibitem{Badger:2004uk}
S.D.~Badger and E.W.N.~Glover, \emph{{Two loop splitting functions in QCD}},
  \href{https://doi.org/10.1088/1126-6708/2004/07/040}{\emph{JHEP} {\bfseries
  07} (2004) 040} [\href{https://arxiv.org/abs/hep-ph/0405236}{{\ttfamily
  hep-ph/0405236}}].

\bibitem{Duhr:2014nda}
C.~Duhr, T.~Gehrmann and M.~Jaquier, \emph{{Two-loop splitting amplitudes and
  the single-real contribution to inclusive Higgs production at N$^3$LO}},
  \href{https://doi.org/10.1007/JHEP02(2015)077}{\emph{JHEP} {\bfseries 02}
  (2015) 077} [\href{https://arxiv.org/abs/1411.3587}{{\ttfamily 1411.3587}}].

\bibitem{Duhr:2013msa}
C.~Duhr and T.~Gehrmann, \emph{{The two-loop soft current in dimensional
  regularization}},
  \href{https://doi.org/10.1016/j.physletb.2013.10.063}{\emph{Phys. Lett. B}
  {\bfseries 727} (2013) 452}
  [\href{https://arxiv.org/abs/1309.4393}{{\ttfamily 1309.4393}}].

\bibitem{Li:2013lsa}
Y.~Li and H.X.~Zhu, \emph{{Single soft gluon emission at two loops}},
  \href{https://doi.org/10.1007/JHEP11(2013)080}{\emph{JHEP} {\bfseries 11}
  (2013) 080} [\href{https://arxiv.org/abs/1309.4391}{{\ttfamily 1309.4391}}].

\bibitem{Dixon:2019lnw}
L.J.~Dixon, E.~Herrmann, K.~Yan and H.X.~Zhu, \emph{{Soft gluon emission at two
  loops in full color}},
  \href{https://doi.org/10.1007/JHEP05(2020)135}{\emph{JHEP} {\bfseries 05}
  (2020) 135} [\href{https://arxiv.org/abs/1912.09370}{{\ttfamily
  1912.09370}}].

\bibitem{Catani:2003vu}
S.~Catani, D.~de~Florian and G.~Rodrigo, \emph{{The Triple collinear limit of
  one loop QCD amplitudes}},
  \href{https://doi.org/10.1016/j.physletb.2004.02.039}{\emph{Phys. Lett. B}
  {\bfseries 586} (2004) 323}
  [\href{https://arxiv.org/abs/hep-ph/0312067}{{\ttfamily hep-ph/0312067}}].

\bibitem{Czakon:2022fqi}
M.~Czakon and S.~Sapeta, \emph{{Complete collection of one-loop
  triple-collinear splitting operators for dimensionally-regulated QCD}},
  \href{https://doi.org/10.1007/JHEP07(2022)052}{\emph{JHEP} {\bfseries 07}
  (2022) 052} [\href{https://arxiv.org/abs/2204.11801}{{\ttfamily
  2204.11801}}].

\bibitem{Catani:2021kcy}
S.~Catani and L.~Cieri, \emph{{Multiple soft radiation at one-loop order and
  the emission of a soft quark\textendash{}antiquark pair}},
  \href{https://doi.org/10.1140/epjc/s10052-022-10001-z}{\emph{Eur. Phys. J. C}
  {\bfseries 82} (2022) 97} [\href{https://arxiv.org/abs/2108.13309}{{\ttfamily
  2108.13309}}].

\bibitem{Zhu:2020ftr}
Y.J.~Zhu, \emph{{Double soft current at one-loop in QCD}},
  \href{https://arxiv.org/abs/2009.08919}{{\ttfamily 2009.08919}}.

\bibitem{Czakon:2022dwk}
M.~Czakon, F.~Eschment and T.~Schellenberger, \emph{{Revisiting the double-soft
  asymptotics of one-loop amplitudes in massless QCD}},
  \href{https://doi.org/10.1007/JHEP04(2023)065}{\emph{JHEP} {\bfseries 04}
  (2023) 065} [\href{https://arxiv.org/abs/2211.06465}{{\ttfamily
  2211.06465}}].

\bibitem{DelDuca:2019ggv}
V.~Del~Duca, C.~Duhr, R.~Haindl, A.~Lazopoulos and M.~Michel, \emph{{Tree-level
  splitting amplitudes for a quark into four collinear partons}},
  \href{https://doi.org/10.1007/JHEP02(2020)189}{\emph{JHEP} {\bfseries 02}
  (2020) 189} [\href{https://arxiv.org/abs/1912.06425}{{\ttfamily
  1912.06425}}].

\bibitem{DelDuca:2020vst}
V.~Del~Duca, C.~Duhr, R.~Haindl, A.~Lazopoulos and M.~Michel, \emph{{Tree-level
  splitting amplitudes for a gluon into four collinear partons}},
  \href{https://doi.org/10.1007/JHEP10(2020)093}{\emph{JHEP} {\bfseries 10}
  (2020) 093} [\href{https://arxiv.org/abs/2007.05345}{{\ttfamily
  2007.05345}}].

\bibitem{Catani:2019nqv}
S.~Catani, D.~Colferai and A.~Torrini, \emph{{Triple (and quadruple) soft-gluon
  radiation in QCD hard scattering}},
  \href{https://doi.org/10.1007/JHEP01(2020)118}{\emph{JHEP} {\bfseries 01}
  (2020) 118} [\href{https://arxiv.org/abs/1908.01616}{{\ttfamily
  1908.01616}}].

\bibitem{DelDuca:2022noh}
V.~Del~Duca, C.~Duhr, R.~Haindl and Z.~Liu, \emph{{Tree-level soft emission of
  a quark pair in association with a gluon}},
  \href{https://doi.org/10.1007/JHEP01(2023)040}{\emph{JHEP} {\bfseries 01}
  (2023) 040} [\href{https://arxiv.org/abs/2206.01584}{{\ttfamily
  2206.01584}}].

\bibitem{Catani:2022hkb}
S.~Catani, L.~Cieri, D.~Colferai and F.~Coradeschi, \emph{{Soft
  gluon\textendash{}quark\textendash{}antiquark emission in QCD hard
  scattering}},
  \href{https://doi.org/10.1140/epjc/s10052-022-11141-y}{\emph{Eur. Phys. J. C}
  {\bfseries 83} (2023) 38} [\href{https://arxiv.org/abs/2210.09397}{{\ttfamily
  2210.09397}}].

\bibitem{Gehrmann-DeRidder:2005btv}
A.~Gehrmann-De~Ridder, T.~Gehrmann and E.W.N.~Glover, \emph{{Antenna
  subtraction at NNLO}},
  \href{https://doi.org/10.1088/1126-6708/2005/09/056}{\emph{JHEP} {\bfseries
  09} (2005) 056} [\href{https://arxiv.org/abs/hep-ph/0505111}{{\ttfamily
  hep-ph/0505111}}].

\bibitem{Currie:2013vh}
J.~Currie, E.W.N.~Glover and S.~Wells, \emph{{Infrared Structure at NNLO Using
  Antenna Subtraction}},
  \href{https://doi.org/10.1007/JHEP04(2013)066}{\emph{JHEP} {\bfseries 04}
  (2013) 066} [\href{https://arxiv.org/abs/1301.4693}{{\ttfamily 1301.4693}}].

\bibitem{Gehrmann-DeRidder:2004ttg}
A.~Gehrmann-De~Ridder, T.~Gehrmann and E.W.N.~Glover, \emph{{Infrared structure
  of e+ e- ---\ensuremath{>} 2 jets at NNLO}},
  \href{https://doi.org/10.1016/j.nuclphysb.2004.05.017}{\emph{Nucl. Phys. B}
  {\bfseries 691} (2004) 195}
  [\href{https://arxiv.org/abs/hep-ph/0403057}{{\ttfamily hep-ph/0403057}}].

\bibitem{Gehrmann-DeRidder:2005alt}
A.~Gehrmann-De~Ridder, T.~Gehrmann and E.W.N.~Glover, \emph{{Gluon-gluon
  antenna functions from Higgs boson decay}},
  \href{https://doi.org/10.1016/j.physletb.2005.03.003}{\emph{Phys. Lett. B}
  {\bfseries 612} (2005) 49}
  [\href{https://arxiv.org/abs/hep-ph/0502110}{{\ttfamily hep-ph/0502110}}].

\bibitem{Gehrmann-DeRidder:2005svg}
A.~Gehrmann-De~Ridder, T.~Gehrmann and E.W.N.~Glover, \emph{{Quark-gluon
  antenna functions from neutralino decay}},
  \href{https://doi.org/10.1016/j.physletb.2005.02.039}{\emph{Phys. Lett. B}
  {\bfseries 612} (2005) 36}
  [\href{https://arxiv.org/abs/hep-ph/0501291}{{\ttfamily hep-ph/0501291}}].

\bibitem{Jakubcik:2022zdi}
P.~Jakub\v{c}\'\i{}k, M.~Marcoli and G.~Stagnitto, \emph{{The parton-level
  structure of e$^{+}$e$^{-}$ to 2 jets at N$^{3}$LO}},
  \href{https://doi.org/10.1007/JHEP01(2023)168}{\emph{JHEP} {\bfseries 01}
  (2023) 168} [\href{https://arxiv.org/abs/2211.08446}{{\ttfamily
  2211.08446}}].

\bibitem{Chen:2023fba}
X.~Chen, P.~Jakub\v{c}\'\i{}k, M.~Marcoli and G.~Stagnitto, \emph{{The
  parton-level structure of Higgs decays to hadrons at N$^{3}$LO}},
  \href{https://doi.org/10.1007/JHEP06(2023)185}{\emph{JHEP} {\bfseries 06}
  (2023) 185} [\href{https://arxiv.org/abs/2304.11180}{{\ttfamily
  2304.11180}}].

\bibitem{Cutkosky:1960sp}
R.E.~Cutkosky, \emph{{Singularities and discontinuities of Feynman
  amplitudes}}, \href{https://doi.org/10.1063/1.1703676}{\emph{J. Math. Phys.}
  {\bfseries 1} (1960) 429}.

\bibitem{Anastasiou:2002yz}
C.~Anastasiou and K.~Melnikov, \emph{{Higgs boson production at hadron
  colliders in NNLO QCD}},
  \href{https://doi.org/10.1016/S0550-3213(02)00837-4}{\emph{Nucl. Phys. B}
  {\bfseries 646} (2002) 220}
  [\href{https://arxiv.org/abs/hep-ph/0207004}{{\ttfamily hep-ph/0207004}}].

\bibitem{Anastasiou:2002wq}
C.~Anastasiou and K.~Melnikov, \emph{{Pseudoscalar Higgs boson production at
  hadron colliders in NNLO QCD}},
  \href{https://doi.org/10.1103/PhysRevD.67.037501}{\emph{Phys. Rev. D}
  {\bfseries 67} (2003) 037501}
  [\href{https://arxiv.org/abs/hep-ph/0208115}{{\ttfamily hep-ph/0208115}}].

\bibitem{Anastasiou:2003yy}
C.~Anastasiou, L.J.~Dixon, K.~Melnikov and F.~Petriello, \emph{{Dilepton
  rapidity distribution in the Drell-Yan process at NNLO in QCD}},
  \href{https://doi.org/10.1103/PhysRevLett.91.182002}{\emph{Phys. Rev. Lett.}
  {\bfseries 91} (2003) 182002}
  [\href{https://arxiv.org/abs/hep-ph/0306192}{{\ttfamily hep-ph/0306192}}].

\bibitem{Anastasiou:2003ds}
C.~Anastasiou, L.J.~Dixon, K.~Melnikov and F.~Petriello, \emph{{High precision
  QCD at hadron colliders: Electroweak gauge boson rapidity distributions at
  NNLO}}, \href{https://doi.org/10.1103/PhysRevD.69.094008}{\emph{Phys. Rev. D}
  {\bfseries 69} (2004) 094008}
  [\href{https://arxiv.org/abs/hep-ph/0312266}{{\ttfamily hep-ph/0312266}}].

\bibitem{Becher:2009cu}
T.~Becher and M.~Neubert, \emph{{Infrared singularities of scattering
  amplitudes in perturbative QCD}},
  \href{https://doi.org/10.1103/PhysRevLett.102.162001}{\emph{Phys. Rev. Lett.}
  {\bfseries 102} (2009) 162001}
  [\href{https://arxiv.org/abs/0901.0722}{{\ttfamily 0901.0722}}].

\bibitem{Gardi:2009qi}
E.~Gardi and L.~Magnea, \emph{{Factorization constraints for soft anomalous
  dimensions in QCD scattering amplitudes}},
  \href{https://doi.org/10.1088/1126-6708/2009/03/079}{\emph{JHEP} {\bfseries
  03} (2009) 079} [\href{https://arxiv.org/abs/0901.1091}{{\ttfamily
  0901.1091}}].

\bibitem{Haber:1988px}
H.E.~Haber and D.~Wyler, \emph{{RADIATIVE NEUTRALINO DECAY}},
  \href{https://doi.org/10.1016/0550-3213(89)90143-0}{\emph{Nucl. Phys. B}
  {\bfseries 323} (1989) 267}.

\bibitem{Kuroda:1999ks}
M.~Kuroda, \emph{{Complete Lagrangian of MSSM}},
  \href{https://arxiv.org/abs/hep-ph/9902340}{{\ttfamily hep-ph/9902340}}.

\bibitem{Nogueira:1991ex}
P.~Nogueira, \emph{{Automatic Feynman graph generation}},
  \href{https://doi.org/10.1006/jcph.1993.1074}{\emph{J. Comput. Phys.}
  {\bfseries 105} (1993) 279}.

\bibitem{vonManteuffel:2012np}
A.~von Manteuffel and C.~Studerus, \emph{{Reduze 2 - Distributed Feynman
  Integral Reduction}},  \href{https://arxiv.org/abs/1201.4330}{{\ttfamily
  1201.4330}}.

\bibitem{Vermaseren:2000nd}
J.A.M.~Vermaseren, \emph{{New features of FORM}},
  \href{https://arxiv.org/abs/math-ph/0010025}{{\ttfamily math-ph/0010025}}.

\bibitem{Gehrmann-DeRidder:2003pne}
A.~Gehrmann-De~Ridder, T.~Gehrmann and G.~Heinrich, \emph{{Four particle phase
  space integrals in massless QCD}},
  \href{https://doi.org/10.1016/j.nuclphysb.2004.01.023}{\emph{Nucl. Phys. B}
  {\bfseries 682} (2004) 265}
  [\href{https://arxiv.org/abs/hep-ph/0311276}{{\ttfamily hep-ph/0311276}}].

\bibitem{Gituliar:2018bcr}
O.~Gituliar, V.~Magerya and A.~Pikelner, \emph{{Five-Particle Phase-Space
  Integrals in QCD}},
  \href{https://doi.org/10.1007/JHEP06(2018)099}{\emph{JHEP} {\bfseries 06}
  (2018) 099} [\href{https://arxiv.org/abs/1803.09084}{{\ttfamily
  1803.09084}}].

\bibitem{Gehrmann:2010ue}
T.~Gehrmann, E.W.N.~Glover, T.~Huber, N.~Ikizlerli and C.~Studerus,
  \emph{{Calculation of the quark and gluon form factors to three loops in
  QCD}}, \href{https://doi.org/10.1007/JHEP06(2010)094}{\emph{JHEP} {\bfseries
  06} (2010) 094} [\href{https://arxiv.org/abs/1004.3653}{{\ttfamily
  1004.3653}}].

\bibitem{Magerya:2019cvz}
V.~Magerya and A.~Pikelner, \emph{{Cutting massless four-loop propagators}},
  \href{https://doi.org/10.1007/JHEP12(2019)026}{\emph{JHEP} {\bfseries 12}
  (2019) 026} [\href{https://arxiv.org/abs/1910.07522}{{\ttfamily
  1910.07522}}].

\bibitem{Clavelli:1996pz}
L.~Clavelli, P.W.~Coulter and L.R.~Surguladze, \emph{{Gluino contribution to
  the three loop beta function in the minimal supersymmetric standard model}},
  \href{https://doi.org/10.1103/PhysRevD.55.4268}{\emph{Phys. Rev. D}
  {\bfseries 55} (1997) 4268}
  [\href{https://arxiv.org/abs/hep-ph/9611355}{{\ttfamily hep-ph/9611355}}].

\bibitem{Ueda:2016yjm}
T.~Ueda, B.~Ruijl and J.A.M.~Vermaseren, \emph{{Forcer: a FORM program for
  4-loop massless propagators}},
  \href{https://doi.org/10.22323/1.260.0070}{\emph{PoS} {\bfseries LL2016}
  (2016) 070} [\href{https://arxiv.org/abs/1607.07318}{{\ttfamily
  1607.07318}}].

\bibitem{Lee:2011jt}
R.N.~Lee, A.V.~Smirnov and V.A.~Smirnov, \emph{{Master Integrals for Four-Loop
  Massless Propagators up to Transcendentality Weight Twelve}},
  \href{https://doi.org/10.1016/j.nuclphysb.2011.11.005}{\emph{Nucl. Phys. B}
  {\bfseries 856} (2012) 95} [\href{https://arxiv.org/abs/1108.0732}{{\ttfamily
  1108.0732}}].

\bibitem{Baikov:2010hf}
P.A.~Baikov and K.G.~Chetyrkin, \emph{{Four Loop Massless Propagators: An
  Algebraic Evaluation of All Master Integrals}},
  \href{https://doi.org/10.1016/j.nuclphysb.2010.05.004}{\emph{Nucl. Phys. B}
  {\bfseries 837} (2010) 186}
  [\href{https://arxiv.org/abs/1004.1153}{{\ttfamily 1004.1153}}].

\bibitem{Chetyrkin:1981qh}
K.G.~Chetyrkin and F.V.~Tkachov, \emph{{Integration by Parts: The Algorithm to
  Calculate beta Functions in 4 Loops}},
  \href{https://doi.org/10.1016/0550-3213(81)90199-1}{\emph{Nucl. Phys. B}
  {\bfseries 192} (1981) 159}.

\bibitem{Catani:1998bh}
S.~Catani, \emph{{The Singular behavior of QCD amplitudes at two loop order}},
  \href{https://doi.org/10.1016/S0370-2693(98)00332-3}{\emph{Phys. Lett. B}
  {\bfseries 427} (1998) 161}
  [\href{https://arxiv.org/abs/hep-ph/9802439}{{\ttfamily hep-ph/9802439}}].

\bibitem{Becher:2009qa}
T.~Becher and M.~Neubert, \emph{{On the Structure of Infrared Singularities of
  Gauge-Theory Amplitudes}},
  \href{https://doi.org/10.1088/1126-6708/2009/06/081}{\emph{JHEP} {\bfseries
  06} (2009) 081} [\href{https://arxiv.org/abs/0903.1126}{{\ttfamily
  0903.1126}}].

\bibitem{korchemksy:1987}
G.~Korchemsky and A.~Radyushkin, \emph{Renormalization of the wilson loops
  beyond the leading order},
  \href{https://doi.org/https://doi.org/10.1016/0550-3213(87)90277-X}{\emph{Nuclear
  Physics B} {\bfseries 283} (1987) 342}.

\bibitem{Collins:1984xc}
J.C.~Collins, \emph{{Renormalization}: {An Introduction to Renormalization, The
  Renormalization Group, and the Operator Product Expansion}}, vol.~26 of
  \emph{Cambridge Monographs on Mathematical Physics}, Cambridge University
  Press, Cambridge (1986),
  \href{https://doi.org/10.1017/CBO9780511622656}{10.1017/CBO9780511622656}.

\bibitem{Bern:2002zk}
Z.~Bern, A.~De~Freitas, L.J.~Dixon and H.L.~Wong, \emph{{Supersymmetric
  regularization, two loop QCD amplitudes and coupling shifts}},
  \href{https://doi.org/10.1103/PhysRevD.66.085002}{\emph{Phys. Rev. D}
  {\bfseries 66} (2002) 085002}
  [\href{https://arxiv.org/abs/hep-ph/0202271}{{\ttfamily hep-ph/0202271}}].

\bibitem{Siegel:1979wq}
W.~Siegel, \emph{{Supersymmetric Dimensional Regularization via Dimensional
  Reduction}}, \href{https://doi.org/10.1016/0370-2693(79)90282-X}{\emph{Phys.
  Lett. B} {\bfseries 84} (1979) 193}.

\bibitem{Grozin:2015kna}
A.~Grozin, J.M.~Henn, G.P.~Korchemsky and P.~Marquard, \emph{{The three-loop
  cusp anomalous dimension in QCD and its supersymmetric extensions}},
  \href{https://doi.org/10.1007/JHEP01(2016)140}{\emph{JHEP} {\bfseries 01}
  (2016) 140} [\href{https://arxiv.org/abs/1510.07803}{{\ttfamily
  1510.07803}}].

\bibitem{matsuura:1988sm}
T.~Matsuura, S.C.~van~der Marck and W.L.~van Neerven, \emph{{The calculation of
  the second order soft and virtual contributions to the Drell-Yan
  cross-cection}},
  \href{https://doi.org/10.1016/0550-3213(89)90620-2}{\emph{Nucl. Phys. B}
  {\bfseries 319} (1989) 570}.

\bibitem{vanNeerven:1985xr}
W.~van Neerven, \emph{{Dimensional Regularization of Mass and Infrared
  Singularities in Two Loop On-shell Vertex Functions}},
  \href{https://doi.org/10.1016/0550-3213(86)90165-3}{\emph{Nucl. Phys.}
  {\bfseries B268} (1986) 453}.

\bibitem{Harlander:2000}
R.V.~Harlander, \emph{{Virtual corrections to g g $to$ H to two loops in the
  heavy top limit}},
  \href{https://doi.org/10.1016/S0370-2693(00)01042-X}{\emph{Phys. Lett. B}
  {\bfseries 492} (2000) 74}
  [\href{https://arxiv.org/abs/hep-ph/0007289}{{\ttfamily hep-ph/0007289}}].

\bibitem{Bern:2003ck}
Z.~Bern, A.~De~Freitas and L.J.~Dixon, \emph{{Two loop helicity amplitudes for
  quark gluon scattering in QCD and gluino gluon scattering in supersymmetric
  Yang-Mills theory}},
  \href{https://doi.org/10.1088/1126-6708/2003/06/028}{\emph{JHEP} {\bfseries
  06} (2003) 028} [\href{https://arxiv.org/abs/hep-ph/0304168}{{\ttfamily
  hep-ph/0304168}}].

\bibitem{Kosower:1997zr}
D.A.~Kosower, \emph{{Antenna factorization of gauge theory amplitudes}},
  \href{https://doi.org/10.1103/PhysRevD.57.5410}{\emph{Phys. Rev. D}
  {\bfseries 57} (1998) 5410}
  [\href{https://arxiv.org/abs/hep-ph/9710213}{{\ttfamily hep-ph/9710213}}].

\bibitem{Campbell:1998nn}
J.M.~Campbell, M.A.~Cullen and E.W.N.~Glover, \emph{{Four jet event shapes in
  electron - positron annihilation}},
  \href{https://doi.org/10.1007/s100529900034}{\emph{Eur. Phys. J. C}
  {\bfseries 9} (1999) 245}
  [\href{https://arxiv.org/abs/hep-ph/9809429}{{\ttfamily hep-ph/9809429}}].

\bibitem{Marcoli:2023xtc}
M.~Marcoli, \emph{{N$^3$LO Antenna Functions for Final-State Radiation}},  in
  \emph{Proceedings for RADCOR 2023}, 2023
  [\href{https://arxiv.org/abs/2310.07934}{{\ttfamily 2310.07934}}].

\bibitem{Gehrmann-DeRidder:2007foh}
A.~Gehrmann-De~Ridder, T.~Gehrmann, E.W.N.~Glover and G.~Heinrich,
  \emph{{Infrared structure of e+ e- ---\ensuremath{>} 3 jets at NNLO}},
  \href{https://doi.org/10.1088/1126-6708/2007/11/058}{\emph{JHEP} {\bfseries
  11} (2007) 058} [\href{https://arxiv.org/abs/0710.0346}{{\ttfamily
  0710.0346}}].

\bibitem{MMTHESIS}
M.~Marcoli, \emph{{The Colourful Antenna Subtraction Method for
  Next-to-Next-to-Leading-Order Calculations in QCD: Formulation and
  Automation}}, Ph.D. thesis, Zurich U., 2023.

\bibitem{Braun-White:2023sgd}
O.~Braun-White, N.~Glover and C.T.~Preuss, \emph{{A general algorithm to build
  real-radiation antenna functions for higher-order calculations}},
  \href{https://doi.org/10.1007/JHEP06(2023)065}{\emph{JHEP} {\bfseries 06}
  (2023) 065} [\href{https://arxiv.org/abs/2302.12787}{{\ttfamily
  2302.12787}}].

\bibitem{Braun-White:2023zwd}
O.~Braun-White, N.~Glover and C.T.~Preuss, \emph{{A general algorithm to build
  mixed real and virtual antenna functions for higher-order calculations}},
  \href{https://arxiv.org/abs/2307.14999}{{\ttfamily 2307.14999}}.

\bibitem{Fox:2023bma}
E.~Fox and N.~Glover, \emph{{Initial-Final and Initial-Initial antenna
  functions for real radiation at next-to-leading order}},
  \href{https://arxiv.org/abs/2308.10829}{{\ttfamily 2308.10829}}.

\bibitem{Garland:2001tf}
L.W.~Garland, T.~Gehrmann, E.W.N.~Glover, A.~Koukoutsakis and E.~Remiddi,
  \emph{{The Two loop QCD matrix element for e+ e- ---\ensuremath{>} 3 jets}},
  \href{https://doi.org/10.1016/S0550-3213(02)00057-3}{\emph{Nucl. Phys. B}
  {\bfseries 627} (2002) 107}
  [\href{https://arxiv.org/abs/hep-ph/0112081}{{\ttfamily hep-ph/0112081}}].

\bibitem{Gehrmann:2023jyv}
T.~Gehrmann, P.~Jakub\v{c}\'\i{}k, C.C.~Mella, N.~Syrrakos and L.~Tancredi,
  \emph{{Planar three-loop QCD helicity amplitudes for $V$+jet production at
  hadron colliders}},  \href{https://arxiv.org/abs/2307.15405}{{\ttfamily
  2307.15405}}.

\bibitem{Abreu:2021asb}
S.~Abreu, F.~Febres~Cordero, H.~Ita, M.~Klinkert, B.~Page and V.~Sotnikov,
  \emph{{Leading-color two-loop amplitudes for four partons and a W boson in
  QCD}}, \href{https://doi.org/10.1007/JHEP04(2022)042}{\emph{JHEP} {\bfseries
  04} (2022) 042} [\href{https://arxiv.org/abs/2110.07541}{{\ttfamily
  2110.07541}}].

\bibitem{Weinzierl:2008iv}
S.~Weinzierl, \emph{{NNLO corrections to 3-jet observables in electron-positron
  annihilation}},
  \href{https://doi.org/10.1103/PhysRevLett.101.162001}{\emph{Phys. Rev. Lett.}
  {\bfseries 101} (2008) 162001}
  [\href{https://arxiv.org/abs/0807.3241}{{\ttfamily 0807.3241}}].

\bibitem{DelDuca:2016ily}
V.~Del~Duca, C.~Duhr, A.~Kardos, G.~Somogyi, Z.~Sz\H{o}r, Z.~Tr\'ocs\'anyi
  et~al., \emph{{Jet production in the CoLoRFulNNLO method: event shapes in
  electron-positron collisions}},
  \href{https://doi.org/10.1103/PhysRevD.94.074019}{\emph{Phys. Rev. D}
  {\bfseries 94} (2016) 074019}
  [\href{https://arxiv.org/abs/1606.03453}{{\ttfamily 1606.03453}}].

\end{thebibliography}\endgroup

\end{document}